\documentclass[paper,twoside,letterpaper]{JHEP}
\usepackage{epsfig,amsfonts,amssymb,amsbsy,amsbsy}
\def\ccommap{\raise 2pt\hbox{,}}

\def\MQ{$\cal M_{\rm q}$}
\def\sign{\mathop{\rm sign}\nolimits}
\def\tr{\mathop{\rm tr}\nolimits}
\def\im{\mathop{\rm Im}\nolimits}
\def\li{\mathop{\rm li}\nolimits}

\def\HQ{{\cal H}_{\rm q}}
\def\Sph#1{{\rm S}^{#1}}
\title{A model for gauge theories with Higgs fields} 
\author{Frank Ferrari\thanks{On leave from Centre National de
la Recherche Scientifique, \'Ecole Normale Sup\'erieure, Paris.}\\    
Joseph Henry Laboratories,\\ Princeton University, Princeton, New
Jersey 08544, USA\\
\email{fferrari@feynman.princeton.edu}}
\preprint{PUPT-1962\\LPTENS-00/28\\ \hepth{0102041}}
\abstract{We discuss in details a simple, purely bosonic, 
quantum field theory belonging to larger class of models with the
following properties:\hfill\\

\noindent a) They are asymptotically free, with a dynamically
generated mass scale.\hfill\\
\noindent b) They have a space of parameters which gets quantum corrections 
drastically modifying the classical
singularity structure. The quantum theory can 
have massless solitons, Argyres-Douglas-like CFTs, exhibit confinement, 
etc\dots\hfill\\
\noindent c) The physics can, to a large extent, be worked out in
models with a large number of supersymmetries as well as in
purely bosonic ones. In the former case, exact BPS mass formulas can be 
derived, brane constructions and embedding in M theory do exist.\hfill\\
\noindent d) The models have an interesting $1/N$ expansion, and it 
is possible to define a double scaling limit in the sense of 
the ``old'' matrix models when approaching the singularities in parameter 
space.\hfill\\

These properties make these theories very good toy models for four 
dimensional gauge theories with Higgs fields, and provide a framework 
where the effects of breaking supersymmetry can be explicitly studied. In 
our model, we work out in details the quantum space of parameters.
We obtain the non-local lagrangian description of the 
Argyres-Douglas-like CFT, and show that 
it admits a strongly coupled fixed point. We also explicitly demonstrate 
property d). The possibility of defining such double scaling limits
was not anticipated on the gauge theory side, and could be 
of interest to understand the gauge theory/string theory 
correspondence.}
\keywords{Nonperturbative Effects, 1/N Expansion, Sigma Models}
\dedicated{To Cl\'elia.}
\begin{document}
%
%

%
%
\section{General presentation}
\subsection{Motivations}
The present paper, which extends and provides full details on
a previous work \cite{cousins}, grew up from the
contradictory feelings one might have with regard to
the dramatic progresses the 
theory of strongly coupled supersymmetric
gauge fields and strings have undergone in the 
past six years (for reviews, see \cite{reviewsSW}, \cite{reviewsbranes},
\cite{reviewsAdS}). Though it is likely that the general intuition gained 
in studying supersymmetric examples will be useful to eventually understand 
more realistic theories, the specific methods and the analytic solutions 
of the supersymmetric theories will probably be irrelevant for solving 
genuinely non-supersymmetric models. This is due to the fact that those 
theories cannot be viewed as perturbation of supersymmetric theories (at 
least if the supersymmetric theories have at least eight supercharges), as 
early \cite{SoftlyB} as well as more recent \cite{PolStra} works tend to
demonstrate. For example, a gravity approximation to the 
hypothetical string theory description of gauge theories, which can yield 
useful insights in supersymmetric or nearly supersymmetric models,
cannot apply to QCD since the 
hadron spectrum is string-like. These limitations have led me to try to 
find a simple framework where the effects of breaking supersymmetry could 
be analyzed. The present paper is devoted to a detailed study
of the simplest non-trivial,
non-supersymmetric, model belonging to a large class of theories
which are {\it simple enough to be tractable even in their
strongly coupled, non-supersymmetric, 
regime, but complex enough so that many interesting questions about gauge 
theories have a 
counterpart in the simple models}. More precisely, and as will be explained
in the following, our simple models have all of the following general
properties:\\
\noindent a) The models are tractable from the non-supersymmetric versions
to the supersymmetric ones. In the latter case exact results can be
obtained (BPS mass formulas in particular) in strict parallel with what is
known for four dimensional supersymmetric gauge theories. Formulas can
actually quantitatively coincide in these cases. Asymptotically free as
well as conformal field theories can be studied.\\
\noindent b) The models have an analogue of a moduli space, with
generically both weakly coupled and strongly coupled regions. Strong
quantum corrections then drastically modify the classical structure. At
weak
coupling we can have solitons playing the r\^ole of magnetic monopoles or
dyons, and these can become massless at strong coupling singularities.
Argyres-Douglas -like CFTs \cite{AD} can appear at strong coupling. All
these phenomena can occur and be studied
in both supersymmetric and non-supersymmetric theories.\\
\noindent c) The supersymmetric versions of the models admit
constructions in terms of branes, and they can be solved via M theory.\\
\noindent d) The models have a non-trivial $1/N$ expansion \`a
la 't Hooft \cite{thooftN}. The large $N$ limit can be unconventional, as
for $N=2$ super Yang-Mills \cite{SD}.\\
\noindent We will illustrate a), b) and d) in this article; c) is already
known, as we will review below. Two additional properties would also be
desirable,\\
\noindent e) The supersymmetric versions of the models can be geometrically
engineered as in \cite{KLM}.\\
\noindent f) The models are dual \`a la Maldacena
to some kind of string theory.\\
\noindent Though I am not aware of any explicit construction, it is very
likely that e) is true, as explained later. As for f),
it is plausible that it could be true in view of c) and d), but we will have
unfortunately nothing to say about this fascinating possibility in this
paper.

A fundamental question of principle, that we would also like to address, is 
whether it is possible to prove, or at least to get a good general 
understanding, that gauge theories (supersymmetric or 
not) or other four dimensional field theories can have a description in 
terms of string theories. The modern starting point is a conjecture 
\cite{Malda} motivated by the relationship between supersymmetric D-branes and 
solitons in closed string theories. It is not clear whether this intuitive 
understanding of the gauge theory/string theory correspondence makes
sense in the general non-supersymmetric case. Interestingly, the results 
of the present paper suggest a way to understand the possible 
proliferation of dualities between four dimensional field theories and 
string theories. The idea is to show that double scaling limits
\cite{BK} can be 
defined in the vicinity of the singularities of the moduli (or parameter) 
spaces of the gauge theories when a non-trivial interacting physics 
develops at low energies. The double scaled theory is then a string theory
\cite{kaza} that can be shown to describe the interacting low energy 
degrees of freedom. We will explicitly demonstrate that such double 
scaling limits can be defined for the
model we consider in this paper. These double scaling limits have 
interesting properties, and we will try to provide a more detailed 
study in a forthcoming publication \cite{ds}.
\subsection{A family of toy models for gauge theories with Higgs fields}
The fact that good toy models exist for four dimensional gauge theories is 
of course not new. A quarter of a century
ago, Polyakov showed in \cite{polnlsm} that two dimensional 
non-linear $\sigma$ models can be asymptotically free, undergo dimensional 
transmutation, and develop infrared slavery, the landmarks of interesting 
four dimensional gauge theories. Based on this idea, many interesting 
results were then obtained (see e.g.~\cite{nlsm}). 
The models we will propose are only 
modest extensions of the original non-linear $\sigma$ models, the main new 
input being a way to mimic the presence of Higgs fields, which are 
conspicuous in modern studies. It seems that either it was not known that 
the standard non-linear $\sigma$ 
models could be modified in a way that would 
make them suitable to compare with modern gauge theory studies, or it was 
not known that with these modifications the models would still be tractable 
enough in the non-supersymmetric cases. It is of course disappointing that 
we have to restrict ourselves to two-dimensional models, but, even at the 
turn of the millenium, this is still the price to pay to discuss 
the effect of breaking supersymmetry in strongly coupled theories.
I hope that the results of \cite{cousins} and
of this paper will convince the reader that the models we are proposing 
constitute a nice, if modest, playground to study this fundamental question.

We now turn to describe the basic idea which led to the construction of 
our models.
One important peculiarity of supersymmetric gauge theories \`a la 
Seiberg-Witten \cite{SW} is that they have massless scalar fields and a 
continuous moduli space of vacua. This is heavily used in the discussion 
of the theories (as well as in many other supersymmetric systems), and may 
appear as being an obvious and impassable obstacle in trying to find 
non-supersymmetric analogues. In non-supersymmetric theories, any vacuum 
degeneracy that may be present classically is generically lifted quantum 
mechanically. In two dimensions, the situation is even worse: 
even supersymmetric theories cannot have a continuous moduli space, because 
the strong infrared fluctuations always make the vacuum wave functional to 
spread over the whole would-be moduli space (a discrete set may remain 
\cite{discms}). However, it turns out that this is
only an outward problem, for the 
following reason. In four dimensions, the moduli space is parametrized by 
Higgs vacuum expectation values (or more precisely by the vevs of
independent gauge invariant combinations of Higgs fields). The physics is 
interesting because the masses of the gauge bosons, which govern the low 
energy coupling, depend on the Higgs vevs via the usual Higgs mechanism.
Moving on the moduli space is then equivalent to varying the low energy 
coupling, and a very interesting physics is associated to the
transition from weak coupling to strong coupling: appearance of strong 
coupling singularities \cite{SW}, non-trivial CFTs \cite{AD}, rearrangement 
of the spectrum of stable states \cite{FB}, etc\dots All this physics is 
largely independent of the fact that the Higgs potential has flat 
directions, but strongly depends on our ability to vary the low energy 
coupling. We are thus led to the conclusion that a good non-supersymmetric 
analogue of the moduli space of supersymmetric theories could be a space of 
parameters on which the low energy coupling depends. This is too vague, 
since there are a priori many ways to vary the low energy coupling, for 
example by putting the theory on a sphere of varying radius or considering a 
finite temperature. To stick as close as possible to the 
supersymmetric gauge theory case,
we will in general consider parameters that correspond to giving 
masses to the fields that contributes with a minus sign
to the $\beta$ function. In the case of the non-linear $\sigma$ model with 
target space the $N-1$ sphere $\Sph{N-1}$ that we will consider in the 
present paper, this simply amounts to giving a mass to the $N-1$
would-be Goldstone bosons. These mass parameters play the r\^ole 
of Higgs vevs, and span a space of parameters $\cal M$. One of our main 
goal is to compute the quantum corrections to $\cal M$, in the same sense 
as quantum corrections to the moduli space of supersymmetric gauge 
theories were computed in \cite{SW} or \cite{Seiberg}.

Though the very simple idea presented above certainly suggests that the 
space of mass parameters in non-linear $\sigma$ models is an interesting 
object to consider, the reader may not be convinced that it is really a 
good analogue of the moduli spaces of supersymmetric gauge theories. The 
present author himself actually became convinced that it is the case only 
after he became aware of papers by Hanany and Hori \cite{braneN2} and Dorey 
and collaborators \cite{susymodels}, where the ${\cal N}=2$
supersymmetric ${\mathbb C}P^{N-1}$ non-linear $\sigma$ model with mass terms
is discussed (in this context, the mass terms are called ``twisted masses,''
and are in one to one correspondence with the holomorphic isometries of 
the target K\" ahler manifold \cite{massN2}).
In \cite{braneN2}, a brane construction of this supersymmetric model is given, 
and M theory is used to obtain non-perturbative results, in strict parallel 
to what was done in the case of ${\cal N}=2$ super Yang-Mills 
\cite{Witbrane}. The striking, in some sense quantitative,
similarity with ${\cal N}=2$ super Yang-Mills 
was then further discussed in \cite{susymodels}.
The property making brane constructions possible is that the mass 
parameters can be interpreted in the ${\mathbb C}P^{N-1}$ model as vector 
multiplets vevs in a gauged linear formulation of the model,
and thus literally
parametrize a Coulomb branch. The works \cite{braneN2,susymodels} 
thus clearly demonstrate that an excellent toy model for ${\cal N}=2$ super 
Yang-Mills in four dimensions is the ${\cal N}=2$ ${\mathbb C}P^{N-1}$ 
non-linear $\sigma$ model with mass terms
in two dimensions (in spite of the fact that ${\cal N}=2$ in four 
dimensions corresponds to eight supercharges, whereas ${\cal N}=2$ in two 
dimensions corresponds to four supercharges).  
In retrospect, this analogy is not too surprising, and there are actually 
many independent arguments in favor of this correspondence.
For example, both types of theories are known to have
similar non-renormalization theorems (\cite{Snren}, \cite{AGF}), and
both admit topological twists (\cite{topoa}, \cite{topob}). 
I think that this
latter property is particularly significant, because the exact results \`a
la Seiberg-Witten (\cite{SW}, \cite{reviewsSW}) are likely to have a
semi-topological origin, very much like the exact results one can obtain in
two dimensions \cite{CV}. Another important fact is that
effective superpotentials in two dimensions are
very similar to effective prepotentials in four dimensions. This was
emphasized for example in \cite{lerche} where two dimensional
superpotentials were geometrically engineered using singular
Calabi-Yau fourfolds and mirror symmetry,
in the same spirit as four dimensional prepotentials can be engineered
using singular Calabi-Yau threefolds and mirror symmetry \cite{KLM}.
Though the models considered in \cite{lerche} were different from the one
we are proposing in this paper, a similar construction is probably
possible. 

We believe that this analogy between supersymmetric 
theories in four and two dimensions can be fruitfully 
extended to the non-supersymmetric cases, and used to study the effects of 
breaking supersymmetry. 
We present below a short dictionary for the correspondence between four
dimensional and two dimensional models. Some of the entries will be
exemplified in later sections.

\TABULAR{|c|c|}{\hline
$4\, {\rm D}$ gauge theories & $2\, {\rm D}$ non-linear $\sigma$ models \\
\hline Gauge bosons & Goldstone bosons\\
Gauge group & Isometry group\\
Number of colors  & Dimension of the compact target space\\
Gauge coupling constant & Inverse radius of the target manifold\\
Higgs fields & Mass terms\\
Moduli space $\cal M$& Space of mass parameters $\cal M$\\
Monopoles and dyons & Kinks\\
S duality & Kramers-Wannier duality\\
Argyres-Douglas CFT & Ginzburg-Landau CFT\\
Eight supercharges & Four supercharges\\
Effective prepotential $\cal F$ & Effective superpotential $W$\\
String description & Branched polymer description\\
\hline }{A gauge theory/$\sigma$ model dictionary}
\vfill\eject

\subsection{Plan of the paper}
We give a general discussion of mass terms, including their 
renormalization properties, in Section 2.1. We work out a few simple 
examples when the target space is the sphere $\Sph{N-1}$, including 
a model that we will call after C. Neumann and
on which we will focus. We also briefly discuss the 
supersymmetric generalizations.
In Section 2.2, the semiclassical properties of the Neumann model are 
explored, including the singularity structure of the
classical space of parameters ${\cal M}_{\rm cl}$, bound states and
solitons.
Section 3 is devoted to the $1/N$ expansion. Emphasis is put on the 
peculiarities introduced by the fact that the dimension of $\cal M$ is of 
order $N$, and on general limitations of the $1/N$ expansion in our model.
We compute the quantum corrections to the metric as well as the mass of 
some stable particles.
We then show how to go beyond the standard
$1/N$ expansion, and we obtain a provisional picture of 
the singularity structure of the quantum space of parameters ${\cal 
M}_{\rm q}$ (also simply denoted by $\cal M$ in the following).
In Section 4 we work out the full structure of $\cal M$, in the large $N$ 
limit.
In Section 5 we analyze the theory in the vicinity of singularities on 
$\cal M$, for any finite $N$. We show that the non-trivial infrared 
physics is most naturally described in terms of a non-local lagrangian 
analogous to a theory of electric and magnetic charges. This brings the 
analogy with an Argyres-Douglas theory to a climax. Due to the simplicity 
of two dimensional theories, we are able to show that the non-local theory 
has a strongly coupled fixed point, and we recover the results of Sections 
3 and 4 independently of a large $N$ approximation.
In Section 6, we briefly show that double scaling limits,
in the sense of \cite{BK}, can be defined
as the singularities on $\cal M$ are approached.
This has never been studied in the context of two dimensional
non-linear $\sigma$ models, and our results suggest very interesting
possibilities for four dimensional gauge theories that have not been
anticipated. Moreover, these double scaling limits have interesting 
properties, which we will discuss elsewhere \cite{ds}.

We have also included three appendices. In Appendix A, we find the most 
general static finite energy solutions to the field equations
of our model at arbitrary $N$. A notable result 
is that we obtain {\it static} solutions describing several standard
sine-Gordon solitons at arbitrary distances from each other. In Appendix B
we derive several simple formulas used in the main text to study the large 
$N$ limit of the theory. Finally in
Appendix C we compute the $1/N$ corrections to 
the equation of the critical hypersurface obtained in Section 3. The 
calculation illustrates some generic properties of the $1/N$ 
expansion, as well as some subtleties associated with infrared divergences 
that can plague this expansion in our model. The result gives a consistency 
check of the simple calculations of Section 3.

To the opposite of \cite{cousins}, we provide in this work detailed 
derivations and elementary discussions of various points. 
This explains in part the length of the paper.

\section{The Neumann model}
\subsection{Mass terms in the non-linear sigma model}
In the following, the target space of our non-linear $\sigma$ 
models is the $N-1$ sphere $\Sph{N-1}$. The discussion could be 
easily generalized to sigma models on symmetric spaces, for example.
At the technical level, the renormalization properties 
of the two dimensional non-linear sigma model are non-trivial 
\cite{BreGuiZin}, and their renormalizability can be proven with the help
of a sort of Zinn-Justin equation, very much like the
renormalizability of gauge theories is analyzed (see e.g.~\cite{ZJbook}).
In the latter case, it is not straightforward to add mass terms for the
gauge bosons that do not spoil the renormalizability of the model.
To do so, it is
necessary to introduce scalar fields, the Higgs bosons, and the gauge
bosons masses are then largely determined by the transformation properties of
the Higgs fields under the gauge group $G$ through the Higgs mechanism. 
The situation is similar for mass terms
in non-linear sigma models: they generally
spoil the renormalizability of the theory, and from this point of view 
they are most naturally characterized
by their transformation properties under the isometry group $G={\rm O}(N)$ 
of the target space. The explicit form of the mass terms in
the lagrangian is then determined unambiguously by $G$ invariance and
power counting.
\subsubsection{Renormalization theory}
Working in the euclidean, the lagrangian without mass terms is
\begin{equation}
\label{Lnlk}
L_{\rm kin} = {1\over 2} \, \sum_{i,j=1}^{N-1}g_{ij}\,
\partial_{\alpha}\Phi_i \partial_{\alpha}\Phi_j,
\end{equation}
where $g_{ij}$ is the standard ${\rm O}(N)$
invariant metric on a sphere of radius $1/g^2$, $g$ being the dimensionless 
coupling constant analogous to the gauge coupling constant.
By taking the $\Phi_i$s to be the usual cartesian coordinates, we have 
\begin{equation}
\label{metric}
g_{ij}=\delta_{ij} + {g^2 \Phi_i \Phi_j \over 1 - g^2 \sum_{i=1}^{N-1}
\Phi_i^2}\cdotp
\end{equation}
The model is defined by a path integral over the fields $\Phi_i$, $1\leq 
i\leq N-1$, with an
${\rm O}(N)$ invariant measure. We have interactions of the form 
$g^{2n}\, (\partial\Phi)^2\, \Phi^{2n}$ for all $n\geq 1$.
A mass term is defined to be
a couple $({\boldsymbol\Gamma},d)$ where $\boldsymbol\Gamma$ is an 
irreducible
representation of ${\rm SO}(N)$ and $d$ is the canonical dimension of the 
mass parameters. In two dimensions, it is consistent to have $d=1$ or $d=2$.
In the latter case, the parameters actually correspond to mass squared.
Different choices of $\boldsymbol\Gamma$
and $d$ do not necessarily correspond to independent mass terms,
as we will see. If $d=1$ the mass
parameters are denoted by ${\mathbf m}\equiv m_{i_1\cdots i_p}$,
and if $d=2$ by ${\mathbf h}\equiv h_{i_1\cdots i_p}$.
The $i_j$s are ${\rm SO}(N)$ indices, and
the tensors $\mathbf m$ or $\mathbf h$ 
satisfies some constraints depending on $\boldsymbol\Gamma$.

Instead of working with the lagrangian (\ref{Lnlk}), which is possible
but awkward, we will introduce a Lagrange multiplier field $\alpha$
and work in a representation where the ${\rm O}(N)$ symmetry is linearly
realized:  
\begin{equation}
\label{Llk}
L_{\rm kin} =  {1\over 2}\, \partial_{\alpha}
{\boldsymbol\Phi}\partial_{\alpha}{\boldsymbol\Phi}
+ {1\over 2}\, \alpha\, \left({\boldsymbol\Phi}^2 - {1\over g^2}\right),
\end{equation}
with ${\boldsymbol\Phi}=(\Phi_1 ,\ldots ,\Phi_N)$. 
Eliminating $\Phi_N$ from (\ref{Llk})
by using the constraint enforced by $\alpha$, one recovers the non-linear
lagrangian (\ref{Lnlk}). We will regulate the theory by a 
simple momentum cutoff $\Lambda_{0}$, which is manifestly ${\rm O}(N)$ 
invariant. Introducing $\mathbf m$ or $\mathbf h$
into the game, the action must then
be taken to be the most general relativistic local functional 
invariant under ${\rm O}(N)$ (by varying both the fields and the
parameters) and compatible with power counting.
The canonical dimensions of the fields are
\begin{equation}
\label{candim}
[\Phi] =0,\quad [\alpha] =2.
\end{equation}
As we will further discussed in Section 3.1,
these dimensions are relevant to study the UV finiteness of both ordinary
perturbation theory and of the $1/N$ expansion 
because the model is asymptotically free.
Due to its canonical dimension, $\alpha$ 
can only appear in a term of the form
$F({\boldsymbol\Phi}^2)\,\alpha$, where $F$ is an arbitrary function.
By integrating over $\alpha$, we get the quantum constraint
\begin{equation}
\label{qconst}
F({\boldsymbol\Phi}^2) =0,
\end{equation}
which can always be solved in the UV as
\begin{equation}
\label{gren}
{\boldsymbol\Phi}^2 = {1\over ZZ_{g}  g^2}\raise 2pt\hbox{,}
\end{equation}
where $ZZ_{g}$ is a renormalization constant. This shows that, as long 
as we integrate over $\alpha$ without introducing a source for this 
field, $F$ can be taken without loss of generality to be
proportional to ${\boldsymbol\Phi}^2- 1/(ZZ_{g}  g^2)$, and 
${\boldsymbol\Phi}^2$ can be replaced by its constant value in the other 
terms of the lagrangian. The subtleties associated with the renormalization 
of $\alpha$-dependent quantities will only show up in Appendix C, and is 
discussed there. Without any mass term, the only non-zero dimension two 
operator that remains is proportional to $\partial_{\alpha} {\boldsymbol\Phi}
\partial_{\alpha} {\boldsymbol\Phi}$, and from this we deduce the form of 
the renormalized lagrangian \cite{BreGuiZin}
\begin{equation}
\label{renlk}
L_{\rm kin} = {Z\over 2}\, \partial_{\alpha}
{\boldsymbol\Phi}\partial_{\alpha}{\boldsymbol\Phi} 
+ {Z\over 2}\, \alpha\, \left({\boldsymbol\Phi}^2 - {1\over Z 
Z_{g}g^2}\right).
\end{equation}
$\boldsymbol\Phi$ and $g$ are now the renormalized fields and coupling 
constant, related to the bare quantities by
\begin{equation}
\label{barevsrena}
{\boldsymbol\Phi} = {{\boldsymbol\Phi_{0}}\over\sqrt{Z}}
\raise 2pt\hbox{,}\quad g = {g_{0}\over\sqrt{Z_{g}}}\cdotp
\end{equation}
In the following, we investigate the simplest mass terms and find 
the scalar potential $V$ they correspond to.
\paragraph{Singlet}
Because of the constraint (\ref{gren}), singlet mass terms are trivial and 
simply correspond to adding a constant to the lagrangian.
\paragraph{Vector $h_{i}$ of dimension two}
This term corresponds to a standard magnetic field,
\begin{equation}
\label{massv2}
V_{(v,2)}= - Z_{(v,2)}\, {\mathbf h}{\boldsymbol\Phi} = -Z_{(v,2)}\, h\Phi_{N},
\end{equation}
where we have used the ${\rm O}(N)$ symmetry to align the magnetic field 
with the $N$th direction. This is a linear symmetry breaking term, and thus
$Z_{(v,2)}=1$. By eliminating $\Phi_{N}$ we have
\begin{equation}
\label{potv2}
V_{(v,2)}= - Z_{(v,2)}\, {h\over g}\, 
\sqrt{1-g^{2}\sum_{i=1}^{N-1}\Phi_{i}^{2}}.
\end{equation}
In addition to giving a mass $\sqrt {Z_{(v,2)}gh}$ to the 
would-be Goldstone bosons, this term also produces an infinite sum of new 
interactions. This model can easily be studied in the large $N$ limit, 
but it is not particularly interesting.
\paragraph{Vector $m_{i}$ of dimension one}
The only new invariant operators of dimension 2 are 
$({\mathbf m}{\boldsymbol\Phi})^{2}$ and ${\mathbf m}^{2}$.
One has also the dimension one operator ${\mathbf m}{\boldsymbol\Phi}$. The 
potential is then
\begin{eqnarray}
\label{massv1}
V_{(v,1)}&=& - {Z_{(v,1)}\over 2}\, ({\mathbf m}{\boldsymbol\Phi})^{2} - 
Z'_{(v,1)}\, M {\mathbf m}{\boldsymbol\Phi}- Z''_{(v,1)}{\mathbf 
m}^{2}\nonumber\\
&=&-{Z_{(v,1)}\over 2}\, m^{2}\Phi_{N}^{2} - Z'_{(v,1)}\, M m \Phi_{N} -
Z''_{(v,1)} m^{2}.\\ \nonumber
\end{eqnarray}
$M$ is a new mass parameter that generically enters the problem. However, 
it can be consistently set to zero thanks to the symmetry 
$\Phi_{N}\mapsto -\Phi_{N}$. $Z''_{(v,1)}$ renormalizes the vacuum energy.
\paragraph{Antisymmetric tensor of dimension two}
No new operator can be constructed,
\begin{equation}
\label{massa2}
V_{(a,2)}=0.
\end{equation}
\paragraph{Traceless symmetric tensor $h_{ij}$ of dimension two}
This corresponds to
\begin{equation}
\label{masss2}
V_{(s,2)}= -{Z_{(s,2)}\over 2}\, {\boldsymbol\Phi}{\mathbf 
h}{\boldsymbol\Phi} = -{Z_{(s,2)}\over 2}\, \sum_{i=1}^{N}h_{i}\Phi_{i}^{2}.
\end{equation}
By diagonalizing the matrix $h_{ij}$, we see that we have $N-1$ 
independent parameters, since the $h_{i}$s satisfy
$\sum_{i=1}^{N}h_{i}=0$. This tracelessness condition can actually
be waived without changing the physics, since the trace part 
of $\mathbf h$ corresponds to adding a constant to the 
lagrangian. This model, that we will call the Neumann model for reasons to 
become clear later in this Section, is the one on which we want to focus
in this paper. We will usually use the $N-1$ independent variables
\begin{equation}
\label{defv}
v_{i}=h_{N}-h_{i},\quad 1\leq i\leq N-1.
\end{equation}
Note that in the particular case $v_{1}=\cdots =v_{N-1}=m^{2}$ we recover 
the model for a dimension one vector mass parameter with $M=0$. By eliminating 
$\Phi_{N}$ we find, up to a constant term, the very simple potential 
\begin{equation}
\label{pots2}
V = {Z_{(s,2)}\over 2}\, \sum_{i=1}^{N-1}v_{i}\Phi_{i}^{2}.
\end{equation}
\paragraph{Antisymmetric tensor $a_{ij}$ of dimension one}
Dimension two operators can be constructed by using $t_{ijkl}=a_{ij}a_{kl}$.
This tensor decomposes into four irreducible representations of ${\rm 
SO}(N)$, amongst which only two can be used to construct invariants with 
the help of the $\Phi_{i}$s: the trivial representation $\tr {\mathbf 
a}^{2}$, and $t_{ikjk}$. We thus get a special case of the previous model, 
for which
\begin{equation}
\label{hsusy}
h_{ij}=-\sum_{k=1}^{N} a_{ik}a_{jk}.
\end{equation}
This special case is particularly significant, however, because it is on 
this form that the Neumann model can be supersymmetrized \cite{Ferrarisusy}.

The Neumann model is singled out by the fact that it is the most general 
model with only quadratic terms in the fields $\Phi$ when the Lagrange 
multiplier $\alpha$ is introduced. This makes the analysis of the large 
$N$ limit very simple as we will see in Section 3. 
More general models can nevertheless
be studied in the large $N$ limit, with the ideas 
presented in the present paper, but at the expense of introducing 
additional auxiliary fields (the large $N$ limit can be subtle in some
cases, though). In this sense the Neumann model is ``minimal,'' 
since it can be studied by introducing the minimal number of auxiliary 
fields.

It is natural to consider another simple model,
\paragraph{Traceless symmetric tensor $m_{ij}$ of dimension 1}
To work out the potential term, we must decompose $t_{ijkl}=m_{ij}m_{kl}$ 
into irreducible representations of ${\rm SO}(N)$. Four such 
representations occur, three of which can be used to construct three 
independent invariants, including the trivial representation which 
renormalizes the vacuum energy. This implies that, 
in addition to the mass parameters themselves, the model depends on a 
new dimensionless coupling constant $G$. Generically, we will also have 
the dimension one operator ${\boldsymbol\Phi}{\mathbf m}{\boldsymbol\Phi}$ 
and a new mass parameter $M$. The potential is
\begin{eqnarray}
\label{masss1}
V_{(s,1)}&=& {Z_{(s,1)}\over 2}\, {\boldsymbol\Phi}{\mathbf m}^{2}
{\boldsymbol\Phi} - {Z'_{(s,1)}G^{2}\over 2}\, ( {\boldsymbol\Phi}{\mathbf m}
{\boldsymbol\Phi})^{2} - Z''_{(s,1)}\, 
M{\boldsymbol\Phi}{\mathbf m}{\boldsymbol\Phi} -
Z'''_{(s,1)}\, \tr {\mathbf m}^{2}\nonumber\\ 
&=&{Z_{(s,1)}\over 2}\, \sum_{i=1}^{N}m_{i}^{2}\Phi_{i}^{2} - 
{Z'_{(s,1)}G^{2}\over 2}\, \left( \sum_{i=1}^{N} 
m_{i}\Phi_{i}^{2}\right)^{2}\nonumber\\
&&\qquad\qquad\qquad\qquad\qquad
- Z''_{(s,1)}\, M\sum_{i=1}^{N}m_{i}\Phi_{i}^{2} -
Z'''_{(s,1)}\, \sum_{i=1}^{N}m_{i}^{2}.\\
\nonumber
\end{eqnarray}
$M$ can be set to zero thanks to the symmetry $m_{i}\mapsto -m_{i}$. $G$ 
can also be consistently
set to zero, since we then recover a special case of the Neumann 
model, for which
\begin{equation}
\label{hsusy2}
h_{ij}=-\sum_{k=1}^{N}m_{ik}m_{jk}.
\end{equation}
Considering $G\not =0$ is however interesting, since the model can be 
supersymmetrized for $G=g$ \cite{Ferrarisusy}
(in the non-supersymmetric version, it is not 
consistent to set $G=g$, and $G$ and $g$ are then two independent coupling 
constants, see below). We thus have two natural supersymmetric versions of 
the Neumann model, with $\mathbf h$ taking the special form 
(\ref{hsusy}), or (\ref{hsusy2}) with $G=g$. In the latter case, the
scalar potential can be written most 
elegantly by introducing a new auxiliary field $\sigma$, which turns out 
to be in the same supersymmetry multiplet as the Lagrange multiplier 
$\alpha$,
\begin{equation}
\label{susypot}
V_{(s,1)}={Z_{\rm SUSY}\over 2}\, \sum_{i=1}^{N}(\sigma + m_{i})^{2}\, \Phi_{i}^{2}.
\end{equation}
This is the ${\cal N}=1$ supersymmetric version of the ${\cal N}=2$
potential considered in \cite{braneN2} and \cite{susymodels}, which is
\begin{equation}
\label{susy2pot}
V_{{\cal N}=2}=
{1\over 2}\, \sum_{i=1}^{N}|\sigma + m_{i}|^{2}\, |\Phi_{i}|^{2},
\end{equation}
where $\sigma$ and the $\Phi_{i}$s are now complex fields.
The ${\cal N}=2$
theory with the potential (\ref{susy2pot}) shows quantitative 
similarities with ${\cal N}=2$ super Yang-Mills in four dimensions 
\cite{braneN2,susymodels}, as we have already pointed out in Section 1.
We thus see that the Neumann model can be viewed as a bosonic version 
of the supersymmetric theories studied in \cite{braneN2,susymodels}.   
\subsubsection{Renormalization constants}
One-loop formulas for the various renormalization constants can be easily 
obtained with the background field method and using Riemann normal 
coordinates, which is very elementary in the case of the sphere.
We recover the well known formulas
\begin{eqnarray}
\label{Zren}
Z_{g} &=& 1 + {N-2\over 2\pi}\,g^{2}\,\ln {\mu\over\Lambda_{0}} + {\cal 
O}(g^{4}),\\
Z &=& 1 + {g^{2}\over 2\pi}\,\ln {\mu\over\Lambda_{0}} + {\cal 
O}(g^{4}),\\ \nonumber
\end{eqnarray}
and we also obtain
\begin{equation}
\label{Zs2}
Z_{(s,2)} = 1 - {g^{2}\over 2\pi}\,\ln {\mu\over\Lambda_{0}}+ {\cal 
O}(g^{4}).
\end{equation}
$\Lambda _{0}$ is the UV cutoff and $\mu$ an arbitrary renormalization 
scale. The RG functions of the Neumann model are then (we have indicated 
$\beta$ up to two loops \cite{loop}, since this is needed later)
\begin{eqnarray}
\beta (g^{2}) &=& {\partial g^{2}\over\partial\ln\mu} =
-{N-2\over 2\pi}\, g^{4} -{N-2\over (2\pi)^{2}}\, g^{6} +
{\cal O}(g^{8}),\label{beta1l}\\
\gamma (g^{2}) &=& {\partial\ln Z\over\partial\ln\mu} = {g^{2}\over 2\pi} +
{\cal O}(g^{4}),\label{gamma}\\
\sigma (g^{2})&=&{\partial\ln Z_{(s,2)}\over\partial\ln\mu} = -{g^{2}\over 2\pi} +
{\cal O}(g^{4}).\label{sigma}\\ \nonumber
\end{eqnarray}
The theory is asymptotically free for $N\geq 3$ \cite{polnlsm},
with a dynamically generated mass 
scale $\Lambda$ defined at one loop by the equation for the coupling at 
scale $\mu$,
\begin{equation}
\label{running}
{1\over g^{2}(\mu)} = {N-2\over 2\pi}\,\ln {\mu\over\Lambda}\cdotp
\end{equation}
We have also computed
\begin{eqnarray}
Z_{(s,1)} &=& 1 - {g^{2}+4G^{2}\over 2\pi}\,\ln 
{\mu\over\Lambda_{0}}+{\cal O}(g^{4}),\label{Zs1}\\
Z'_{(s,1)} &=& 1 - {3g^2\over\pi} + {\cal O}(g^{4}),\label{Zps1}\\
\nonumber
\end{eqnarray}
from which we can deduce the $\beta$ function for the coupling $G$,
\begin{equation}
\label{betaG}
\beta_{G}(G^{2},g^{2}) = -{G^{2}(2G^{2}-3g^{2})\over\pi}
+ {\cal O}(g^{4}).
\end{equation}
This shows unambiguously that $G$ is a new coupling, independent of $g$,
and thus that introducing mass terms for the Goldstone bosons will 
generically introduce new dimensionless coupling constants into the 
theory. This is another common point with gauge theories, where 
in addition to the gauge coupling constant we have the 
couplings in the Higgs potential. However, in two dimensions, the new 
couplings cannot spoil asymptotic freedom: in spite of possible plus signs 
in $\beta$ functions like (\ref{betaG}), the physical
coupling is really something 
like $G^{2}m^{2}$ where $m$ is some mass parameter, and is always 
irrelevant in the UV.
\subsection{The Neumann model}
From now on, we focus exclusively on the Neumann model, whose minkowskian
classical lagrangian and field equations are
\begin{equation}
\label{neumannlag}
L= {1\over 2}\, \sum_{i=1}^{N} \Bigl( 
\partial_{\mu}\Phi_{i}\partial^{\mu}\Phi_{i} + h_{i}\Phi_{i}^{2} \Bigr)
-{\alpha\over 2}\, \left( \sum_{i=1}^{N} \Phi_{i}^{2} - {1\over 
g^{2}}\right),
\end{equation}
\begin{equation}
\label{fieldeq}
\partial_{\mu}\partial^{\mu}\Phi_{i} = (h_{i}-\alpha)\, \Phi_{i}.
\end{equation}
The space of parameters $\cal M$ is spanned by the $N-1$ real independent 
variables 
\begin{equation}
\label{vdef}
v_{i}=h_{N}-h_{i}.
\end{equation}
A singularity on $\cal 
M$ is a point where some of the degrees of freedom become massless. The 
set of all the singularities on $\cal M$ is called
the critical hypersurface ${\cal H}\subset {\cal M}$. 
We will distinguish between the classical critical 
hypersurface ${\cal H}_{\rm cl}$ and the quantum critical hypersurface 
${\cal H}_{\rm q}$. We will also speak loosely of a classical ${\cal 
M}_{\rm cl}$ and quantum ${\cal M}_{q}$
space of parameters, meaning $\cal M$ equipped with ${\cal H}_{\rm cl}$ 
or ${\cal H}_{q}$ respectively. 
For generic values of the parameters $v_{i}$, the model has a symmetry
${\mathbb Z}_{2(1)}\times \cdots \times{\mathbb Z}_{2(N)}$ with
\begin{equation}
\label{Z2s}
{\mathbb Z}_{2(i)} : \Phi_{j} \longmapsto (-1)^{\delta_{ij}}\,\Phi_{j}.
\end{equation}
When the $v_{i}$s coincide for $p$ disjoint subsets of
indices $I_{1},\ldots ,I_{p}$, of respective
cardinality $k_{1},\ldots ,k_{p}$, the symmetry group is enlarged to
$\times _{i\notin I_{1}\bigcup\cdots\bigcup I_{p}} {\mathbb Z}_{2(i)}\times 
{\rm O}(k_{1})\times\cdots\times {\rm O}(k_{p})$.

The Neumann model, in addition of being a natural extension of the 
standard non-linear $\sigma$ model, is also a generalization of the 
sine-Gordon model. Indeed the classical potential
\begin{equation}
\label{Vcl}
V_{\rm cl} = -{1\over 2}\, \sum_{i=1}^{N} h_{i}\Phi_{i}^{2}
\end{equation}
reduces for $N=2$ to the sine-Gordon potential
\begin{equation}
\label{SGpot}
V_{\rm sG}= -{v_{1}\over 4g^{2}}\, \cos 2\theta
\end{equation}
by writing
\begin{equation}
\label{sGpara}
\Phi_{1}={\sin\theta\over g}\raise 2pt\hbox{,}\quad
\Phi_{2}={\cos\theta\over g}\cdotp
\end{equation}
In the case $N=3$, our model is also related, through Haldane's
map \cite{hald}, to the anisotropic XYZ quantum large integer
spin chain in one dimension, or equivalently to the anisotropic classical 
Heisenberg model in two dimensions. This analogy is quite helpful 
to understand the physics of the model. The model for $N=3$ has instantons,
and we could introduce in this case a non-zero $\theta$ angle. The physics
would then strongly depend on $\theta$. 
We will not do that in the following, however, since we want to study the
models for arbitrary $N$ in a unified way. Considering a non-zero $\theta$
would be more natural in the context of a work on
the ${\mathbb C}P^{N}$ model with mass terms. 

The model defined by (\ref{neumannlag}) is also the field theoretic 
generalization of a famous integrable system in classical mechanics 
corresponding to the 
motion of a particle on a sphere with a quadratic potential,
\begin{equation}
\label{solitonequa}
{d^{2}\Phi_{i}\over dx^{2}} = (\alpha -h_{i})\, \Phi_{i},\quad
\sum_{i=1}^{N}\Phi_{i}^{2} = {1\over g^{2}}\cdotp
\end{equation} From
This problem was first studied by C. Neumann 150 years ago (for $N=3$), 
hence the name ``Neumann'' for the model (\ref{neumannlag}).
The time independent solutions to the field theory 
satisfy (\ref{solitonequa}). In Appendix A we use integrability to 
find the most general solitonic solutions of the Neumann model, 
generalizing the famous sine-Gordon solitons.
\subsection{The weakly coupled theory}
At weak coupling, the classical vacua of the theory correspond to the
minima of the potential (\ref{Vcl})
defined on the sphere $\sum_{i=1}^{N}\Phi_{i}^{2} = 1/g^{2}$. If we 
restrict ourselves, without loss of generality, to the region ${\cal 
M}_{N}$ of parameter space defined by $v_{i}\geq 0$,
then we have generically two equivalent vacua related by the
spontaneously broken ${\mathbb Z}_{2(N)}$ symmetry, such that
\begin{equation}
\label{vaccl}
\langle \Phi_{i}\rangle = \pm {\delta_{iN}\over g}\cdotp
\end{equation}
By expanding around one of these vacua, we read from the lagrangian that 
the $N-1$ ``Goldstone bosons'' $\Phi_{i}$, $1\leq i\leq N-1$, have masses
\begin{equation}
\label{mass}
m_{i}=\sqrt{v_{i}}.
\end{equation}
This relation together with (\ref{running}) shows that the coupling is weak 
in the region
\begin{equation}
\label{wccond}
v_{i} \gg \Lambda^{2}\quad {\rm for}\quad 1\leq i\leq N-1.
\end{equation}
It is actually sufficient to have
$N-2$ heavy fields for the physics to be weakly interacting, for example
\begin{equation}
\label{wccond2}
v_{i} \gg \Lambda^{2}\quad {\rm for}\quad 1\leq i\leq N-2.
\end{equation}
The low energy theory is then a sine-Gordon model whose 
elementary field has a mass $\sqrt{v_{N-1}}$ and whose non-running coupling 
is small.

The classical hypersurface of singularities ${\cal H}_{\rm cl}$ in ${\cal 
M}_{N}$ is trivially determined by (\ref{mass}) to be the union of the 
hyperplanes $v_{i}=0$. In addition to ${\cal M}_{N}$, there are $N-1$ 
other regions ${\cal M}_{j}\subset {\cal M}$ defined by
\begin{equation}
\label{Mis}
{\cal M}_{j}= \{ (v_{1},\ldots ,v_{N-1}) \mid v_{i}^{(j)}= v_{i}-v_{j}\geq 
0 \},
\end{equation}
or equivalently by 
$h_{j}=\max_{i}h_{i}$ in ${\cal M}_{j}$. ${\cal H}_{\rm cl}\cap {\cal 
M}_{j}$ is then the union of the hyperplanes $v_{i}^{(j)}=0$
in ${\cal M}_{j}$. We have reprensented
${\cal M}_{\rm cl}$ for the cases $N=3$ and $N=4$ in Figure 1.

\EPSFIGURE{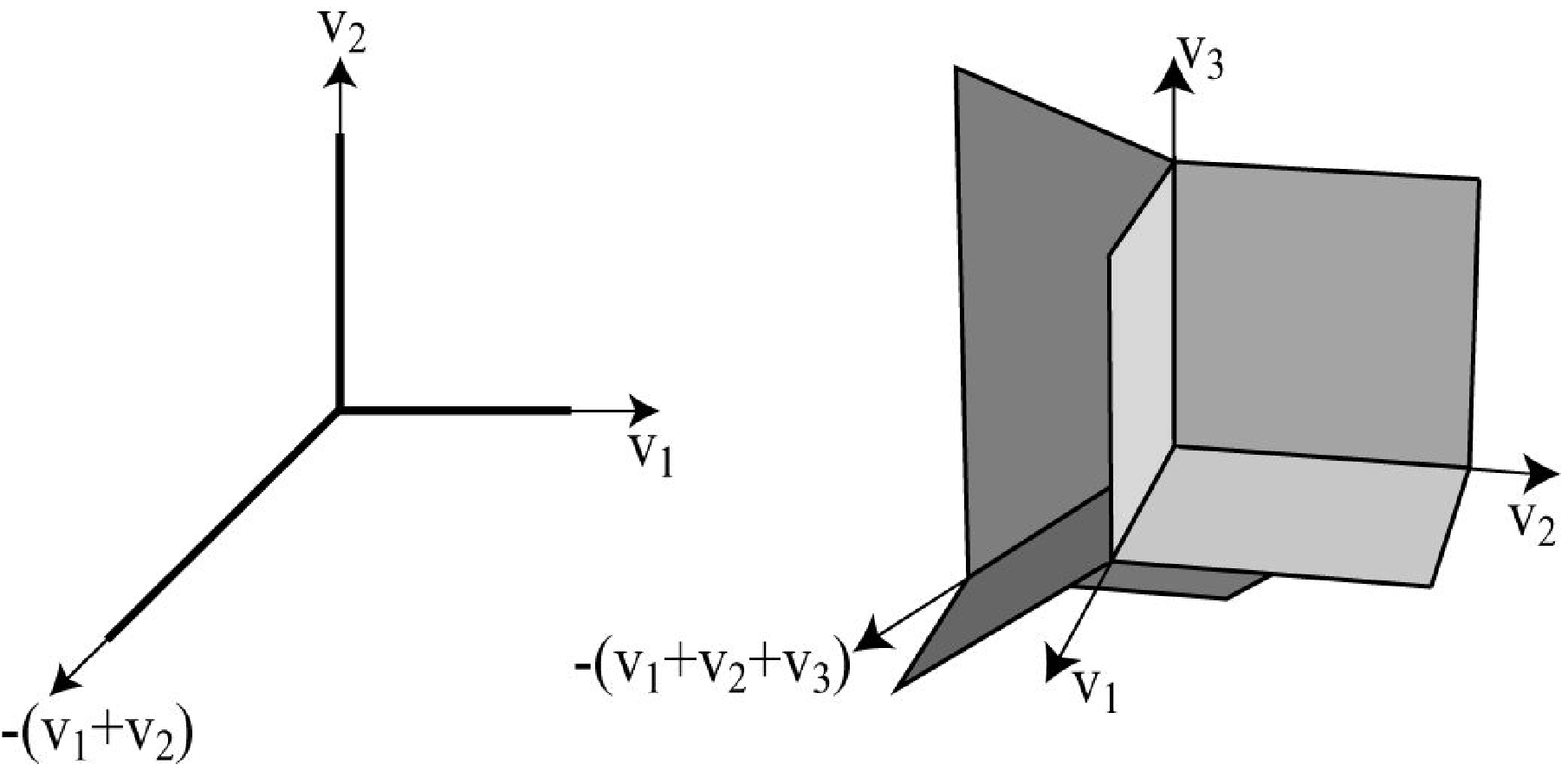,width=15cm}{The 
classical space of parameters ${\cal 
M}_{\rm cl}$ in the cases $N=3$ (left) and $N=4$ (right).
\label{Mclassical}}
\subsubsection{Integrability?}
In view of the relation of our model with the Neumann integrable system, the
${\rm O}(N)$ non-linear $\sigma$ model and the sine-Gordon model,
the reader may wonder whether it could be integrable itself 
or not. However, the simple arguments that can be used to prove
integrability when $v_{1}=\cdots=v_{N-1}=0$ or $N=2$ no longer work,
because the model combines in general
the complications due to both an explicit mass term and a non-zero $\beta$
function. In other words, there are enough 
operators to spoil the higher conservation laws
of the standard non-linear $\sigma$ model. Integrability can actually
be easily 
disproved in the weakly coupled region of parameter space, for example by 
computing the $S$ matrix element for the scattering of elementary particles,
\begin{equation}
\label{scattering}
(i,{\mathbf q}_{1}) + (j,{\mathbf q}_{2}) \longrightarrow 
(k,{\mathbf p}_{1})+(l,{\mathbf p}_{2}),
\end{equation}
where ${\mathbf q}_{1}=(\omega _{i,q_{1}},q_{1}), {\mathbf q}_{2}$ and
${\mathbf p}_{1}, {\mathbf p}_{2}$ are the incoming and outgoing 
two-momenta respectively. In the Born approximation, which is valid at weak 
coupling, we have
\begin{eqnarray}
\langle k,p_{1};l,p_{2}\mid S -1 \mid i,q_{1};j,q_{2} \rangle &=&
\delta ^{(2)}\bigl( {\mathbf p}_{1}+{\mathbf p}_{2}-{\mathbf q}_{1}-
{\mathbf q}_{2}\bigr)\nonumber\\
&&\qquad\quad {ig^{2}\over 16\pi ^{2}}\, 
\Bigl( s\, \delta_{ij}\delta_{kl} + t\,\delta_{ik}\delta_{jl} + 
u\,\delta_{il}\delta_{jk}\Bigr)\label{Smatrix}\\ \nonumber
\end{eqnarray}
where $s=({\mathbf q}_{1}+{\mathbf q}_{2})^{2}$, $t=({\mathbf 
q}_{1}-{\mathbf p}_{1})^{2}$ and $u=({\mathbf q}_{1}-{\mathbf p}_{2})^{2}$ 
are the usual Mandelstam variables, and with the normalization
\begin{equation}
\label{norminout}
\langle j,p \mid i,q \rangle = \omega_{i,q}\, \delta _{ij}\delta (q-p) =
\sqrt{q^{2}+m_{i}^{2}}\, \delta _{ij}\delta (q-p).
\end{equation}
It is clear from (\ref{Smatrix}) that processes where the number of 
particles of a given mass changes are allowed, which proves that the model 
cannot be integrable in this regime \cite{integrable}. In the ${\rm 
O}(N-1)$ symmetric case $v_{i}=\cdots =v_{N-1}$ where all the elementary 
particles have the same mass, the $S$ matrix element can be written
\begin{eqnarray}
\langle k,p_{1};l,p_{2}\mid S -1 \mid i,q_{1};j,q_{2} \rangle &=&
\omega_{q_{1}}\delta (q_{1}-p_{1})\, \omega_{q_{2}}\delta (q_{2}-p_{2})
\nonumber\\
&&\hskip -5em
\bigl( \delta_{ij}\delta_{kl}\, S_{1} + \delta_{ik}\delta_{jl}\, S_{2}
+\delta_{il}\delta_{jk}\, S_{3}\bigr) + (k\leftrightarrow l, 
p_{1}\leftrightarrow p_{2})\label{Smatrix2},\\ \nonumber
\end{eqnarray}
with
\begin{equation}
\label{Sis}
S_{1}= {ig^{2}\over 8\pi^{2}}\, {s\over\sqrt{-su}}\raise 2pt\hbox{,}\quad
S_{2}= 1,\quad S_{3}= {ig^{2}\over 8\pi^{2}}\, {u\over\sqrt{-su}}\cdotp
\end{equation}
On this form, it is clear that the Yang-Baxter equation is violated
\cite{integrable}, and 
thus the theory cannot be integrable even in the most symmetric case.
The above reasoning does not exclude that the model could be 
integrable for some special values of the parameters at strong 
coupling, but apart from the case $v_{1}=\cdots =v_{N-1}=0$
we think it is very unlikely.
This is confirmed by the analysis in the large $N$ approximation, see 
Section 3.
\subsubsection{Bound states}
From what is known for the sine-Gordon model, we can suspect that we have
bound states of the elementary particles at 
weak coupling. This is a priori possible because the interactions between 
the $\Phi_{i}$s are attractive thanks to the derivatives in 
the interaction terms (see (\ref{Lnlk}) and (\ref{metric})). 
At weak coupling, we can use a non-relativistic approximation
to investigate the bound state spectrum. The 
quantum mechanical attractive potential in the two-particle subspace can be 
straightforwardly deduced from the $S$ matrix element (\ref{Smatrix}) and 
is simply
\begin{equation}
\label{nonrelpot}
V(X)= -{1\over 2}\, g^{2}\, \delta(X)\otimes J
\end{equation}
where $J$ acts in internal space,
\begin{equation}
\label{Jdef}
\langle kl\mid J\mid ij\rangle =\delta_{ij}\delta_{kl}.
\end{equation}
We see that processes like $(ii)\rightarrow (jj)$ with $i\not =j$ are 
possible, as long as the kinematical non-relativistic constraint 
$m_{i}=m_{j}$ is satisfied. In general, stable bound states will exist 
between particles of the lowest mass. Let us study the two-particle bound 
states in the symmetric case $v_{1}=\cdots =v_{N-1}=v$. This amounts to 
solving the Shr\" odinger equation for two particles of mass $\sqrt{v}$      
interacting through (\ref{nonrelpot}). The wavefunctions $\psi 
_{ij}(x_{1},x_{2})$ must satisfy 
$\psi_{ii}(x_{1},x_{2})=\psi_{ii}(x_{2},x_{1})$ due to bose statistics. It 
is very elementary to see that there is a unique bound state which is a 
singlet of ${\rm O}(N-1)$ ($\psi_{ij}\propto\delta_{ij}$) and whose mass is
\begin{equation}
\label{massbs}
m_{\rm b}=2\sqrt{v}\, \Bigl( 1-{1\over 32}\, (N-1)^{2}g^{4} \Bigr).
\end{equation}
The symmetric mixing between the $N-1$ $\Phi_{i}$--$\Phi_{i}$ two-particle 
states making up the bound state dramatically stabilizes it. This is 
particularly significant for the physics of the model, as we will see later. 
In particular, in the large $N$ limit \`a la 't Hooft \cite{thooftN} 
\begin{equation}
\label{largeNthooft}
N\rightarrow\infty ,\quad g\rightarrow 0,\quad g^{2}N = {\rm constant,}
\end{equation}
the interactions between the fields $\Phi_{i}$ are of order $1/N$, but they 
can nevertheless 
form a bound state whose binding energy is of order $N^{0}$, as 
(\ref{massbs}) shows. We will discuss the properties of this bound state 
further in Section 3. It is also natural to look for multiparticle bound states, 
which we know must exist in the limit where the model approach the 
sine-Gordon model (when one of the $v_{i}$s is much smaller than the 
others). This is non-trivial, because the $J$ matrices corresponding to 
different pairs of particles don't always commute, and we have not tried to 
study the corresponding Schr\" odinger equation. Note that these 
multiparticle bound states will be unstable in the field theory,
which is unlike the case of the 
sine-Gordon model, but their lifetime can be large. In the case where the 
$v_{i}$s are not all equal, we will still have a stable bound state in 
which the pair $\Phi$-$\Phi$ corresponding to the particle of lowest mass 
dominate. We will see how (\ref{massbs}) generalizes in this case in 
Section 3 in the large $N$ limit. 
\subsubsection{Solitons}
The most general time independent, finite energy solutions to the field
equations (\ref{fieldeq}) are derived in Appendix A. They satisfy
(\ref{solitonequa}) and must tend towards one of the vacua (\ref{vaccl})
of our theory.
Let us assume for the moment that $v_{1}>v_{2}>\cdots v_{N-1}$.
The only
solutions corresponding to stable particles are then found by restricting the 
fields to $\Phi_{1}=\cdots =\Phi_{N-2}=0$, and correspond to the standard 
sine-Gordon soliton and anti-soliton of mass
\begin{equation}
\label{massesol}
m_{\rm sol}= {2\sqrt{\mathstrut v_{N-1}}\over g^{2}}\cdotp
\end{equation}
For example the soliton solution joining the North pole of the sphere 
$\Sph{N-1}$ at $x=-\infty$ to the South pole at $x=+\infty$ and centered 
at $x=x_{c}$ is given by
\begin{equation}
\label{solution}
\Phi_{N-1}={\sin\theta\over g}\raise 2pt\hbox{,}\quad
\Phi_{N}={\cos\theta\over g}\raise 2pt\hbox{,}
\end{equation}
with
\begin{equation}
\label{solution2}
\tan {\theta\over 2} = e^{\sqrt{v_{N-1}} (x-x_{c})}.
\end{equation}
There are also solutions corresponding to sine-Gordon solitons of masses
\begin{equation}
\label{massgensol}
m_{\rm i,sol}= {2\sqrt{\mathstrut v_{i}}\over g^{2}}\raise 2pt\hbox{,}
\end{equation}
when the only varying fields are taken to be $\Phi_{N}$ and $\Phi_{i}$.
The particles associated with these solutions are unstable and would decay 
to the stable solution (\ref{solution}), (\ref{solution2}) by emitting 
elementary quanta.

There are also more general solutions, which are worked out in details in 
Appendix A, describing a succession of sine-Gordon kinks and anti-kinks of 
different masses. 
The force between these kinks magically cancel at 
the classical level, a consequence of the integrability of the equations
(\ref{solitonequa}). It is very unlikely that this property is maintained 
at the quantum level. The static solution for the case 
$N=3$ is depicted in Figure 2.

\EPSFIGURE{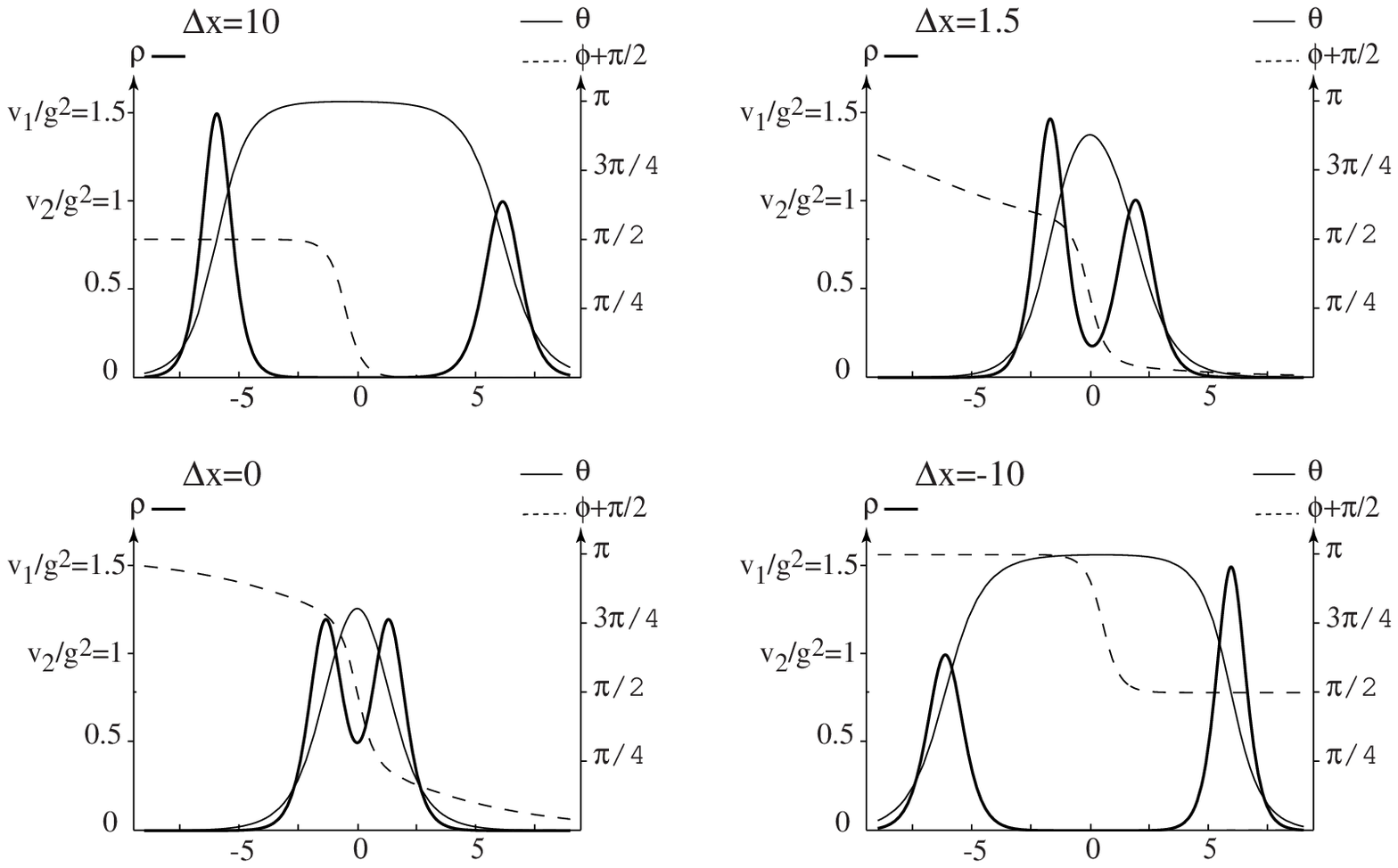,width=15cm}{
A static solution 
for the case $N=3$, $v_{1}=1.5$, $v_{2}=1$, and $g=1$, describing two kinks 
of respective masses $2\sqrt{v_{1}}/g^{2}$ and $2\sqrt{v_{2}}/g^{2}$,
at different values of the distance $\Delta x_{c} =x_{c2}-x_{c1}$ between
their centers. The energy density $\rho$ is represented 
in thick solid line, and the spherical angles $\theta$ and $\phi$
in thin solid line and dashed line respectively.
The maximum energy density of a given isolated kink 
is $v_{i}/g^{2}$. The formulas used to obtain the figure are given in 
Appendix A.3.\label{solprofiles}}

When several of the $v_{i}$s coincide, the soliton solutions have 
collective coordinates correponding to the enhanced ${\rm O}(p)$ 
symmetries. Let us discuss briefly the most symmetric case $v_{1}=\cdots 
=v_{N-1}=v$. The general solution $\Phi_{i,{\rm sol}}(x-x_{c};\xi_{\alpha})$
is obtained from (\ref{solution}) and 
(\ref{solution2}) by applying an arbitrary ${\rm SO}(N-1)$ rotation 
$R(\xi_{\alpha})$. The angles $\xi_{\alpha}$
can be taken to parametrize the coset ${\rm 
SO}(N-1)/{\rm SO}(N-2)=\Sph{N-2}$. The 
quantization in the one-soliton sector, in the moduli space (low energy) 
approximation, then proceeds in the usual way, by restricting the fields 
to be of the form
\begin{equation}
\label{modspace}
\Phi_{i}(x,t)=\Phi_{i,{\rm sol}}\bigl( x-x_{c}(t);\xi^{\alpha}(t)\bigr),
\end{equation}
and replacing this ansatz in the lagrangian (\ref{neumannlag})
to find the quantum
mechanics governing the collective coordinates $\xi^{\alpha}(t)$ and
$x_{c}(t)$. The resulting lagrangian turns out to be
\begin{equation}
\label{Lmoduli}
L_{\rm moduli}= - m_{\rm sol} + {1\over 2}\, m_{\rm sol}\, \dot x_{c}^{2} +
{m_{\rm sol}\over 2 v}\, g_{\alpha\beta}\, \dot\xi^{\alpha}\dot\xi^{\beta},
\end{equation}
where $g_{\alpha\beta}$ is the $\rm {O}(N-1)$ invariant metric on the 
sphere $\Sph{N-2}$ of radius unity. The Schr\"odinger equation 
corresponding to (\ref{Lmoduli}) involves the Beltrami
laplacian on the sphere
$\Sph{N-1}$, whose eigenvalues and eigenvectors are well-known. We get a
tower of states of mass
\begin{equation}
\label{masstower}
m_{{\rm sol},J}= m_{\rm sol} + {v\over 2m_{\rm sol}}\, J(J+N-3)
\end{equation}
filling multiplets of ${\rm SO}(N-1)$ corresponding to the completely 
symmetric and traceless tensors of rank $J$.
The low-lying states must be stable at weak coupling,
since then $v/m_{\rm sol}=\sqrt{v}\, 
g^{2}/2\ll \sqrt{v}$, the mass of the elementary particles.

We also have soliton/anti-soliton bound states described by the breather 
solution of the sine-Gordon model. It could be interesting to investigate, 
for example in the case $v_{1}=\cdots =v_{N-1}$, the possible relationship 
between these two-solitons bound states and the multiparticles bound states 
formed by the $\Phi_{i}$s. We have not performed this analysis, however; 
for our purposes, only the stable $\Phi$--$\Phi$ singlet will be relevant.

\section{The 1/N expansion}
\subsection{General properties and limitations of the large N approximation}
The large $N$ expansion \` a la 't Hooft (\ref{largeNthooft})
is one of the main non-perturbative tool at our 
disposal. The basic idea (for reviews see for example \cite{revlargeN}) is 
to integrate {\it exactly} over the fields $\Phi_i$ in (\ref{neumannlag}), 
which yields an effective action proportional to $N$. 
The $1/N$ expansion is then nothing but a
loop expansion for this effective action, and it is non-perturbative 
with respect to the other parameters of the theory. For example, the 
connected vacuum amplitude can be written in the Neumann model as
\begin{equation}
\label{vacexp}
W = \sum_{l=0}^{\infty} N^{1-l}\, W_l(v_i/\Lambda).
\end{equation}
A $2n$-point function would have an additional factor of $1/N^{n-1}$.
Though it is very useful, the $1/N$ expansion has
nevertheless two important limitations that will bother us in
the following, and that are likely to show up in gauge theories as well.

The first limitation comes from the fact that our space of parameters 
$\cal M$ is
$N-1$ dimensional. In the large $N$ limit, instead of working with a
discrete set of dimensionless parameters $x_i=v_i/\Lambda$, one should
really introduce a density
\begin{equation}
\label{denpar}
r(x) = {1\over N-1}\, \sum_{i=1}^{N-1} \delta (x-x_i),
\end{equation}
and consider only smooth enough functions $r(x)$. What this means in
practice is that we cannot study in this approximation scheme
non-generic parameters where a small number of masses are adjusted to take
particular values. To be concrete, let us suppose that we want to investigate
the region in parameter space where one of the $v_i$s, say $v_1$, goes to
zero. In the $1/N$ expansion, the $v_i$s typically appear through
combinations of the form
\begin{equation}
\label{defV}
{1\over V} = {1\over N-1}\, \sum_{i=1}^{N-1} {1\over v_i}\cdotp
\end{equation}
We see that, though $V$ is generically of order $N^0$, it is of order $N$
when $v_1\rightarrow 0$. This means that the $1/N$ counting is completely
modified in this limit, and the standard expansion can no longer be trusted.
This problem is of course a serious drawback, since we know in particular
from gauge
theory studies that interesting phenomena are usually associated with fine
tuning of parameters (``critical points''). Fortunately, there are two ways
out of this problem. First, one can consider only $p$ dimensional subspaces
${\cal S}_p\subset {\cal M}$, 
with $p\ll N$, for example by choosing $v_1=\cdots 
v_{[(N-1)/p]} = w_1$, $v_{[(N-1)/p]+1} =\cdots = v_{[2(N-1)/p]} = w_2$,
etc\dots The physics on ${\cal S}_p$, including special points,
can then be studied reliably in the large $N$ expansion.
This trick is useful in the supersymmetric
generalizations of the Neumann model \cite{Ferrarisusy}.
In the present case,
we need to do better, however, because interesting physics can be
associated with only {\it one} parameter going to a special value, like
$v_1\rightarrow 0$. The idea is then to treat exactly
the small number of degrees of freedom that play a special r\^ole in the
particular limit we are interested in. For example, in the limit
$v_1\rightarrow 0$ with all the other $v_i$s positive,
we will integrate over the $N-2$ fields $\Phi_2,\ldots ,\Phi_{N-1}$ in
(\ref{neumannlag}), while keeping explicitly the fields $\Phi_1$ and $\Phi_N$
together with $\alpha$. More generally, it is actually very convenient to 
always keep explicitly the field $\Phi_N$ when working on ${\cal M}_{N} = \{ 
v_{i}\geq 0 \}$ 
for example, since this field is singled out by the fact that
$\langle\Phi_N \rangle \not = 0$ at weak coupling in this region.

The second limitation of the $1/N$ expansion that we will encounter comes 
from the fact that the terms in the expansion can behave badly for some 
values of the parameters. For example, it could be that some 
of the coefficients $W_{l}(v_{i}/\Lambda )$ in (\ref{vacexp}) are singular for 
some (generic)
values of the $v_{i}$s. Typically, these coefficients can be expanded 
in powers of the coupling constant
at weak coupling, but when the coupling grows the series might become 
singular.
Beyond that point, the coefficients might again be expandable in terms of 
some ``dual'' coupling. At the singular points,
the standard $1/N$ expansion fails completely. We 
will see in Section 3.3 that the physics can nevertheless
be extracted in the large $N$ limit, by resumming in some sense
the $1/N$ series. We will also see in Section 6 that in our model the 
singular behaviour of the coefficients of the $1/N$ expansion is very 
specific, and that this has some very interesting consequences.
The same kind of phenomena are likely to occur
in gauge theories as well,
for example in the vicinity of an Argyres-Douglas point.
\paragraph{Renormalization}
At the technical level, the $1/N$ expansion needs to be renormalized.
It is actually easy to find the divergent part of the renormalization
constants that make the theory finite to all orders in $1/N$. 
This is possible because the theory is asymptotically free, and the $1/N$
expansion is non-perturbative in $g$. The divergent contributions can then
be obtained by summing up ordinary perturbation theory by using the
renormalization group. As is well known,
these divergent terms are completely determined by
the two-loop $\beta$ function and the one-loop $\gamma$ and $\sigma$
functions, all given in (\ref{beta1l}), (\ref{gamma}) and (\ref{sigma}).
More precisely, we have the exact formulas
\begin{eqnarray}
{1\over g_0^2} &=& {N-2\over 2\pi}\,\ln {\Lambda_0 \over\Lambda} +
{1\over 2\pi}\, \ln\ln {\Lambda_0 \over\Lambda} + {\rm finite\ terms},
\label{renNg} \\ 
\ln Z &=& -{1\over N-2}\, \ln\ln {\Lambda_0 \over\Lambda}
+ {\rm finite\ terms},\label{renNZ} \\
\ln Z_{(s,2)} &=& {1\over N-2}\, \ln\ln {\Lambda_0 \over\Lambda}
+ {\rm finite\ terms},\label{renNZs2}\\ \nonumber
\end{eqnarray}
which yield
\begin{eqnarray}
Z &=& 1- {1\over N}\, \ln\ln {\Lambda_0 \over\Lambda} + {k\over N}+
{\cal O}(1/N^2),\label{renNz2}\\
Z_{(s,2)} &=& 1 + {1\over N}\, \ln\ln {\Lambda_0
\over\Lambda} + {k'\over N}+{\cal O}(1/N^2),\label{renNzs22}\\ \nonumber
\end{eqnarray}
where $k$ and $k'$ are finite constants.
The finite terms in $\ln Z$ and $\ln Z_{(s,2)}$
will contribute to the divergent terms in $Z$ and $Z_{(s,2)}$ at order
$1/N^2$ and higher, however.
\subsection{Massless states}
To keep the formulas as simple as possible, let us begin by restricting 
ourselves to the ${\cal O}(N-1)$ symmetric case $v_{1}=\cdots 
=v_{N-1}=v\geq 0$. We can then take without loss of generality $h_{N}=v$ and
$h_{i}=0$ for $1\leq i\leq N-1$.
By integrating over the $N-1$ fields $\Phi_{i}$, $1\leq 
i\leq N-1$, in (\ref{neumannlag}), and Wick rotating to the euclidean, we 
obtain, in the large $N$ limit, the non-local effective action
\begin{equation}
\label{Seff1}
S_{\rm eff}[\alpha ,\Phi_{N}] = N\, s_{\rm eff}[\alpha ,\varphi]
\end{equation}
with
\begin{equation}
\label{phidef}
\Phi_{N} = \sqrt{N}\,\varphi
\end{equation}
and
\begin{equation}
\label{seff1}
s_{\rm eff}[\alpha ,\varphi] = \int\! d^{2}x \left(
{1\over 2}\, \partial_{\alpha}\varphi \partial_{\alpha}\varphi +
{\alpha -v\over 2}\, \varphi^{2} - {\alpha\over 2Ng^{2}} \right)
+ s[\alpha].
\end{equation}
The functional $s[\alpha]$ is defined and studied in details in Appendix B, 
to which we refer the reader, and $g$ is the renormalized coupling
(\ref{running}). The effective potential corresponding to 
$s_{\rm eff}$ is
\begin{equation}
\label{veff1}
v_{\rm eff}(\alpha ,\varphi ) = {\alpha -v\over 2}\, \varphi^{2} -
{\alpha\over 8\pi}\, \ln {\alpha\over e\Lambda^{2}}\raise 2pt\hbox{,}
\end{equation}
and the saddle point equations governing the physics in the limit 
$N\rightarrow\infty$ are
\begin{equation}
\label{saddle}
\left\{ \matrix{
(\alpha_{*} -v)\, \varphi_{*} &=& 0 \cr
4\pi \varphi_{*}^{2} &=&
\displaystyle \ln {\alpha_{*}\over\Lambda^{2}}\cdotp\cr}\right.
\end{equation}
At weak coupling $v\gg\Lambda^{2}$, the ${\mathbb Z}_{2(N)}$ 
symmetry is spontaneously broken and $\varphi _{*}\not =0$ (\ref{vaccl}). 
We will 
loosely call the ``weak coupling" region in parameter space the whole 
region where ${\mathbb Z}_{2(N)}$ is broken and where we have
\begin{equation}
\label{wcsad}
{\rm Weak\ coupling:}\quad\left\{ \matrix{
\alpha_{*} &=&v\hfill\cr
\varphi_{*}^{2} &=&\displaystyle {1\over 4\pi}\, \ln 
{v\over\Lambda^{2}}\cdotp\cr}\right.
\end{equation}
Equations (\ref{wcsad}) give
\begin{equation}
\label{vacq}
\langle\Phi_{N}^{2}\rangle = {1\over g^{2}(\mu =\sqrt{v})}
={1\over g_{\rm eff}^{2}}\raise 2pt\hbox{,}
\end{equation}
which is simply the classical formula (\ref{vaccl}) where 
the bare coupling has 
been replaced by the effective low energy coupling $g_{\rm eff}$,
the running coupling (\ref{running}) evaluated at the 
scale $\sqrt{v}$ corresponding to the mass of the elementary particles.
On the other hand, when $v=0$, we know that $\langle\Phi_N\rangle =0$ as a
consequence of the ${\rm SO}(N)$ symmetry. Thus, there must be a phase 
transition at strong coupling for some $v=v_{\rm c}\sim\Lambda^2$,
where the ${\mathbb Z}_{2(N)}$ is restored. We would like to
understand the nature of this transition,
and whether $v_{\rm c}=0$ or $v_{\rm c}>0$. 
Since it occurs at strong coupling in a non-abelian system, the standard
intuition would suggest that a mass gap is formed and thus that any phase
transition must be first order. This would be the case for example 
if $v_{\rm c}=0$. However, we are going to show that this is not what
happens: the transition is second order, and we have massless states at
strong coupling. To do so,
we evaluate the stability of the saddle point (\ref{wcsad}). 
The effective potential for $\varphi$ is obtained by 
integrating out $\alpha$ from (\ref{veff1}),
\begin{equation}
\label{veffphi}
v_{\rm eff}(\varphi) = {\Lambda^{2}\over 8\pi}\, \Bigl(
e^{4\pi\varphi^{2}} - 4\pi\varphi^{2}\, {v\over\Lambda^{2}}\Bigr)\cdotp
\end{equation}
This shows that we have a second order transition at $v=v_{\rm
c}=\Lambda^2$, between a broken symmetry phase $v>\Lambda^{2}$ and a phase
$v<\Lambda^{2}$ where the ${\mathbb Z}_{2(N)}$ is restored.
This is the ``strong coupling" region of 
parameter space, where the saddle point is
\begin{equation}
\label{scsad}
{\rm Strong\ coupling:}\quad\left\{ \matrix{
\alpha_{*}&=&\Lambda^{2}\cr
\varphi_{*}&=&0.\cr}\right.
\end{equation}
It is interesting to investigate in details what happens at the point 
$v=\Lambda^{2}$. In the $1/N$ expansion, the effective 
coupling constant $g_{\rm eff}$ goes to 
infinity at that point, which is directly related to the fact that we have
massless degrees of freedom. This phenomenon is also known to occur in
{\it supersymmetric} gauge theories when a magnetic monopole 
becomes massless. 
Actually we will see in Section 4 and 5 that effects which
are non-perturbative in $1/N$ make the effective
coupling constant finite at the transition
point $v=\Lambda^2$. The critical theory will be argued to be
very similar to an Argyres-Douglas CFT \cite{AD}, where both magnetic
monopoles (topologically stable solitons) and electrically charged
particles (created by the elementary fields) can become massless.
In our case, it is clear that the stable ${\rm SO}(N-1)$ singlet
soliton found at the end of Section 2.3.3 must be massless when
$v=\Lambda^{2}$. This is simply due to the fact that
the two degenerate minima of the potential
(\ref{veffphi}) merge at that point, and any solitonic solution to the
effective action must then become trivial. To find out whether any
perturbative state could become massless together with the soliton, it is
natural to first look at the mass of the elementary fields $\Phi_i$. 
One can calculate straightforwardly the two point functions 
in the leading $N\rightarrow\infty$ approximation. In momentum 
space they are simply
\begin{equation}
\label{twopoint}
\langle\Phi_{i}\Phi_{j}\rangle (p) = {\delta_{ij}\over p^{2}+\alpha_{*}}
\raise 2pt\hbox{,}
\end{equation}
which shows that the fields $\Phi_{i}$, $1\leq i\leq N-1$, create free 
particles of masses $m_{i}=\sqrt{\alpha_{*}}$. The $m_{i}$s thus take 
their classical values $m_{i}=\sqrt{v}$ as long as $v>\Lambda^{2}$, and 
then take the constant value $m_{i}=\Lambda$ when $v<\Lambda^{2}$: at 
strong coupling, a mass gap is created for these particles, as the standard 
lore suggests. To find massless perturbative states 
at $v=\Lambda^{2}$, we thus have 
to look more carefully at the menu of stable particles described in Section 
2.3. As explained there, though the interactions between 
the elementary particles are of order $1/N$, a bound state whose binding 
energy is of order $N^{0}$ can be formed because of a dramatic 
stabilization due to the mixing between the different flavors. This bound 
state is created by the operator $\sum_{i=1}^{N-1}\Phi_{i}^{2}$, but can 
equivalently be associated with
\begin{equation}
\label{phidefe}
\varphi={1\over \sqrt{Ng^{2}}}\,
\sqrt{1-g^{2}\sum_{i=1}^{N-1}\Phi_{i}^{2}}, 
\end{equation}
and thus its mass 
must appear as a pole in the $\langle\varphi\varphi\rangle$ correlation 
function. By expanding
\begin{equation}
\label{phiexp}
\varphi = \varphi_{*} + {\chi\over\sqrt{N}}\raise 2pt\hbox{,}\quad
\alpha = \alpha_{*} + {\beta\over\sqrt{N}}\raise 2pt\hbox{,}
\end{equation}
we immediately obtain from (\ref{seff1}) the leading term in the large $N$ 
expansion,
\begin{equation}
\label{twoptchi}
\langle\chi\chi\rangle (p) = {\tilde s^{(2)}(p^{2};\alpha_{*})\over
(p^{2}+\alpha_{*}-v)\, \tilde s^{(2)}(p^{2};\alpha_{*}) - \varphi_{*}^{2}}
\cdotp
\end{equation}
The function $\tilde s^{(2)}$ is defined in Appendix B.
The bound state mass $m_{\rm b}$ thus satisfies the equation
\begin{equation}
\label{mbeq}
m_{\rm b}^{2} = \alpha_{*}-v- {\varphi_{*}^{2}\over
\tilde s^{(2)}(p^{2}=-m_{\rm b}^{2};\alpha_{*})}\cdotp
\end{equation}
At strong coupling the solution is simply
\begin{equation}
\label{mbsc}
m_{\rm b}=\sqrt{\Lambda^{2} -v}, \quad {\rm for}\quad  v\leq\Lambda^{2}.
\end{equation}
We see that when $v=0$ and we have the full ${\rm O}(N)$ symmetry,
$m_{\rm b}=\Lambda =m_{i}$, and the spectrum consists in a
${\rm O}(N)$ vector \cite{integrable}
made up by the $N-1$ elementary particles and the $\varphi$ field.
We also discover that when $v=\Lambda^{2}$
the mass of the bound state vanishes!

When $v>\Lambda^{2}$, we have to solve
\begin{equation}
\label{mbwc}
\sqrt{{4v\over m_{\rm b}^{2}} -1}\, \ln {v\over\Lambda^{2}} =\pi -
2\, \arctan\sqrt{{4v\over m_{\rm b}^{2}} -1} \cdotp
\end{equation}
When $v\gg\Lambda^{2}$, this yields
\begin{equation}
\label{mbseries}
m_{\rm b} = 2\sqrt{v}\, \Bigl( 1- {1\over 32}\, \bigl( Ng^{2}\bigr)^{2}
+ {1\over 32\pi}\, \bigl( Ng^{2}\bigr) ^{3} + 
{\cal O}\bigl(\bigl( Ng^{2}\bigr) ^{4}, 1/N\bigr) \Bigr),
\end{equation}
in agreement with (\ref{massbs}). In the vicinity of $v=\Lambda^{2}$, 
(\ref{mbsc}) and (\ref{mbwc}) predicts
\begin{eqnarray}
m_{\rm b}^{2} & \simeq & \Lambda^{2}\, 
\Bigl( 1- {v\over\Lambda^{2}} \Bigr)\quad 
{\rm for}\quad v\rightarrow\Lambda^{2},\ v<\Lambda^{2},\nonumber \\
m_{\rm b}^{2} & \simeq & 2\Lambda^{2}\, \Bigl( {v\over\Lambda^{2}} -1\Bigr)
\quad {\rm for}\quad v\rightarrow\Lambda^{2},\ v>\Lambda^{2}.
\label{wrong}\\ \nonumber
\end{eqnarray}
\EPSFIGURE{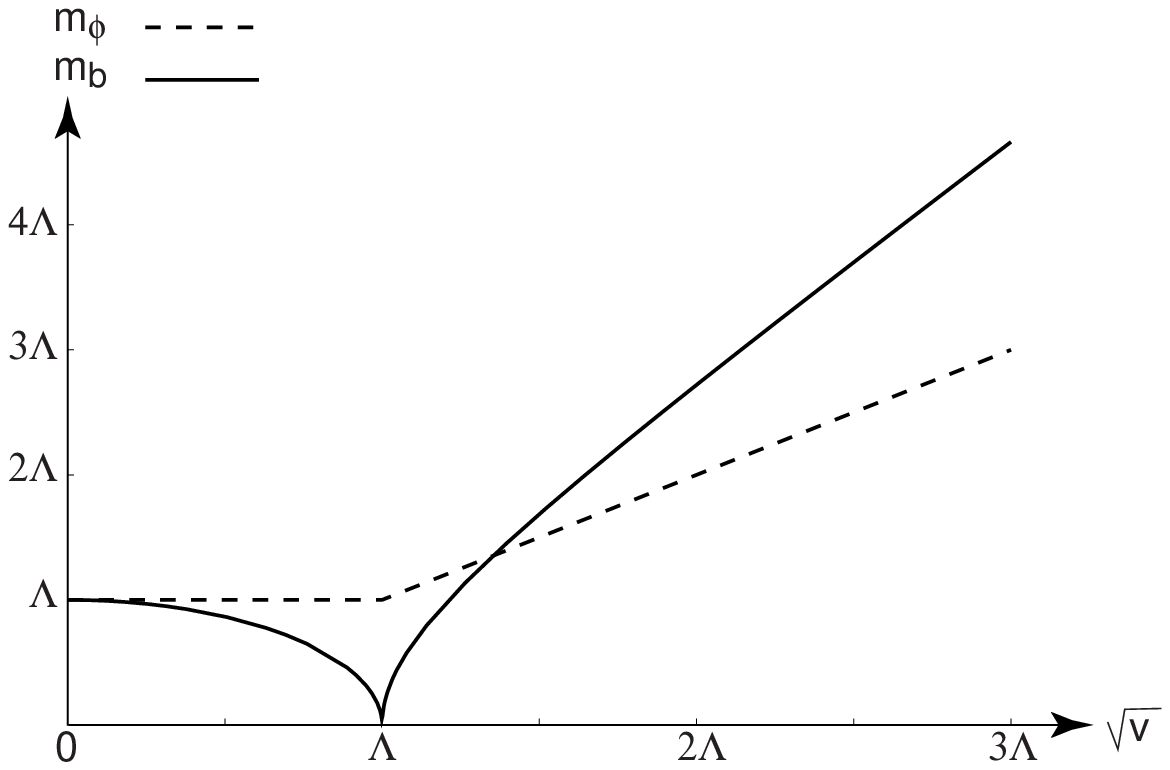,width=12cm}{
The mass of the elementary particles created by the $\Phi_{i}$s, $1\leq 
i\leq N-1$ (dashed line) and of the $\Phi$-$\Phi$ bound state (solid line),
as a function of $\sqrt{v}=\sqrt{v_{1}}=\cdots =\sqrt{v_{N-1}}$,
in the leading $1/N$ approximation.
\label{massgraph}}

We have depicted in Figure 3 the masses of the elementary particles and of 
the bound state as a function of $\sqrt{v}$, as given by the leading term in the 
$1/N$ expansion. As we will see in section 3.4, this leading 
approximation is actually incorrect near $v=\Lambda^{2}$, and in 
particular the formulas (\ref{wrong}) are wrong. However, the most 
important fact that the soliton and the bound state are massless will 
remain, and we will see that the correct form of (\ref{wrong}) is simply
\begin{equation}
\label{right}
m_{\rm b}^{2}\propto m_{\rm sol}^{2} \propto 
\Lambda^{2}\Bigl( 1-{v\over\Lambda^{2}}
\Bigr) ^{2} \quad {\rm for\ } v\rightarrow\Lambda^{2} .
\end{equation}
\subsection{Correlators and quantum corrected metric}
We continue to restrict ourselves to the symmetric case $v_1=\cdots
=v_{N-1}$ in this subsection. This makes the formulas simpler, but do not
affect the physics we want to discuss. Our aim is to obtain the large $N$
formulas for the $S$ matrix, generalizing (\ref{Sis}), and to compute the
quantum corrections to the metric on the sphere target space.
The generating functional $Z[J]$ for the correlation functions is given by
a path integral
\begin{eqnarray}
Z[J] &=& \int\! {\cal D}\chi {\cal D}\beta \exp\Bigl[
-(N-1)\, s_{\rm eff} [\alpha_* + \beta /\sqrt{N}, \varphi_* + \chi
/\sqrt{N}] \nonumber \\
&&\hskip 10em
+ {1\over 2}\, \int\! d^2 x \sum_{i=1}^{N-1}
J_i \, {1\over -\partial ^2 + \alpha_* + \beta /\sqrt{N} }\, J_i \Bigr] 
\Biggm/ \nonumber \\
&& \int\! {\cal D}\chi {\cal D}\beta \exp\Bigl[
-(N-1)\, s_{\rm eff} [\alpha_* + \beta /\sqrt{N}, \varphi_* + \chi
/\sqrt{N}]\Bigr] \cdotp \label{Zfunc}\\ \nonumber
\end{eqnarray}
The large $N$ Feynman diagrams contributing to the four point function
$\langle \Phi _{i_1}\Phi _{i_2}\Phi_{i_3}\Phi_{i_4}\rangle$ are indicated
in Figure 4. The wavy lines correspond to the $\beta$ field propagator
\begin{equation}
\label{betaprop}
\langle\beta\beta\rangle (p) = { p^2 + \alpha_* -v \over
(p^2 + \alpha_* -v) \tilde s^{(2)} (p^2,\alpha_*) -\varphi_*^2}\cdotp
\end{equation}
\EPSFIGURE{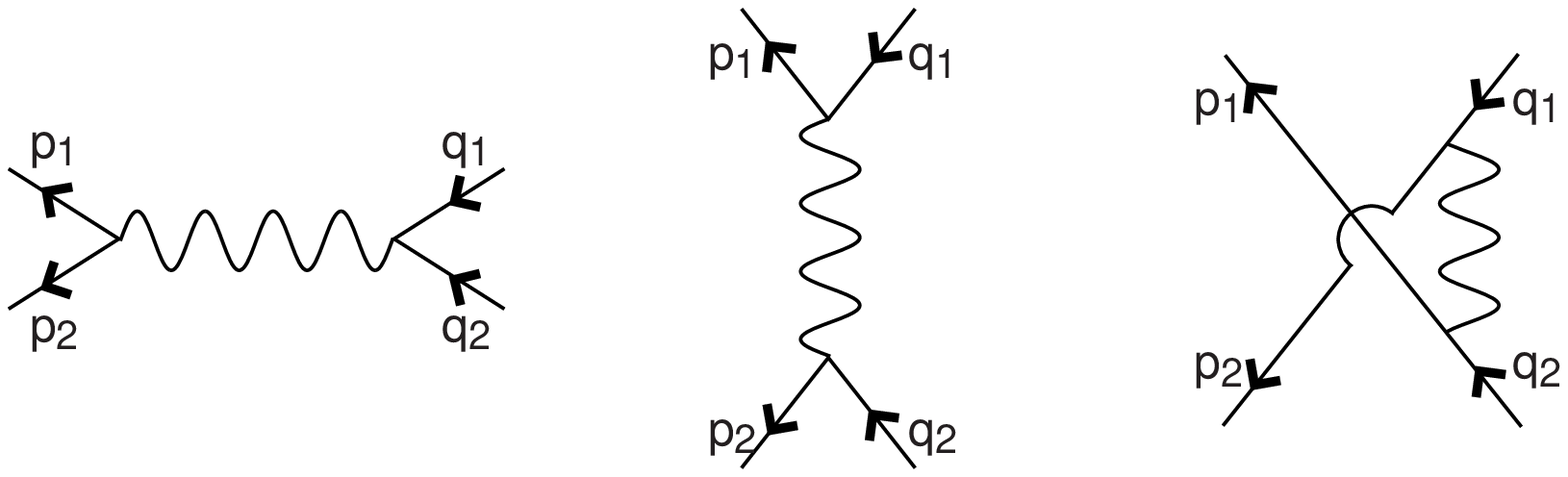,width=15cm}{Large $N$ Feynman diagrams for the
four-point function. The wavy line corresponds to the $\beta$ field 
propagator, which is itself a sum of bubble diagrams in ordinary 
perturbation theory.\label{feyn1}}

The functions $S_1$, $S_2$ and $S_3$ defined by (\ref{Smatrix2}) can then
be deduced from the four-point function. At weak coupling ($v>\Lambda^2$),
we have
\begin{equation}
\label{Sis2wc}
S_1 = {i g_{\rm eff}^2 \over 8\pi ^2}\, 
{s\over\sqrt{-su}} \,
{1\over 1 + Ng_{\rm eff}^2 s\, \tilde s^{(2)} (p^2 = -s;v)}
\raise 2pt\hbox{,}\quad S_2 =1,\quad S_3 =S_1 (s\mapsto u),
\end{equation}
or equivalently in terms of the rapidity $\theta >0$ such that
$s=4v\cosh ^2 (\theta /2)$ and $u= -4v\sinh^2 (\theta /2)$,
\begin{equation}
\label{Sis2wc2}
S_1 (\theta) = {i g_{\rm eff}^2 \over 8\pi ^2}\,
{1\over\displaystyle\tanh (\theta /2) } \,
{1\over\displaystyle 1 + Ng_{\rm eff}^2\,
{\displaystyle \theta - i\pi \over \displaystyle 4\pi\tanh (\theta /2)
}}\raise 2pt\hbox{,}\quad S_2(\theta)=1,\quad S_3(\theta) = S_1(i\pi -
\theta).
\end{equation}
$g_{\rm eff}^2$ is defined as usual to be the low energy coupling, 
i.e.~the running coupling (\ref{running}) evaluated at $\mu =\sqrt{v}$.
These formulas illustrate nicely how the $1/N$ expansion works and produces
non-perturbative results. The genuinely non-perturbative information 
lies on the full $\theta$ dependence of the denominators, not on the fact
that $g_{\rm eff}$ appears in these denominators. In particular, when
$\theta\rightarrow\infty$, the non-perturbative corrections become
dominant, even if $v\gg\Lambda^2$, but due to asymptotic freedom, we know
that in this regime the $S$ matrix can also be deduced from
perturbation theory by using the renormalization group.
Indeed we have
\begin{equation}
\label{Shighener}
S_1(\theta) \mathop{\hbox{=}}_{\theta\rightarrow\infty} 
{1\over 2\pi N\theta} = {ig^2 (\mu =\sqrt{s})\over 8\pi^2}
\end{equation}
which is the perturbative result (\ref{Sis}) at high energy, but with the
coupling $g$ evaluated at the center of mass energy, which is different
from the low energy coupling. In the strong coupling region $v<\Lambda^2$,
we have
\begin{equation}
\label{Sis2sc}
S_1 = {i\over 8\pi^2 N\sqrt{-su}}\, {1\over\tilde s^{(2)}(p^2=-s;\Lambda^2)}
\raise 2pt\hbox{,}\quad S_2 = 1- {i\Lambda^2\over \pi N\sqrt{-su}}
\raise 2pt\hbox{,}\quad S_3 = S_1 (s\mapsto u),
\end{equation}
or equivalently
\begin{equation}
\label{Sis2sc2}
S_1(\theta) = {i\over 2\pi N}\, {1\over \theta -i\pi}\raise 2pt\hbox{,}
\quad
S_2(\theta) = 1- {i\over 2\pi N \sinh\theta }\raise 2pt\hbox{,}\quad
S_3(\theta) = -{i\over 2\pi N\theta}\cdotp
\end{equation}
The functions $S_i$s are independent of $v$ in this regime, but this is
true only in the leading $1/N$ approximation.
The formulas (\ref{Sis2sc}) show that the $\Phi_i$s, $1\leq i\leq N-1$,
no longer form a two
particles bound state for $v<\Lambda^2$, and thus that the field $\Phi_N$
must be considered as an independent degree of freedom in this region.  
Clearly the geometric interpretation of the $\Phi_i$s as living on a sphere
cannot be valid for $v<\Lambda^2$. To better understand what is really
going on, it is natural to compute the quantum corrections to the metric on
the target space sphere as a function of $v$. This amounts to computing the
coefficient of $p^2$ in a low energy expansion of all
the connected, one-particle irreducible,
$2n$-point functions $\langle\Phi_{i_1}\cdots\Phi_{i_{2n}}\rangle$.
The ${\rm SO}(N-1)$ symmetry restricts a priori the metric to be of the form
\begin{equation}
\label{metricgen}
g_{ij} = f_{1}\Bigl(\sum_{k=1}^{N-1}\Phi_k^2\Bigr)\, \delta_{ij} + 
f_{2}\Bigl(\sum_{k=1}^{N-1}\Phi_k^2\Bigr)\, \Phi_i \Phi_j 
\end{equation}
for unknown functions $f_{1}$ and $f_{2}$. Actually, we are going to show 
that in the leading $1/N$ approximation and for $v>\Lambda^{2}$ 
\begin{equation}
\label{metricq}
g_{ij}=\delta_{ij} + {\Phi_{i}\Phi_{j}\over\displaystyle {N\over 4\pi}\, 
\ln {v\over\Lambda^{2}} - \sum_{k=1}^{N-1}\Phi_{k}^{2}}\raise 2pt\hbox{,}
\end{equation}
which is simply the ${\rm SO}(N)$ invariant metric on a sphere of radius 
$1/g_{\rm eff}$. This very simple result is equivalent to the statement 
that the $p^{2}$ terms in the low energy expansion of the $2n$ point 
functions is simply given by the perturbative,
tree level result, with the bare coupling 
being replaced by the low energy coupling $g_{\rm eff}$. This 
simplification occurs because the low energy expansion of the $\beta$ 
field propagator (\ref{betaprop}) is $\langle\beta\beta\rangle (p) = 
-Ng_{\rm eff}^{2}\, p^{2} + {\cal O}(p^{4})$, and thus only one such 
propagator can appear in a diagram contributing to the metric. A connected 
diagram of this type for the $2n$ points function 
must then be entirely constructed from one $\langle\beta\beta\rangle$ 
propagator, $n-2$ vertices $\beta\chi^{2} /\sqrt{N}$ coming from the 
expansion of $s_{\rm eff}$ (\ref{seff1},\ref{phiexp}), and $2n-4$ 
contractions between $\beta$ and $\chi$ which each give a factor of 
$\sqrt{Ng_{\rm eff}^{2}}$, in addition to the $n$ $\Phi\Phi\beta /\sqrt{N}$
vertices. Such diagrams at low energies depend 
on $N$ and $g_{\rm eff}$ only through a multiplicative factor 
$Ng_{\rm eff}^{2}\, N^{-(n-2)/2}\, (Ng_{\rm eff}^{2})^{(2n-4)/2}\, 
N^{-n/2}=N^{1-n}\, (Ng_{\rm eff}^{2})^{n-1} = g_{\rm eff}^{2(n-1)}$. This 
dependence on the coupling is the same as the one found
in a tree level perturbative calculation. The dependence in the momenta 
must then also coincide, since the large $N$ result must 
reduce to the tree level result when $g_{\rm eff}\ll 1$. This proves 
(\ref{metricq}).

\EPSFIGURE{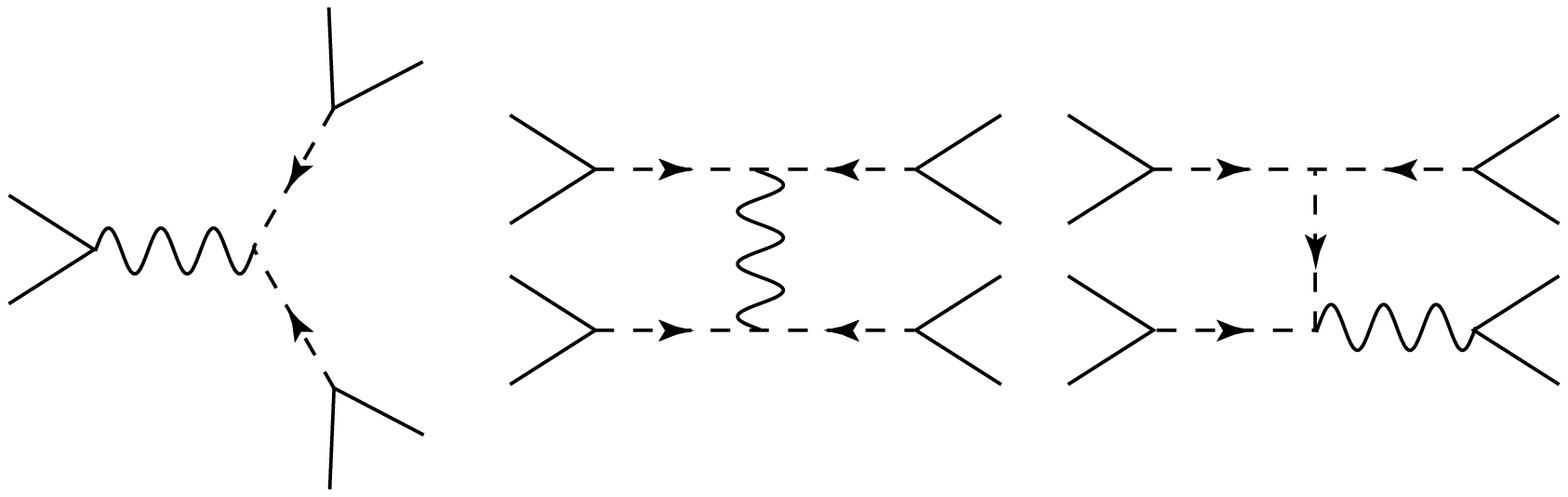,width=15cm}{Typical diagrams relevant to the 
calculation of the metric and contributing to the 
6-point (left) and 8-point (center and right)
functions. The $\langle\beta\beta\rangle$ propagator 
is represented by a wavy line, and the $\langle\beta\chi\rangle$ 
contraction by a dashed line with an arrow
pointing toward the $\chi$ insertion.\label{feyn2}}

We thus see that the effective target space sphere 
shrinks symmetrically to a point when $v\rightarrow \Lambda^{2}$, 
$v>\Lambda^{2}$.
The geometrical interpretation of the $\sigma$ model
is thus lost in the strong coupling
region $v<\Lambda^{2}$, which corresponds 
in some sense to a continuation of the sphere to negative radius. 
Similar continuations have already been 
seen in the context of supersymmetric models \cite{CVnegr}, and when the
$\sigma$ model is conformal they actually play an 
important r\^ole in understanding the topology changing
transitions in string theory (see for example \cite{reviewgreene}).
We will further discuss the consequences of 
this interpretation in Section 5.

\EPSFIGURE{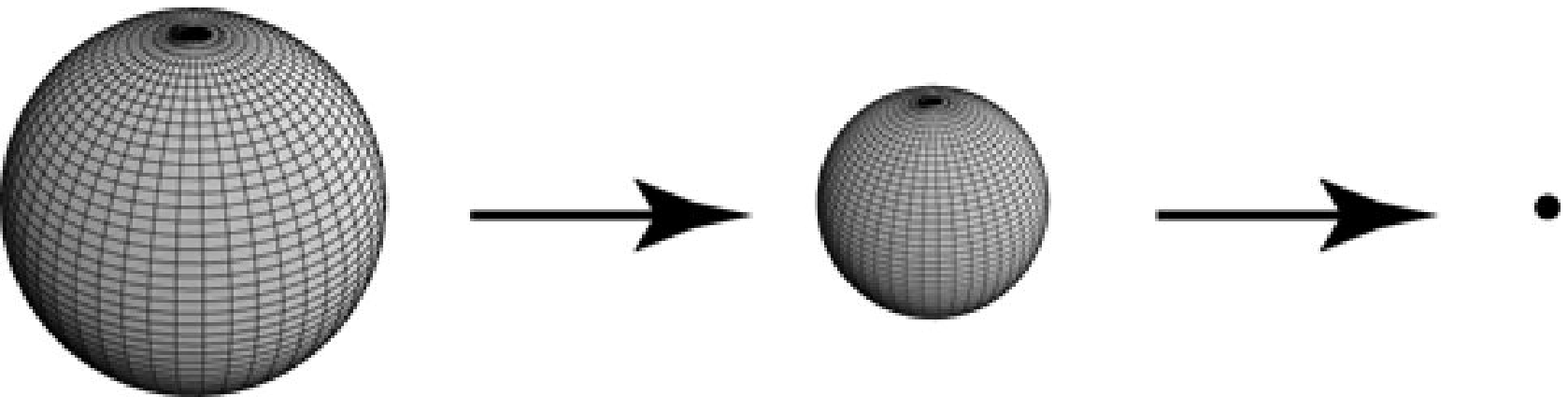,width=15cm}{The effective target space sphere 
shrinks symmetrically to a point 
when $v\rightarrow\Lambda^{2}$ in the leading $1/N$ approximation.
At strong coupling, the geometrical interpretation of the $\sigma$ model 
is lost. This is reminiscent of the stringy geometry described in the 
context of supersymmetric non-linear $\sigma$ model.\label{shrink}}
\subsection{The critical hypersurface $\HQ$}
Let us now discuss the general case where the $v_{i}\geq 0$
can be distinct. Formulas (\ref{seff1})---(\ref{wcsad}) and (\ref{scsad}) 
are replaced respectively by
\begin{equation}
\label{seff2}
s_{\rm eff}[\alpha ,\varphi] = \int\! d^{2}x \left(
{1\over 2}\, \partial_{\alpha}\varphi \partial_{\alpha}\varphi +
{\alpha -h_{N}\over 2}\, \varphi^{2} - {\alpha\over 2Ng^{2}} \right)
+{1\over N}\, \sum_{i=1}^{N-1} s[\alpha -h_{i}],
\end{equation}
\begin{equation}
\label{veff2}
v_{\rm eff}(\alpha ,\varphi ) = {\alpha -h_{N}\over 2}\, \varphi^{2} -
{1 \over 8\pi N}\, \sum_{i=1}^{N-1} (\alpha -h_{i})\,
\ln {\alpha -h_{i}\over e\Lambda^{2}}\raise 2pt\hbox{,}
\end{equation}
\begin{equation}
\label{saddle2}
\left\{ \matrix{
(\alpha_{*} -h_{N})\, \varphi_{*} &=& 0\hfill\cr
4\pi \varphi_{*}^{2} &=&\displaystyle {1\over N}\,\sum_{i=1}^{N-1}
\ln {\alpha_{*}-h_{i}\over\Lambda^{2}}\raise 2pt\hbox{,}\cr}\right.
\end{equation}
\begin{eqnarray}
{\rm Weak\ coupling:}&\quad &\left\{ \matrix{\displaystyle
\alpha_{*} &=& h_{N}\hfill\cr\displaystyle
\varphi_{*}^{2} &=&\displaystyle {1\over 4\pi N}\,\sum_{i=1}^{N-1}
\ln {v_{i}\over\Lambda^{2}}\raise 2pt\hbox{,}
\cr}\right.\label{wcsad2}\\
{\rm Strong\ coupling:}&\quad &\left\{ \matrix{\displaystyle
\sum_{i=1}^{N-1} \ln {\alpha_{*}-h_{i}\over\Lambda^{2}} &=& 0\cr
\displaystyle\varphi_{*} &=& 0.\cr}\right.\label{scsad2}\\ \nonumber
\end{eqnarray}
Note that the various sums over $i$ should be more rigorously replaced by 
integrals with the help of (\ref{denpar}). This will always be understood 
in the following.
The border between the weak coupling region $\langle\Phi_{N}\rangle\not =0$
and the strong coupling region $\langle\Phi_{N}\rangle =0$ is a 
codimension one hypersurface one which both the
soliton and the bound state become massless. The equation for this quantum 
hypersurface of singularity ${\cal H}_{\rm q}$ in ${\cal M}_{N}$
is immediately deduced from (\ref{wcsad2}) and (\ref{scsad2}),
\begin{equation}
\label{Heq1}
{\cal H}_{\rm q}\cap {\cal M}_{N}\ :\quad 
\prod_{i=1}^{N-1}v_{i} =\Lambda^{2(N-1)}.
\end{equation}
The critical point $v_{1}=\cdots =v_{N-1}=v_{\rm c}=\Lambda^{2}$ discussed
previously is of course on ${\cal 
H}_{\rm q}$. ${\cal H}_{\rm q}$ has also other sheets in the 
other, physically equivalent, regions ${\cal M}_{i}$ of parameter space 
(\ref{Mis}), whose equations are simply
\begin{equation}
\label{Heqgen1}
{\cal H}_{\rm q}\cap {\cal M}_{i}\ :\quad 
\prod_{j\not =i} v^{(i)}_{j} =\Lambda^{2(N-1)}.
\end{equation}
We thus get a preliminary picture (it is only partially correct as will be 
discussed in Section 4) of the quantum space of
parameters, in the large $N$ limit.

\EPSFIGURE{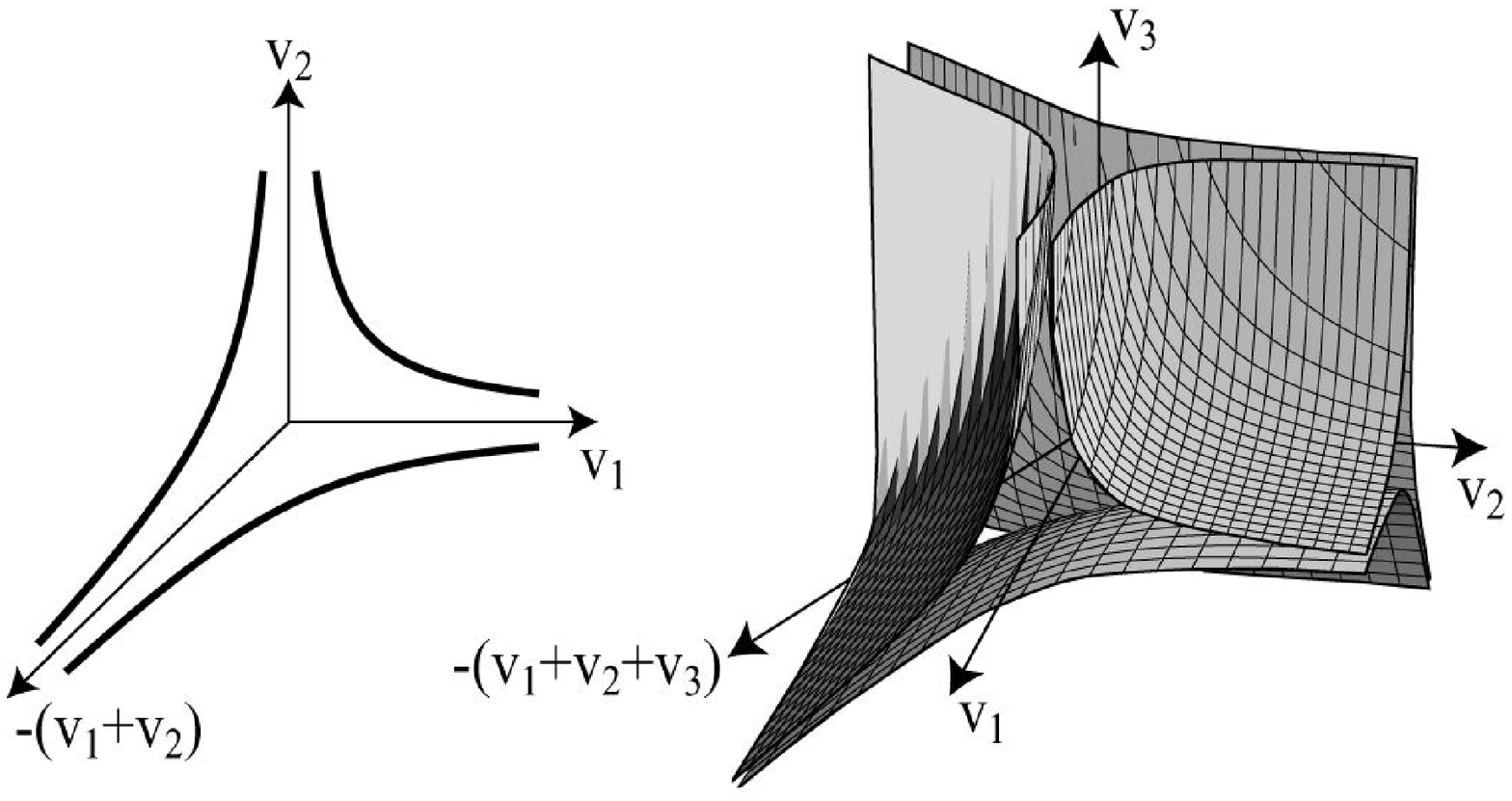,width=15cm}{
A preliminary picture of the quantum space of parameters ${\cal M}_{\rm q}$ 
in the cases $N=3$ (left) and $N=4$ (right).
\label{hquantum1}}

It is very likely that the qualitative 
picture is still valid even for values of $N$ as small as $N=3$ (we will 
give strong evidence that this is the case in Section 5), and we have used 
equations (\ref{Heq1}) and (\ref{Heqgen1}) in the cases $N=3$
and $N=4$ to make Figure~7. This figure should be compared with the 
classical case given in Figure 1. One might have expected that the 
singularities on parameter space would have disappeared in the quantum 
case, a mass gap being created, and the ${\mathbb Z}_{2(N)}$ symmetry being 
restored through a first order phase transition. We see that on the 
contrary the classical singular hyperplanes split into two in the quantum 
case, and the ${\mathbb Z}_{2(N)}$ symmetry is restored through a second 
order phase transition. We discuss in the next subsection how we can 
understand the nature (Ising) of this transition in the large $N$ 
limit, and we will also discuss it at finite $N$ in Section 5.

We can also write down the formulas generalizing (\ref{mbsc}) and 
(\ref{mbwc}) to the case of arbitrary positive $v_{i}$s. At strong 
coupling, the mass $m_{\rm b}$ satisfies
\begin{equation}
\label{mbsc2}
m_{\rm b} = \sqrt{\alpha_{*} - h_{N}},
\end{equation}
with $\alpha_{*}$ determined by (\ref{scsad2}), and at weak coupling we 
have
\begin{equation}
\label{mbwc2}
\sum_{i=1}^{N-1} \ln {v_{i}\over\Lambda^{2}} = 2\sum_{i=1}^{N-1}
{1\over \sqrt{4v_{i}/m_{\rm b}^{2} -1}}\,
\arctan {1\over\sqrt{4v_{i}/m_{\rm b}^{2} -1}}\, \cdotp
\end{equation}
\vfill\eject

\subsection{Beyond the 1/N expansion: the Ginzburg-Landau description}
In the vicinity of the critical hypersurface ${\cal H}_{\rm q}$, the usual
$1/N$ expansion is inconsistent because of infrared divergences. These
divergences are due to two different effects. First, the $\chi$ field
becomes massless and thus the
$\langle\chi\chi\rangle$ propagator (\ref{twoptchi}) 
goes to $1/p^2$. Consequently the $1/N$ expansion cannot be defined
because of the usual IR divergences associated with massless scalar
propagators in two dimensions. Second, and independently of the fact that
we are in two dimensions, the $1/N$ corrections are calculated by taking
into account interaction terms corresponding to relevant
operators. Such interactions always produce IR divergences in
perturbing around a massless theory. This is strictly analogous to the
textbook discussion of the corrections to the mean field approximation for
the $\phi^4$ theory near dimension 4, except that we are dealing presently
with
the $1/N$ perturbation theory instead of the ordinary perturbation theory.
To analyse the divergences, one could study in details the large $N$
Feynman graphs and isolate the most divergent contributions. Instead, we
will proceed in a slightly more heuristic, but actually equivalent, way. In
a first step, we would like to integrate out the field $\alpha = \alpha_* +
\beta /\sqrt{N}$, 
since the divergences are due to the masslessness of $\chi$. 
At large $N$, this can be done by solving the variational equation
\begin{equation}
\label{vareqalpha}
{\delta s_{\rm eff}[\alpha,\varphi]\over\delta\alpha (x)} =0,
\end{equation}
which at low energy reduces to
\begin{equation}
\label{vareqle}
{\partial v_{\rm eff}\over\partial\alpha} = 0.
\end{equation}
By using (\ref{veff2}) and (\ref{seff2}), we then obtain a low energy
effective action valid in the vicinity of ${\cal H}_{\rm q}$, and
depending on $\chi=\Phi_{N}$ only,
\begin{equation}
\label{seffphi}
S_{\rm eff}[\chi] = \int\! d^2x \, \Bigl( {1\over 2}\,
\partial_{\alpha}\chi \partial_{\alpha}\chi - {\delta v\over 2}\,
\chi^{2} +{\pi V\over N}\,\chi^4 + {\cal O}\bigl(\chi ^{6}/N^{2}\bigr)
\Bigr),
\end{equation}
where we have defined
\begin{equation}
\label{Vdef}
{1\over V} = {1\over N-1}\, \sum_{i=1}^{N-1} {1\over v_{i}}\ccommap
\end{equation}
and
\begin{equation}
\label{defdeltav}
\delta v = {V\over N-1}\, 
\sum_{i=1}^{N-1}\ln {v_{i}\over\Lambda^{2}}\cdotp
\end{equation}
The equation for $\HQ$ is simply $\delta v=0$. Higher derivative
corrections to (\ref{vareqle}) produce terms like $\chi ^2
\partial_{\alpha}\chi \partial_{\alpha}\chi /N$ 
which are irrelevant in the IR. Similarly, we have dropped in
(\ref{seffphi}) irrelevant terms
or order $\chi ^{2k}/N^{k-1}$, $k\geq 3$. In a $1/N$ 
expansion at $\delta v =0$, their contribution is subleading with respect
to the $\chi^{4}/N$ term, to {\it all} orders in $1/N$. Indeed, IR
divergences compensate for the $1/N$ factors coming from the $\chi^{4}/N$ 
interaction, which should thus be considered as giving leading $N^{0}$ 
contributions when $\delta v\rightarrow 0$.
A way to understand this is
to rescale the space-time and momentum variables
\begin{equation}
\label{rescalex}
x=\sqrt{N}\, x'\, ,\quad p= p'/\sqrt{N},
\end{equation}
which eliminates any $N$ dependence in (\ref{seffphi}) on $\HQ$,
\begin{equation}
\label{seffphi2}
S_{\rm eff}[\chi] = \int\! d^2 x' \, \Bigl( {1\over 2}\,
\partial'_{\alpha}\chi \partial'_{\alpha}\chi + \pi V\, \chi^4 + 
{\cal O}\bigl( \chi^{6}/N\bigr) \Bigr).
\end{equation}
The IR properties $p\rightarrow 0$ can now be studied
by working at fixed $p'=p\sqrt{N}$ and using a simple
minded $1/N$ expansion in the $p'$ variables. 
This amounts to perturbing
around the interacting theory (\ref{seffphi2}) where the $\chi^{4}$ term 
is taken into account exactly, and contrary to the
perturbation around the free massless theory this is perfectly
well-defined.

We recognize in (\ref{seffphi}) the Ginzburg-Landau
description of an Ising critical point \cite{LGZad}. The low energy physics
on ${\cal H}_{\rm q}$ ($\delta v=0$) is thus governed by an Ising CFT.
Formula
(\ref{right}) is then simply derived by noting that a deviation from the
critical value $v=\Lambda^2$ corresponds to turning on the energy operator
in the Ising CFT, a dimension one operator. Note that displacements along 
the critical surface correspond to operators that are irrelevant in the 
IR, since the Ising CFT does not have any marginal deformation.
The appearance of a
non-trivial interacting CFT at a point where both a soliton and a
perturbative state become massless is of course very reminiscent of an
Argyres-Douglas CFT \cite{AD}, though in our case it appears to correspond
to a very banal critical theory with a local description (\ref{seffphi}).
Nevertheless, we will see in Section 5 that a natural way to understand the
appearance of this interacting CFT, independently of a large $N$
approximation, is through a description in terms of a more exotic
non-local theory of electric
and magnetic charges (spin waves and vortices). The Ising CFT is then
nothing but a strong coupling fixed point of this non-local theory. This
latter description is more akin to what one may expect in the context of
gauge theories, where simple local description \`a la Ginzburg-Landau may
not exist.

The reader may be concerned at this point by the fact that even though
the standard $1/N$ expansion is not valid in the vicinity of $\HQ$,
the very existence of $\HQ$ itself has
been demonstrated in this simple framework. To investigate the consistency
of the analysis, one should compute the $1/N$ corrections to the equation
for $\HQ$ (\ref{Heq1}), and see if they are small. This calculation, which
illustrates several properties of the $1/N$ expansion, is
presented in Appendix C. The result is that though
we do encounter IR divergences
when computing the corrections to (\ref{Heq1}), they are mild
logarithmic divergences. The only consequence is that the leading
corrections, instead of being of order $1/N$, are of order $(\ln N)/N$. 
This is still a small correction when $N\rightarrow\infty$, as required.

\section{The quantum space of parameters \MQ}

The equation
(\ref{Heq1}) for $\HQ$ is not reliable when a small number of the $v_i$s
goes to zero. Indeed, this equation is obtained in the leading 
$N\rightarrow\infty$ approximation in the form
\begin{equation}
\label{Heq1bis}
{\cal H}_{\rm q}\cap {\cal M}_{N}\ :\quad
{1\over N}\, \sum_{i=1}^{N-1}\ln {v_i\over\Lambda^2} =0.
\end{equation}
Then, when for example $v_{N-1}$ is much smaller than the other $v_i$s, 
the corresponding term in the sum (\ref{Heq1bis}), which is of order $1/N$,
inconsistently becomes larger than the sum of the other terms, which is of
order $N^0$. Naively, one might expect that this problem could be eliminated
by taking into account the $1/N$ corrections to the 
equation for $\HQ$ (\ref{Heq}), but this
is not true: the fact that the large $N$
counting is changed in the limit $v_{N-1}\rightarrow 0$
affects all orders of the $1/N$ expansion used in Section 3.
\subsection{Weak coupling}
Actually, we can readily understand that (\ref{Heq1bis}) is not correct,
even qualitatively, on the whole parameter space. Indeed,
when $v_{N-1}= 0$ and all the other $v_i$s are much larger
than $\Lambda^2$, the low energy physics is governed by a weakly coupled 
${\rm O}(2)$ non-linear $\sigma$ model. This is a {\it massless} phase,
and thus we see that at weak coupling the classical (Figure 1) and 
quantum hypersurface of singularities must coincide. This is unlike the 
prediction based on (\ref{Heq1bis}), see Figure 7.
More precisely, for $v_{N-1}\ll v_i$ and $v_i\gg\Lambda^2$,
$1\leq i\leq N-2$,  
the low energy effective action is reliably determined by a
one-loop perturbative calculation (\ref{beta1l}) to be
\begin{equation}
\label{Slowa}
S_{\rm eff} = {1\over 2 g_{\rm eff}^2}\, \int\! d^2 x\, \Bigl(
\partial_{\alpha}\theta \partial_{\alpha}\theta - {v_{N-1}\over 2}\, \cos
2\theta \Bigr)
\end{equation}
with
\begin{equation}
\label{reff}
{1\over g_{\rm eff}^2} = {1\over 4\pi}\, \ln {\prod_{i=1}^{N-2}
v_{i}\over\Lambda^{2(N-2)}}\cdotp
\end{equation}
As already discussed in Section 2, we get a sine-Gordon model whose coupling
in standard normalizations is
\begin{equation}
\label{betaSG}
\beta_{\rm SG} = 2 g_{\rm eff}.
\end{equation}
It is important to note that, though $\HQ = {\cal H}_{\rm cl}$ at weak
coupling, the classical and quantum low energy physics are drastically
different. Classically, we have a free massless theory independent of
$g_{\rm eff}$. Quantum mechanically, we have the non-trivial CFT of a boson
compactified on a circle of radius $1/g_{\rm eff}$. Even at very small 
coupling,
the physics (anomalous dimensions, etc\dots ) depends on $g_{\rm
eff}$ in a non-trivial way. The fact that the classical and quantum
singularity structure can coincide though the corresponding physics are
different is also seen in supersymmetric gauge theories, for example in
${\cal N}=1$ super Yang-Mills with $N_{\rm f} = N_{\rm c}+1$
\cite{reviewsSW}. 

The description in terms of (\ref{Slowa}) is very useful at small $g_{\rm
eff}$ (for a review on the sine-Gordon model including the original
references, see e.g.~\cite{raja}). We can for example
use (\ref{Slowa}) to compute the mass of the two particle bound state,
\begin{equation}
\label{massbs2}
m_{\rm b} = {4\sqrt{\mathstrut v_{N-1}}\over 
g_{\rm eff}^2}\, \Bigl( 1-{g_{\rm eff}^{2}
\over 2\pi}\Bigr)\, \sin {g_{\rm eff}^2\over
2 \bigl( 1-g_{\rm eff}^2 / 2\pi\bigr)}\cdotp
\end{equation}
This formula should be compared to the analogous formulas (\ref{massbs}) and
(\ref{mbwc2}) which were obtained in different regimes. The mass of the
soliton (\ref{solution2}) is also known,
\begin{equation}
\label{masssolSG}
m_{\rm sol} = {2\sqrt{\mathstrut v_{N-1}}\over g_{\rm eff}^2}\, \Bigl(
1-{g_{\rm eff}^2\over 2\pi}\Bigr)\cdotp
\end{equation}
The action (\ref{Slowa}) also predicts that the bound state becomes unstable
when $g_{\rm eff}^2 \geq 2\pi /3$, and that we have a massless phase for
$g_{\rm eff}^2 > 2\pi$ even when $v_1\not =0$. This is of course
incorrect, and (\ref{Slowa}) is not a good description of the physics at
strong coupling. We know that the non-abelian degrees of freedom
must come into the game and create a mass gap $\Lambda$. Moreover,
the results of Section 3 indicates that for $v_{N-1}\not =0$, there should
be a critical value of $g_{\rm eff}$ at which both the bound state and the
soliton become massless and the low energy theory is an Ising CFT. 
This is consistent with the idea that (\ref{massbs2}) only gives an 
upper bound on the mass of the bound state, which can actually
be significantly lower when the coupling increases due to the
mixing with the other degrees of freedom (see Section 2).
\subsection{Strong coupling}
To investigate the strongly coupled region, the idea is to modify the large
$N$ expansion of Section 3 in order to treat exactly
the fields $\Phi_N$ {\it and}
$\Phi_{N-1}$ is the region of small $v_{N-1}$. This amounts to
integrating out the $N-2$ fields $\Phi_1,\ldots\Phi_{N-2}$ from 
(\ref{neumannlag}), while keeping explicitly $\Phi_N = \sqrt{N}\varphi$ and
$\Phi_{N-1}=\sqrt{N}\varphi '$. 
The effective action, which replaces (\ref{seff2}), is then simply
\begin{eqnarray}
s_{\rm eff}[\alpha ,\varphi,\varphi'] &=& \int\! d^{2}x \left(
{1\over 2}\, \partial_{\alpha}\varphi \partial_{\alpha}\varphi +
{1\over 2}\, \partial_{\alpha}\varphi' \partial_{\alpha}\varphi' +
{\alpha -h_{N}\over 2}\,\varphi^{2}+{\alpha -h_{N-1}\over 2}\,\varphi'^{2}
 - {\alpha\over 2Ng^{2}} \right)\nonumber\\
&&\hskip 19em +{1\over N}\, \sum_{i=1}^{N-2} s[\alpha -h_{i}].\label{seff3}
\\ \nonumber
\end{eqnarray}
We can then repeat straightforwardly the arguments of Sections 3.4 and 3.5. 
The large $N$ low energy effective action at small $v_{N-1}$
coincide with the one-loop result
(\ref{Slowa}) except in a vanishingly small region in parameter space 
corresponding to $\tilde\delta v\ll\Lambda^{2}$ where
\begin{equation}
\label{tdeltav}
\tilde\delta v = {\tilde V\over N-2}\,  \sum_{i=1}^{N-2} \ln 
{v_{i}\over\Lambda^{2}}\ccommap
\end{equation}
with
\begin{equation}
\label{tVdef}
{1\over\tilde V} = {1\over N-2}\, \sum_{i=1}^{N-2} {1\over v_{i}}\cdotp
\end{equation}
In this region, the fluctuations of the radius of the effective target 
space circle become important, and the low energy physics is governed by 
the Ginzburg-Landau action
\begin{equation}
\label{seffAT}
S_{\rm eff}= \int\! d^{2}x\, \Bigl( {1\over 
2}\,\partial_{\alpha}\chi\partial_{\alpha}\chi + {1\over 
2}\,\partial_{\alpha}\chi'\partial_{\alpha}\chi' 
+ {v_{N-1}\over 2}\,\chi'^{2} - 
{\tilde\delta v\over 2}\,\left(\chi^{2} + \chi'^{2}\right)
+ {\pi\tilde V\over N}\, \left( \chi^{2} + 
\chi'^{2}\right)^{2}\Bigr).
\end{equation}
This type of Ginzburg-Landau theory has already been studied \cite{LG}, 
following the ideas of \cite{LGZad}. If $v_{N-1}=0$ and $\tilde\delta v 
>0$, it gives a description of the compactified boson CFT. 
The marginal operator corresponding to a change in the radius is
$\chi^{2}+\chi'^{2}$. We thus get a smooth interpolation with the physics 
described by (\ref{Slowa}). If $v_{N-1}>0$, we have an Ising 
critical point corresponding to the restoration of the ${\mathbb Z}_{2(N)}$
symmetry at $\tilde\delta v=0$, which means that the equation for $\HQ$ is
\begin{equation}
\label{HeqLG}
\HQ\cap {\cal M}_{N}\ :\quad \prod_{i=1}^{N-2} v_{i} 
=\Lambda^{2(N-2)},\quad {\rm for\ } 0<v_{N-1}\ll\Lambda^{2}.
\end{equation}
We thus also obtain a smooth interpolation of the physics described by 
(\ref{seffphi}), but we now see that $\HQ$ must intersect the 
hyperplane $v_{N-1}=0$. 
Similarly, if $v_{N-1}<0$, we have an Ising critical point
corresponding to the restoration of the ${\mathbb Z}_{2(N-1)}$
symmetry at $\tilde\delta v=v_{N-1}$, and thus
\begin{equation}
\label{HeqLG2}
\HQ\cap {\cal M}_{N-1}\ :\quad \prod_{i=1}^{N-2} v_{i} =
\left( \Lambda^{2}e^{v_{N-1}/\tilde V}\right)^{N-2},\quad {\rm for\ } 
-\Lambda^{2}\ll v_{N-1} <0.
\end{equation}
For $v_{N-1}=0$ and $\tilde\delta v=0$, the two Ising CFT found 
for $v_{N-1}>0$ and $v_{N-1}<0$ respectively
are coupled through an ${\rm O}(2)$ 
invariant interaction, giving an Ashkin-Teller critical point. This CFT is 
equivalent to the compactified boson CFT described by (\ref{Slowa}) for 
$v_{N-1}=0$ at $g_{\rm eff}^{2}=\pi /2$, which is itself equivalent to the 
${\mathbb Z}_{2}$ orbifold CFT at the self-dual point \cite{Saleur} (for an 
elementary discussion, see \cite{Ginsparg}). The marginal operator 
corresponding to the variation of the orbifold radius is 
$\chi^{2}\chi'^{2}$ \cite{LG}, but this is not associated to any 
microscopic operator in the Neumann model. For this reason, the 
equivalence with an orbifold theory does not seem to play a particular
r\^ole in our model. What is extremely 
significant, however, is that the value $g_{\rm eff}^{2}=\pi /2$ also 
corresponds to the critical coupling for a Kosterlitz-Thouless phase 
transition \cite{KT}. For $v_{N-1}=0$ and $\tilde\delta v<0$, (\ref{seffAT}) is 
indeed believed \cite{LG} to describe a massive phase, and no longer the 
compactified boson CFT. This is a very interesting physics: the 
creation of the mass gap in our model, which is fundamentally
due to the non-abelian nature of the degrees of freedom,
can be understood as coming from a condensation of vortices in the abelian 
description (\ref{Slowa}). This aspect will be discussed in great details 
in the next Section.

The fact that the transition must occur at $g_{\rm eff}^{2}=\pi /2$, and 
not at $g_{\rm eff}^{2}=\infty$ as predicted by the $N\rightarrow\infty$ 
approximation (\ref{HeqLG}, \ref{reff}), is an exact result that cannot be 
deduced in any simple approximation scheme. Physically, it means that the 
effective target space does not literally shrinks to a point, but that the 
effective radius nevertheless becomes so small that its quantum 
fluctuations are important and any classical geometric 
interpretation of the target space is impossible. This is likely to be 
true anywhere on $\HQ$, and thus for example the endpoint of Figure 6 
should rather be viewed as a very small, but non-vanishing, fluctuating
quantum sphere.
\subsection{The ansatz for the equation of $\HQ$}
\EPSFIGURE{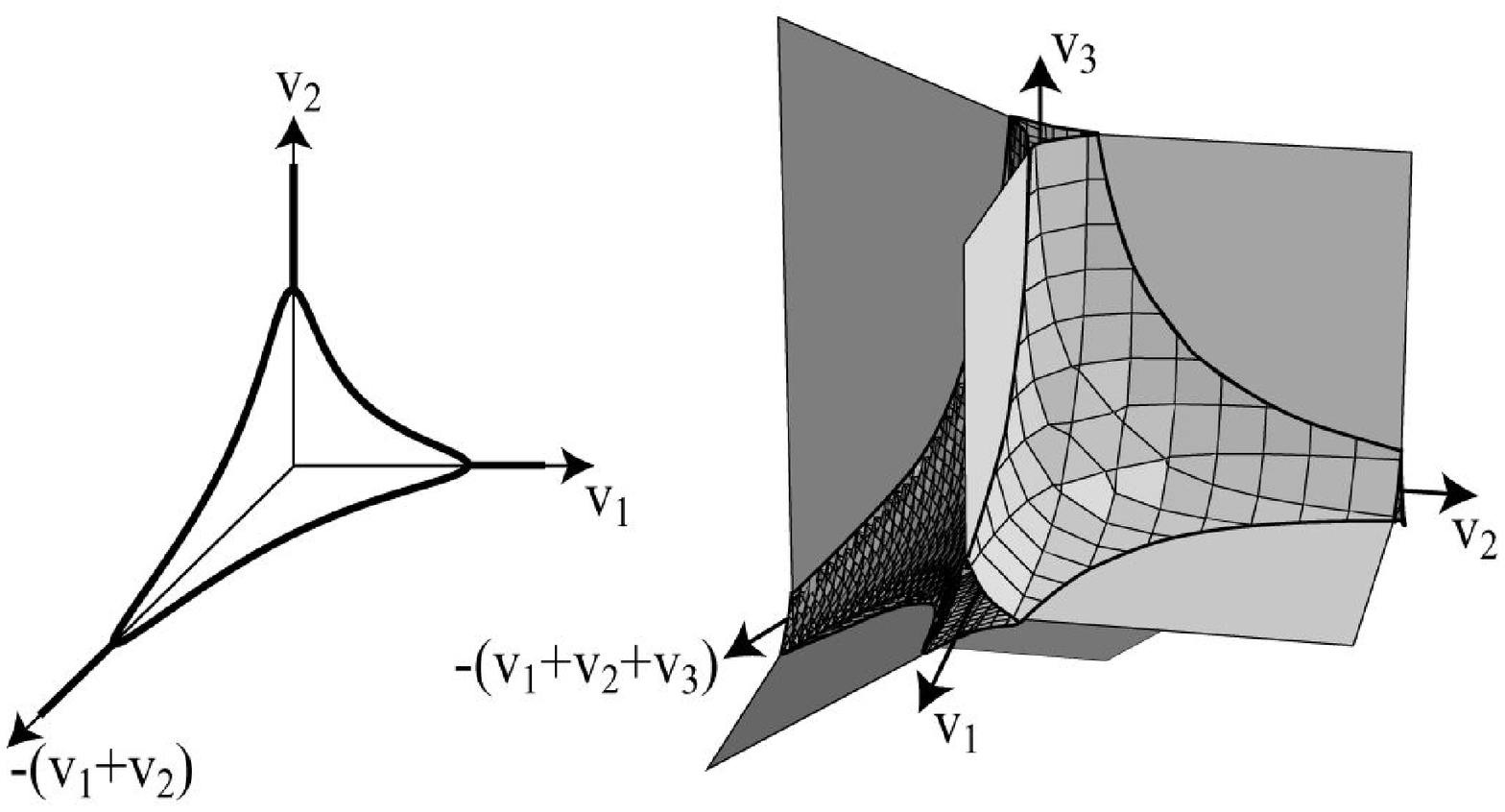,width=15cm}{The quantum space of parameters ${\cal
M}_{\rm q}$ in the cases $N=3$ (left) and $N=4$ (right). We have chosen 
arbitrarily $\tilde\Lambda /\Lambda = 3/2$.
\label{Mquantum}}

We would like to present an equation for $\HQ$ valid in the large $N$ 
limit on the whole parameter space $\cal M$, that interpolates smoothly 
between the different regimes (\ref{Heq1}, \ref{Heqgen1}, \ref{HeqLG},
\ref{HeqLG2}). To get such an equation, 
we clearly need to treat all the regions ${\cal M}_{i}$ (\ref{Mis})
symmetrically. It does not seem to be possible to do this rigorously, for 
reasons explained at lenght above: we need different $1/N$ expansions in 
the different regions ${\cal M}_{i}$, and yet another expansion to 
understand the transition between the different regions. However, there is 
a field, the Lagrange multiplier $\alpha$, that we can treat symmetrically 
on the whole parameter space. In particular, we can compute an 
$N\rightarrow\infty$
effective potential for $\alpha$ by integrating out the $N$ fields 
$\Phi_{i}$ from (\ref{neumannlag}),
\begin{equation}
\label{valpha}
v_{\rm eff}(\alpha) = -{1\over 8\pi N}\, \sum_{i=1}^{N} (\alpha -h_{i})\, 
\ln {\alpha -h_{i}\over e\Lambda^{2}}\cdotp
\end{equation}
The saddle point $\alpha_{*}$ satisfies
\begin{equation}
\label{sadgenalp}
{dv_{\rm eff}(\alpha =\alpha_{*})\over d\alpha} =
-{1\over 8\pi N}\, \sum_{i=1}^{N-1} 
\ln {\alpha_{*} - h_{i}\over\Lambda^{2}} =0.
\end{equation}
It is easy to see that (\ref{sadgenalp}) is consistent with our preceding 
formulas (\ref{wcsad2}, \ref{scsad2}). For example, if we consider the region 
${\cal M}_{N}$ at weak coupling ($\Lambda$ small), (\ref{sadgenalp}) 
implies an expansion $\alpha_{*} = h_{N} + 
\Lambda^{2(N-1)}/\prod_{i=1}^{N-1}v_{i} + \cdots$, and thus 
the corrections to 
(\ref{wcsad2}) go to zero except on a vanishingly small region of ${\cal 
M}_{N}$ when $N\rightarrow\infty$. The advantage of (\ref{sadgenalp}) is 
that it is smooth over the entire $\cal M$. An 
improved equation for $\HQ$ in ${\cal M}_{N}$ is then to use (\ref{saddle2}),
\begin{equation}
\label{provi}
\prod_{i=1}^{N-1}(\alpha_{*}-h_{i})=\Lambda^{2(N-1)},
\end{equation}
but with $\alpha_{*}$ now being determined by (\ref{sadgenalp}).
A natural generalization of (\ref{provi}) on the whole of parameter space is
\begin{equation}
\label{Heqfin}
\HQ\ :\quad
\sum_{i=1}^{N} \prod_{\scriptstyle j=1\atop \scriptstyle j\neq i}^{N}
(\alpha_{*}-h_{j}) = \kappa \Lambda^{2(N-1)} =
\tilde\Lambda^{2(N-1)},
\end{equation}
where we have introduced a ``phenomenological'' constant $\kappa >1$ which,
though it cannot be seen in the $N\rightarrow\infty$ limit,
is necessary to make $\HQ$
intersect with the hyperplanes separating the different regions ${\cal
M}_i$. $\kappa$ can be considered as a non-perturbative 
quantity analogous to the Kosterlitz-Thouless critical coupling
$g_{\rm eff}=\pi /2$ that cannot be calculated in a large $N$ expansion.
The best justification we can find for (\ref{Heqfin}) is by looking a
posteriori to its consequences: by construction, is reproduces the correct
result in the regions ${\cal M}_i$ separately (Section 3.4), 
and it can be checked easily
that it also gives a correct interpolation between the different regions
(Section 4.2).
Though our results have been obtained at large $N$, it is very 
likely that they are qualitatively valid even for $N=3$, and
we have used (\ref{sadgenalp}) and (\ref{Heqfin}) to obtain a sketch
of the quantum space of parameters \MQ\ in the cases $N=3$ and 
$N=4$ in Figure 8.

In our model, it is necessary to cross a surface of singularities to go 
from weak coupling to strong coupling. This is of course not true in
general. For example, 
one may turn on a magnetic field (a $(v,2)$ mass term in the terminology 
of Section 2) in addition to the $(s,2)$ term considered so far. 
This would make the phase transitions first order.

\section{The non-local theory of electric and magnetic charges}
\subsection{General discussion}
The picture of the quantum space of parameters in Figure 8 clearly shows
that we have two qualitatively different regimes in our model. Outside the 
hypersurface of singularities $\HQ$, the usual geometrical 
interpretation of the non-linear $\sigma$ model as a sum over maps from 
${\mathbb R}^{2}$ to $\Sph{N-1}$ is valid. The use of the coordinates on 
the sphere as the fundamental physical degrees of freedom is then 
justified. On the contrary, in the interior of $\HQ$, the 
geometrical interpretation breaks down. It is of course still possible to 
define formally the model as a sum over maps from ${\mathbb R}^{2}$ to 
$\Sph{N-1}$, but the path integral will be dominated by highly singular maps 
from which the image of a smooth target space manifold cannot be 
reconstructed. A very interesting possibility is that there could be a 
(non-local)
change of variables in the path integral to new degrees of freedom that 
would give a natural description of the physics at strong coupling. These 
new degrees of freedom may, or, more probably in our case, may not,
have a geometrical interpretation in terms of 
coordinates on a new target space. In the case of ${\cal N}=2$ 
supersymmetric non-linear $\sigma$ models, this change of variables can be 
argued to exist in many cases and is called the ``mirror map'' 
(see e.g.~\cite{reviewgreene}, \cite{vafamirror}).
In the case of the integrable
${\rm O}(N)$ non-linear $\sigma$ model, which corresponds to the origin of 
our space of parameters (Figure 8), it is actually a long standing 
problem to find these ``good'' degrees of freedom (see e.g.~\cite{polbook}).
When $N=3$, a possibility is to use variables associated with
instantons. In our model, instanton calculations can be done reliably
at weak coupling since the $v_{i}$s provide there an IR cutoff.
However, I think that it is unlikely that instantons, which are smooth 
configurations of the original fields, can account for the physics near 
and in the interior of $\HQ$ (it would be interesting to investigate this 
point, though). Moreover, one would like to have a unified description for 
all $N\geq 3$, and not only for $N=3$, since the physics is qualitatively 
the same for all those cases. It is more natural to suspect
that the relevant degrees of freedom at strong coupling are
more singular configurations. Figure 6 suggests that
configurations of the fields where
\begin{equation}
\label{zerosc}
{\mathbf \Phi} =0
\end{equation}
at some space-time points will be conspicuous. Around such points, 
the angular coordinates on the sphere can vary abruptly. These 
configurations are suppressed classically by the kinetic energy term in the 
action, but at strong coupling we see that they can actually
contribute significantly to the path integral. For example, in 
the case of the two-sphere, we have two coordinates $\theta$ and 
$\phi$ with the identifications
\begin{equation}
\label{identif}
(\theta,\phi)\equiv (\theta +2\pi,\phi),\quad
(\theta,\phi)\equiv (\theta,\phi +2\pi),\quad
(\theta,\phi)\equiv (-\theta,\phi +\pi),
\end{equation}
which suggest that relevant degrees of freedom could be vortices in the 
$\theta$ and $\phi$ variables as well as configurations where $\theta$ has 
square root branch cuts, as in a ${\mathbb Z}_{2}$ orbifold.
Unfortunately,
I do not know how to take into account this three types of defects 
together in an ${\rm O}(3)$ invariant way. Nevertheless, 
we can more modestly look at 
the regime $v_{N-1}\rightarrow 0$ studied in Section 4
where the low energy effective theory is abelian. When $v_{N-1}=0$,
we have demonstrated at large $N$ that the transition from weak (massless 
phase) to strong (massive phase) coupling is of the Kosterlitz-Thouless 
type. It is of course well-known \cite{KT} that such a transition
can be understood as being triggered by a condensation of vortices 
$\theta\equiv\theta + 2\pi$. Let us stress an important non-trivial point 
of principle here. The low energy action at $v_{N-1}=0$,
\begin{equation}
\label{O2}
S_{\rm eff} = {1\over 2g_{\rm eff}^{2}} \int\! d^{2}x\, 
\partial_{\alpha}\theta \partial_{\alpha}\theta \, ,
\end{equation}
does {\it not} by itself predicts a Kosterlitz-Thouless transition at 
$g_{\rm eff}^{2}=\pi /2$, but rather a massless phase at any $g_{\rm 
eff}$. The action (\ref{O2}) is often used as a continuum version of the 
${\rm O}(2)$ spin model defined on a lattice, $g_{\rm eff}^{2}$ playing 
the r\^ole of a temperature. The lattice model is the object of interest in 
condensed matter physics, and was studied by Kosterlitz and Thouless 
\cite{KT}. In the lattice hamiltonian, configurations for which 
$\theta\rightarrow\theta +2\pi$ abruptly are not suppressed;
the suppression is an artefact 
of the continuum description given by (\ref{O2}). By introducing the 
vortices by hand in the continuum formulation (which means that
we choose a non-zero 
fugacity for the singular field configurations corresponding to vortices in 
the definition of the path integral), one can get a good 
modeling of the lattice model \cite{KT}, and in particular predict the 
mass gap at high temperature and the correct behaviour at the transition.
In our case, however, we are not dealing with lattice models, but 
with a bona fide field theory. What has been proven in the large 
$N$ limit in Section 4 is that a non-zero fugacity for the vortices is {\it 
dynamically} generated in the model because of the non-abelian 
(continuous) degrees of freedom. The physical origin of the condensation 
of vortices is thus completely different in our case and in the well-known 
abelian lattice model. Note that the vortices are not instantons of the 
non-abelian model and can exist for all $N$. This suggests that, more 
generally, the ${\rm O}(N)$ model may be naturally described in terms of 
``non-abelian'' generalized vortices (\ref{identif}).

In the following, we are going to study how the standard picture of vortex 
condensation is modified when we go to the regime where $v_{N-1}$ is small 
but non-zero. In Section 3 and 4 we obtained in the large $N$ limit 
a simple Ginzburg-Landau description of the physics on $\HQ$. We do not 
expect to get as simple a description in the present context. In general, 
both weak coupling and strong coupling degrees of freedom can be relevant 
in the vicinity of $\HQ$, where the transition between the 
weak coupling and strong coupling behaviours takes place, giving an 
unusual non-local description of the low energy physics. The same kind on
physics has of course already been studied in condensed matter works on
abelian lattice models. A $\cos p\theta$ potential in the lagrangian is
interpreted in this context as a $p$-fold symmetry breaking term modeling
crystal anisotropy. When $p\geq 4$, a perturbative analysis \`a la
Kosterlitz is possible (\cite{Kada}, \cite{RGCoulomb}), but when $p\leq 3$,
and in particular in the case $p=2$ we are interested in (\ref{Slowa}), 
this is not possible as we will review. Even in those cases, the nature of
the physics is rather clear from the lattice point of view in particular
(see the Appendix of \cite{isinga} or \cite{isingb}). 
The general ideas presented later in this Section
are thus certainly not new, though a
comprehensive analysis of the kind we offer does not seem to have appeared.
Our main goal is really to emphasize the striking similarity between our
non-supersymmetric system and the
physics in the vicinity of an Argyres-Douglas point in supersymmetric gauge
theories. Whether similar things can happen in non-supersymmetric four
dimensional theories is of course still an open problem.

Before we turn to these points, let us make a simple heuristic remark.
In our model, we have a clear interpretation of the fact that a compactified
boson can be equivalent to the ${\rm O}(2)$ symmetric Ashkin-Teller model, 
and that this happens precisely at the Kosterlitz-Thouless coupling. One 
may then wonder whether a natural explanation for the fact that these 
CFTs are also equivalent to the ${\mathbb Z}_{2}$ orbifold at the self-dual 
radius could emerge. A naive argument in our context could be the
following: since the Kosterlitz-Thouless transition is
triggered by the non-abelian degrees of freedom, the angle $\theta$ should
know somehow that is lives on a sphere, and an orbifold-like
identification $\theta\equiv -\theta$ 
should then be implemented (\ref{identif}).
However, the orbifold of the circle theory at the Kosterlitz-Thouless 
radius {\it is not} the orbifold at the self-dual radius. The naive
interpretation thus does no make sense as it stands.
It would be interesting to 
try to include the effect of the other $\Phi$ coordinates on the sphere to 
see if this picture might be made consistent.
\subsection{The non-local description in the abelian regime}
A general configuration of the field $\theta$, including 
vortices, can be written
\begin{equation}
\label{thetavort}
\theta (x) = \theta _{0}(x)+\theta _{v}(x)
\end{equation}
where $\theta_{0}$ is the non-singular part and
\begin{equation}
\label{thetavort2}
\theta_{v}(x) = \sum_{i}n_{i} \arg (x-x_{i})
\end{equation}
corresponds to vortices (``magnetic charges'')
of charge $n_{i}$ centered at the points $x_{i}$.
When the total charge $\sum n_i$ is zero, the action (\ref{Slowa}) can be
finite and is given by
\begin{eqnarray}
S_{\rm eff}(n_i;x_i) &=& {1\over 2g_{\rm eff}^{2}}\, \int\! d^{2}x\,
\partial_{\alpha}\theta_{0}\partial_{\alpha}\theta_{0} - 
{\pi\over g_{\rm eff}^{2}}\, \sum _{i,j} n_{i}n_{j}\, \ln |x_{i}-x_{j}|
\nonumber\\ && \hskip 1em
-{v_{N-1}\over 8g_{\rm eff}^{2}}\,\int\! d^{2}x\, \Biggl[
\prod_{i}\left( {z-z_{i}\over \bar z -\bar z_{i}} \right)^{n_{i}} \,
e^{2i\theta_{0}} + \prod_{i}\left( {\bar z-\bar z_{i}\over 
z-z_{i}}\right)^{n_{i}}\, e^{-2i\theta_{0}} \Biggr].\label{seffvort}\\
\nonumber
\end{eqnarray}
We recognize the standard Coulomb interaction between magnetic charges in
two dimensions, but we have also a complicated interaction term between the
magnetic charges and the fundamental spin waves (``electric charges''). 
We have noted $z_i = x_i^1 + ix_i^2$ and $\bar z_i = x_i^1 -i x_i^2$. 
The charges $n_i=\pm 1$ are the most relevant, as can be checked
straightforwardly along the lines of the calculation presented below. We will
thus write the partition function in term of the dynamically generated
fugacity $F= f \Lambda^2$ ($f$ is a dimensionless constant) for the $n_i=\pm
1$ vortices as
\begin{equation}
\label{partv1}
Z = \sum_{n=1}^{\infty} {f^{2n} \Lambda^{4n}\over (n!)^2}\,
\int\!\prod_{i=1}^{n} \bigl(d^{2}x_{i}d^{2}y_{i}\bigr) \,
\int\! {\cal D}\theta_0 \, e^{-S_{\rm eff}[\theta_0;x_i,y_i]},
\end{equation}
where the $x_i$s and $y_i$s parametrize the positions of the charge $+1$ and
charge $-1$ vortices respectively, and $S_{\rm eff}[\theta_0;x_i,y_i]$
is the action (\ref{seffvort}) for such a configuration. By expanding the
exponential in powers of $v_{N-1}$ and calculating the path integral over
$\theta_0$, we get
\vfill\break
\begin{eqnarray}
Z &=& \sum_{m,n=0}^{\infty} {\bigl(v_{N-1}/ 2g_{\rm eff}^{2}\bigr)^{2m}
\over (m!)^{2}}\,
{\bigl(f\Lambda^{2}\bigr)^{2n}\over (n!)^{2}} \, \int\!
\prod_{p=1}^{m}\bigl(d^{2}u_{p}d^{2}w_{p}\bigr) 
\prod_{i=1}^{n} \bigl(d^{2}x_{i}d^{2}y_{i}\bigr)
\nonumber\\
&&\hskip 6em
\prod_{1\leq p<q\leq m} \left( |u_{p}-u_{q}||w_{p}-w_{q}|\right)^{2g_{\rm 
eff}^{2}/\pi} \, \prod_{1\leq p,q\leq m} |u_{p}-w_{q}|^{-2g_{\rm 
eff}^{2}/\pi} \nonumber\\
&&\hskip 6em
\prod_{1\leq i<j\leq n}\left( |x_{i}- x_{j}||y_{i}-y_{j}|\right)^{2\pi 
/g_{\rm eff}^{2}}\, \prod_{1\leq i,j\leq n} |x_{i}-y_{j}|^{-2\pi /g_{\rm 
eff}^{2}} \nonumber\\
&&\hskip 6em
\prod_{{\scriptstyle 1\leq i\leq n\atop\scriptstyle 1\leq p\leq m}}
{(u_{p}-x_{i})(w_{p}-y_{i})(\bar u_{p}-\bar y_{i})(\bar w_{p}-\bar x_{i}) 
\over 
(\bar u_{p}-\bar x_{i})(\bar w_{p}-\bar y_{i})(u_{p}-y_{i})(w_{p}-x_{i})}
\cdotp\label{Zvort1}\\ \nonumber
\end{eqnarray}
The second line in (\ref{Zvort1}) corresponds to the standard mass expansion
of the original sine-Gordon model (\ref{Slowa}), the third line corresponds
to the interaction between vortices, which is also described by a
sine-Gordon model of the type (\ref{Slowa}) but with a coupling
$g_{\rm eff}' = \pi /g_{\rm eff}$, and the fourth line corresponds to the
interactions between the electric and magnetic charges. This partition
function is reproduced, including the electric/magnetic interactions,
by the perturbative formula
\begin{equation}
\label{Zvort2}
Z = \bigl\langle e^{-S_{\rm int}} \bigr\rangle _{\rm free}
\end{equation}
where the average is computed with the free field weight (\ref{O2}) and
\begin{equation}
\label{Sinteract}
S_{\rm int} = -\int\! d^{2}x\,
\Biggl( {v_{N-1}\over 4g_{\rm eff}^{2}}\, \cos 2\theta + 
2f\Lambda^{2}\, \cos {2\pi\over g_{\rm eff}^2}\,\tilde\theta\Biggr).
\end{equation}
The dual field $\tilde\theta$ is defined by
\begin{equation}
\label{dualfield}
d \tilde\theta = -i*\! d\theta.
\end{equation}
The formula (\ref{Zvort2}) for $Z$ is perturbative, since it involves an
average over free fields. Interestingly,
an off-shell, non-perturbative, formulation of this
type of theories exists \cite{tseyt}. The basic idea is to treat the fields
$\theta$ and $\tilde\theta$ as independent variables, and try to find an
action whose equations of motions give (\ref{dualfield}) when the
interactions are turned off. 
By writing (\ref{dualfield}) in the minkowskian,
\begin{equation}
\label{dualfield2}
\partial_0\tilde\theta = -\partial_1\theta ,\quad
\partial_1\tilde\theta = -\partial_0\theta ,
\end{equation}
we see that $\partial_1\tilde\theta$ is in some sense the canonical
momentum associated with $\theta$. The phase space path integral then
suggests to try the action \cite{tseyt}
\begin{equation}
\label{nonloc}
S_{\rm em} = -{1\over 2g_{\rm eff}^2}\,\int\! d^2x\,
\Bigl( \partial_0\theta\partial_1\tilde\theta + 
\partial_0\tilde\theta\partial_1\theta + 
\bigl( \partial_1\theta\bigr)^2 + \bigl( \partial_1\tilde\theta\bigr)^2
\Bigr) + S_{\rm int}[\theta,\tilde\theta]
\end{equation}
which indeed yields the correct equations of motion.
If $S_{\rm int}$ depends on $\theta$ only, then by integrating out
$\tilde\theta$ we would recover the standard Lorentz invariant
action for a scalar, but in general it is impossible to write
down a manifestly Lorentz invariant action
for both $\theta$ and its dual. Operators associated with a
dyon of electric charge $n_e$ and magnetic charge $n_m$ are
\begin{equation}
\label{opvort}
{\cal O}_{(n_e,n_m)} = e^{\pm i(n_e\theta +
2\pi n_m \tilde\theta /g_{\rm eff})}.
\end{equation}
We thus see that our model (\ref{Sinteract}) corresponds to the case where
both a magnetic charge (the vortex)
$(n_e=0,n_m=1)$ and an electric charge $(n_e=2,n_m=0)$ are included. By
rescaling the fields
\begin{equation}
\label{resfields}
\theta = g_{\rm eff}\, \phi,\quad\tilde\theta = g_{\rm eff}\, \tilde\phi
\end{equation}
and defining
\begin{equation}
\label{memndef}
F_e = {v_{N-1}\over 2g_{\rm eff}^2}\,\ccommap\quad F_m=4f\Lambda^2
\end{equation}
we can bring the action in a form
\begin{equation}
\label{Snloc}
S_{\rm em} = -{1\over 2}\int\! d^2x\,
\Bigl( \partial_0\phi\partial_1\tilde\phi +
\partial_0\tilde\phi\partial_1\phi +
\bigl( \partial_1\phi\bigr)^2 + \bigl( \partial_1\tilde\phi\bigr)^2
+ F_e\, \cos \bigl( 2g_{\rm eff}\phi\bigr) + F_m\,\cos
\bigl( 2\pi\tilde\phi /g_{\rm eff}\bigr)
\Bigr) 
\end{equation}
which is manifestly invariant under a strong/weak coupling $S$ duality,
\begin{equation}
\label{S}
g_{\rm eff} \mathop{\longleftrightarrow}^{\displaystyle S}
{\pi\over g_{\rm eff}}\,\ccommap\quad
F_e \mathop{\longleftrightarrow}^{\displaystyle S} F_m\, ,\quad 
\phi \mathop{\longleftrightarrow}^{\displaystyle S} \tilde\phi.
\end{equation}
Note that $S$ is not the usual $T$ duality that would exchange the unit
magnetic charge with the unit electric charge and $g_{\rm eff}$ with
$2\pi /g_{\rm eff}$, in the same way as the monodromy at an Argyres-Douglas 
point is not the full duality group. 

The quantization of (\ref{Snloc}) can be done in the standard way. 
We have two second class constraints, 
\begin{equation}
\label{constraints}
\Pi = -{1\over 2}\,\partial_1 \tilde\phi ,\quad
\tilde\Pi = -{1\over 2}\, \partial_1 \phi,
\end{equation}
and the Dirac bracket yields the equal time
commutation relation
\begin{equation}
\label{quantization}
\left[ \phi (x^0,x^1),\tilde\phi (x^0,x'^{1})\right] =i\,
\Theta (x^1 - x'^{1})
\end{equation}
where $\Theta$ is the Heavyside step function. We see that the non-locality
of our theory, which describes the interaction of electric and magnetic
degrees of freedom, is simply encoded in this commutation relation.
Let us note finally
that actions describing both electric and magnetic charges in
four dimensions, and that should play a r\^ole in understanding
Argyres-Douglas CFTs, have been constructed along the lines of (\ref{Snloc})
\cite{SSen}.
\subsection{The strongly coupled fixed point \`a la Argyres-Douglas}
The standard way to treat (\ref{Snloc}) would be to use perturbation 
theory in $F_{e}$ and $F_{m}$ (\cite{Kada}, \cite{elecmag}). This makes 
sense when the corresponding operators are nearly marginal. What are we 
expecting in our case? From the results of Section 4, or from the intuition 
gained in lattice model, we would like to show that (\ref{Snloc}) 
has an Ising fixed point. If we have only one fixed point, which is 
likely, it must occur at the $S$-invariant point
\begin{equation}
\label{sdcoupling}
g_{\rm eff}=g_{*}=\sqrt{\pi}, \quad F_e=F_m.
\end{equation}
On the other hand, the perturbative
dimensions of the ``electric'' and ``magnetic'' operators are respectively
\begin{equation}
\label{dimop}
\Delta_{\rm elec} = {g_{\rm eff}^{2}\over\pi}\ccommap\quad
\Delta_{\rm mag} = {\pi\over g_{\rm eff}^{2}}\ccommap
\end{equation}
which would correspond at the fixed point to
\begin{equation}
\label{dimopfp}
\Delta_{\rm elec*} = \Delta_{\rm mag*} = 1,
\end{equation}
values far below the threshold of marginality $\Delta =2$ where perturbation 
theory is valid. This means that the fixed point
we are looking for must be at strong coupling. This is also what is 
expected in four dimensions, and in general it would imply, even in two 
dimensions, that the theory in not tractable. Luckily, in the very 
particular case at hand, we have a powerful tool at our disposal that can 
solve the problem: we can fermionize and replace the scalar $\phi$ by a Dirac
spinor $\psi$. Fermionizing is of course a very natural 
thing to do since we are looking for an Ising critical point. Moreover, 
the self-dual dimensions (\ref{dimopfp}) strongly suggest that the electric 
and magnetic operators can be viewed as fermion mass terms near the fixed 
point. Fermionization is a bit unusual in our case, however, because 
the ``topological'' symmetry $\tilde\phi\rightarrow\tilde\phi + {\rm
constant}$ is broken by the interaction term in (\ref{Snloc}) and thus we do
not have a conserved fermion current $j\propto *d\phi$ (in other words,
the relations (\ref{dualfield}) or (\ref{dualfield2}) are no longer correct
due to the interactions between $\phi$ and $\tilde\phi$). The perturbative
reasoning of Coleman \cite{coleman} can nevertheless be reproduced
straightforwardly, starting from (\ref{Zvort2}). 
In the same way as the sine-Gordon potential introduces
Dirac mass terms, the $\cos (2\pi\tilde\phi /g_{\rm eff})$ term introduces
Majorana mass terms. More precisely, we can identify 
\begin{equation}
\label{ferm1}
\mu\,\cos\bigl( 2g_{\rm eff}\phi\bigr) = -\pi \, \bar\psi\psi ,\quad
\mu\,\cos\bigl( 2\pi\tilde\phi /g_{\rm eff}\bigr) = -\pi\, 
\bar\psi^C \psi,
\end{equation}
where we have introduced an arbitrary renormalization scale $\mu$.
The theory (\ref{Snloc}) is then equivalent to the fermionic theory with
lagrangian
\begin{equation}
\label{fermS1}
S_{\rm ferm} = \int\! d^2x\, \Bigl( i
\bar\psi \gamma^{\mu}\partial_{\mu}\psi +
{\pi F_e\over\mu}\, \bar\psi\psi + {\pi F_m\over\mu}\, \bar\psi^C \psi -
{G\over 2}\, \bar\psi\gamma^{\mu}\psi\, \bar\psi\gamma_{\mu}\psi\Bigr)
\end{equation}
with
\begin{equation}
\label{coupl}
{G\over\pi} = {\pi -g_{\rm eff}^2\over g_{\rm eff}^2}\cdotp
\end{equation}
It is useful to decompose the Dirac fermion $\psi$ in terms of two Majorana
fermions $\lambda=-\lambda^C$ and $\chi=-\chi^C$ such that
\begin{eqnarray}
\psi &=& \pmatrix{\psi_-\cr\psi_+\cr} = {\lambda + i\chi\over\sqrt{2}}=
{1\over\sqrt{2}}\,
\pmatrix{\lambda_- + i\chi_-\cr \lambda_+ +i\chi_+\cr}\nonumber\\
\bar\psi &=& (\bar\psi_{+},\bar\psi_{-}) = {\bar\lambda 
-i\bar\chi\over\sqrt{2}} = {1\over\sqrt{2}}\, (\lambda_{+}-i\chi_{+}, 
\lambda_{-}-i\chi_{-}).\label{decomp}\\ \nonumber
\end{eqnarray}
In terms of these new variables, and by introducing the light-cone
coordinates $x^{\pm} = x^0 \pm x^1$ and
\begin{equation}
\label{newvar}
M = {\pi (F_e - F_m)\over\mu}\ccommap\quad
M'= {\pi (F_e + F_m)\over\mu}\ccommap
\end{equation}
we have
\begin{eqnarray}
\hskip -2em S_{\rm ferm} &=& \int\! d^2x\, \Bigl(
\lambda_+\partial_-\lambda_+ + \chi_+\partial_-\chi_+ -
\lambda_-\partial_+\lambda_- - \chi_-\partial_+\chi_-\nonumber\\
&&\qquad -iM'\, \chi_-\lambda_+ -
iM\, \chi_+\lambda_- -
{\pi (\pi -g_{\rm eff}^2)\over 2g_{\rm eff}^2} \,
\lambda_-\chi_-\lambda_+\chi_+ \Bigr).\label{fermS2}\\ \nonumber
\end{eqnarray}
We thus see that at the self-dual point (\ref{sdcoupling}), for which
$M=0$, we have at low
energy a free Majorana fermion $(\lambda_-,\chi_+)$, which indeed describes
an Ising critical point as was to be shown. The $S$ duality (\ref{S}) acts
as
\begin{equation}
\label{S2}
M\mathop{\longleftrightarrow}^{\displaystyle S} -M,\quad
M'\mathop{\longleftrightarrow}^{\displaystyle S} M',\quad
\chi_{\pm}\mathop{\longleftrightarrow}^{\displaystyle S}\mp\chi_{\pm},\quad
\lambda_{\pm}\mathop{\longleftrightarrow}^{\displaystyle S}\lambda_{\pm},
\end{equation}
and thus reduces at low energy to the Kramers-Wannier duality of the Ising
model.

I would like to conclude this Section by giving a more rigorous derivation
of (\ref{fermS1}) and explaining at the same time
a point that might puzzle the reader. The puzzle is the following.
The duality (\ref{S}) should be valid not only at the fixed point,
but also for any coupling $g_{\rm eff}\not =g_*$. Using (\ref{S}) and
(\ref{S2}), the four-fermions
interaction term in (\ref{fermS1}) naively transforms under $S$ as
\begin{equation}
\label{S3}
G\, \bar\psi\gamma^{\mu}\psi\,\bar\psi\gamma_{\mu}\psi
\mathop{\longleftrightarrow}^{\displaystyle S}
{G\over 1+G/\pi}\, \bar\psi\gamma^{\mu}\psi\,\bar\psi\gamma_{\mu}\psi.
\end{equation}
This is clearly inconsistent, since by 
repeating the transformation (\ref{S3}), we
could make the coupling arbitrarily small. The subtlety comes from the
definition of the fermion current from which the
four-fermions interaction is constructed \cite{current}. 
The canonical definition, that must be used for example in a Hamiltonian
approach, uses a point-splitting regularization {\it at equal time}.
However, such a definition 
does not produce a Lorentz vector. A correct definition of the current must
actually involve both a space-like and a time-like splitting. There is a
one-parameter family of Lorentz covariant
current that can be defined in that way as we will review below.
Products of fields taken at time-like intervals depend on the dynamics of
the theory. The Lorentz invariant
four-fermions interaction must then depend in some hidden
way on the coupling constant $G$, and this invalidates (\ref{S3}).
Since we don't want to repeat the tedious
arguments of \cite{current}, we are going to show much more
straightforwardly how this comes about by giving at the same time
a rigorous non-perturbative derivation of
(\ref{fermS1}) starting from (\ref{Snloc}).
We will consider the hamiltonian in the Schr\"odinger picture,
\begin{equation}
\label{hamil}
H = {1\over 2}\, \Biggl[
\Biggl( {d\phi\over dx}\Biggr)^2 + \Biggl( {d\tilde\phi\over dx}
\Biggr)^2+F_e\, \cos\bigl( 2g_{\rm eff}\phi\bigr) + F_m\,\cos\bigl(
2\pi\tilde\phi /g_{\rm eff}\bigr)\Biggr],
\end{equation}
with the quantization condition (\ref{quantization}), and apply the
bosonization/fermionization formulas on $H$. In the Schr\"odinger picture, 
all operators are of course regularized by point splitting at equal time.
Fermions can be defined by 
using any scalar fields $X$ and $\tilde X$ satisfying the same quantization 
condition as $\phi$ and $\tilde\phi$ (\ref{quantization}) by
\begin{equation}
\label{bosferm}
\psi_- = \sqrt{{\mu\over 2\pi}}\, e^{i\sqrt{\pi}( \tilde X+X)},\quad
\psi_+ = \sqrt{{\mu\over 2\pi}}\, e^{i\sqrt{\pi}( \tilde X-X)}.
\end{equation}
For our purposes, we need
\begin{equation}
\label{Xdef}
X = {g_{\rm eff}\over\sqrt{\pi}}\, \phi\, ,\quad
\tilde X = {\sqrt{\pi}\over g_{\rm eff}}\, \tilde\phi\, .
\end{equation}
The current $j^{\mu} = \bar\psi\gamma^{\mu}\psi$ is
\begin{equation}
\label{currentbos}
j^{0}=i(\psi_{+}\bar\psi_{+}-\psi_{-}\bar\psi_{-}) = {g_{\rm 
eff}\over\pi}\, {d\phi\over dx}\,\ccommap\quad
j^{1}=-i(\psi_{+}\bar\psi_{+} + \psi_{-}\bar\psi_{-}) = {1\over g_{\rm 
eff}}\, {d\tilde\phi\over dx}\, \ccommap
\end{equation}
and this is not a Lorentz vector. The four-fermions interaction is then
defined by
\begin{equation}
\label{ffint}
\bar\psi\gamma^{\mu}\psi\, \bar\psi\gamma_{\mu}\psi \equiv
q_{0}\bigl( j^{0}\bigr)^{2} - q_{1} \bigl( j^{1} \bigr)^{2}\, ,
\end{equation}
where $q_{0}$ and $q_{1}$ are parameters that we will adjust later to make 
the interaction a Lorentz scalar. It is straightforward to check that 
the fermion hamiltonian
\begin{eqnarray}
H_{\rm ferm} &=& \bar\psi_{-} {d\psi_{-}\over dx} + \bar\psi_{+} 
{d\psi_{+}\over dx} + {\pi F_{e}\over\mu}\, \Bigl( \psi_{-}\bar\psi_{+} + 
\psi_{+}\bar\psi_{-}\Bigr) + {\pi F_{m}\over\mu}\, \Bigl( 
\psi_{-}\psi_{+} + \psi_{+}\psi_{-}\Bigr)\nonumber\\
&& \hskip 18em +{G\over 2}\, \Bigl(
q_{0}\bigl( j^{0}\bigr)^{2} - q_{1} \bigl( j^{1} \bigr)^{2} \Bigr)
\label{hferm}\\ \nonumber
\end{eqnarray}
is equivalent to the bosonic hamiltonian
\begin{eqnarray}
H_{\rm bos} &=& {1\over 2}\, \Biggl[ {g_{\rm eff}^{2}\over\pi} \, \Biggl( 
1+{Gq_{0}\over\pi}\Biggr) \Biggl( {d\phi\over dx}\Biggr)^{2} +
{\pi\over g_{\rm eff}^{2}}\, \Biggl( 1-{Gq_{1}\over\pi}\Biggr)
\Biggl( {d\tilde\phi\over dx}\Biggr)^{2} \Biggr]\nonumber\\
&&\hskip 12em
+ F_{e}\, \cos\bigl( 2g_{\rm eff}\phi\bigr) + F_{m}\, \cos\bigl(
2\pi\tilde\phi /g_{\rm eff}\bigr).\label{hb}\\ \nonumber
\end{eqnarray}
The four-fermions interaction will be a Lorentz scalar if the $G\not =0$ 
theory is Lorentz invariant with the same speed of light $c=1$ as the free 
$G=0$ fermionic theory. This in turn is equivalent to the condition that 
$H_{\rm bos}$ coincide with H (\ref{hamil}), which happens for
\begin{equation}
\label{Lorentz}
q_{1} = {g_{\rm eff}^{2}\over\pi}\, q_{0}
\end{equation}
and
\begin{equation}
\label{couplgen}
{G\over\pi} = {1\over q_{0}}\, {\pi - g_{\rm eff}^{2}\over g_{\rm eff}^{2}}
\cdotp
\end{equation}
Equation (\ref{Lorentz}) shows that we have one free parameter $q_{0}$ in 
defining the scalar
four-fermions interaction, and also that this interaction 
depends implicitly on $g_{\rm eff}$ through $q_{1}$. Equation 
(\ref{couplgen}) generalizes the standard Coleman's formula (\ref{coupl}) 
which corresponds to the choice $q_{0}=1$. Of course the physics does not 
depend on $q_{0}$. We can now deduce the correct
duality transformations of the 
four-fermions term. Equations (\ref{ffint}), (\ref{Lorentz}) and 
(\ref{couplgen}) implies
\begin{equation}
\label{ffintdef}
G\, \bar\psi\gamma^{\mu}\psi\, \bar\psi\gamma_{\mu}\psi =
{\pi (\pi -g_{\rm eff}^{2})\over g_{\rm eff}^{2}}\, \Bigl(
\bigl( j^{0} \bigr)^{2} - {g_{\rm eff}^{2}\over\pi}\, \bigl( j^{1}\bigr)^{2}
\Bigr).
\end{equation}
Equations (\ref{currentbos}), (\ref{decomp}) and (\ref{S2}) show that
\begin{equation}
\label{Stranscurrent}
j^{0}\mathop{\longleftrightarrow}^{\displaystyle S} j^{1},
\end{equation}
and thus by using (\ref{S}) we finally deduce that the four-fermions term is 
actually {\it invariant} under the $S$ duality.

It would be interesting to study possible strongly coupled fixed points
in theories of the same kind as (\ref{Snloc}), 
but including other electric and magnetic
charges, multi-components fields, and $\theta$ angles
\cite{phasetheta}. The fact that the Argyres-Douglas CFTs appearing in four
dimensional supersymmetric gauge theories admit an ADE classification
\cite{ADE} suggests that all the minimal unitary CFTs, that also follow the
ADE pattern \cite{zuber}, could be obtained in
this way. A very simple example is the case where the electric charge
$n_e=2$ of our model is replaced by an electric charge $n_e=3$. This yields
a strongly coupled fixed point of the Potts type.

\section{The double scaling limits}

We have discussed three different descriptions of the non-trivial
low energy physics appearing in our model near the singularities of
parameter space. The Ginzburg-Landau description (Sections 3.5 and 4.2) 
and the fermionic description (Section 5.3) are very simple, and can exist
because of peculiarities of two-dimensional physics. The third description
(Section 5.2), in terms of a theory of electric and magnetic charges, is a
priori more generic, and I have emphasized the strong similarity with
Argyres-Douglas CFTs in four dimensions in particular. We would like now to
derive a fourth description, whose counterpart in four dimensions would be a
string theory description of the vicinity of an Argyres-Douglas CFT (or more
generally of the vicinity of generic non-trivial critical points that can be
found in the context of gauge theories with Higgs fields). We will be rather 
brief on this interesting aspect of our model here, since we hope to provide
more details in \cite{ds}. We are simply going to show 
explicitly that a double scaling limit in the sense of the ``old'' matrix 
models \cite{BK} can be defined when we approach the Ising or Ashkin-Teller
singularities on $\HQ$, 
and that the double scaled theory coincide with the low energy theory 
in the vicinity of $\HQ$. This provides a duality between the interacting 
theory describing the low energy degrees of freedom and a theory of 
randomly branched polymers, as explained in \cite{polym} and as will be 
reviewed in \cite{ds}. The discussion is similar to the case of the 
ordinary vector models, as reviewed for example by J.~Zinn-Justin 
in \cite{revlargeN}.

There are three different types of critical behaviour occuring in our 
model (Ising, Ashkin-Teller and compactified boson), and thus a priori 
three different types of double scaling limits can be considered. Let us 
first discuss the vicinity of the Ising critical point. As discussed in 
Section 3.5, when we approach the critical hypersurface $\HQ$, $\delta 
v\rightarrow 0$, the standard $1/N$ expansion in plagued by IR divergences. 
This means that in the large $N$ expansion of physical quantities, like 
(\ref{vacexp}), the coefficients $W_{l}$ will diverge as $\delta 
v\rightarrow 0$. Proving that a double scaling limit can be defined when 
$\delta v\rightarrow 0$ amounts to proving that these divergences have a 
specific form for any $l$ such that they can be compensated for by taking 
$N\rightarrow\infty$ and $\delta v\rightarrow 0$ in a correlated way. 
The double scaled theory then gets contributions from all orders in $1/N$. 
This means in particular that we must take into account the renormalization 
of the original non-linear $\sigma$ model to all orders. Remarkably 
enough, this can be done very simply. The crucial 
point is of course that the divergences in $W_{l}$ are due to IR effects. 
\vfill\eject
We have shown in Section 3.5 that at momenta $p\ll\Lambda$, the physics is 
correctly described by the action (\ref{seffphi}). By using the rescaled 
variables (\ref{rescalex}), we deduce that physical quantities can be 
correctly described by the action
\begin{equation}
\label{sdscale}
S[\chi] = \int\! d^2 x' \, \Bigl( {1\over 2}\,
\partial_{\alpha}'\chi \partial_{\alpha}'\chi - {N \delta v\over 2}\,
\chi^{2} +\pi V\,\chi^4 \Bigr)
\end{equation}
for momenta
\begin{equation}
\label{momentads}
p' \ll \Lambda \sqrt{N}.
\end{equation}
If we now take the limit $N\rightarrow\infty$, the effective UV
cut-off of (\ref{sdscale}), which is of order 
\begin{equation}
\label{effuv}
\Lambda_{0,{\rm eff}} \sim \Lambda\sqrt{N},
\end{equation}
goes to infinity. The action (\ref{sdscale}) thus needs to be renormalized
accordingly, in order to get a finite limit. This is done with the help of 
the standard normal ordering,
\begin{equation}
\label{normalorder}
\chi ^{4} = :\!\chi^{4}\! : + 6\, {1\over 2\pi}\ln {\Lambda_{0,{\rm 
eff}}\over\Lambda}\, \chi^{2} + 3\, \bigl( {1\over 2\pi}\ln
{\Lambda_{0,{\rm eff}}\over\Lambda} \Bigr)^{2}.
\end{equation}
By replacing into (\ref{sdscale}), we see that,
up to some trivial factors, we obtain a finite double scaled partition 
function in the limit
\begin{equation}
\label{defscaI}
N\rightarrow\infty\, ,\quad \delta v\rightarrow 0, \, \quad N\, {\delta 
v\over V} - 3 \ln N = {\rm constant},
\end{equation}
as was to be shown. In momentum dependent quantities, $p'=p\sqrt{N}$ must 
be held fixed.
The same reasoning show that the correct double scaling limit in the 
vicinity of the Ashkin-Teller critical point (\ref{seffAT}) is defined by
\begin{eqnarray}
\label{defscaAT}
&& N\rightarrow\infty\, ,\quad \tilde\delta v\rightarrow 0\, ,\quad 
v_{N-1}\rightarrow 0 \, ,\nonumber \\
&& N\, {\tilde\delta v\over\tilde V} - 4\ln N = {\rm constant},\quad
N\, {\tilde\delta v - v_{N-1}\over \tilde V} - 4\ln N = {\rm constant'}.\\
\nonumber
\end{eqnarray}
The last case that we have to consider is the vicinity of the compactified
boson CFT, for example $v_{\rm N-1}\rightarrow 0$ with $\prod_{i=1}^{N-2}
v_i > \Lambda^{2(N-2)}$. We cannot define a double scaling limit in this
case. Indeed, in the $N\rightarrow\infty$ limit, 
the physics is well described by the action (\ref{Slowa}), 
since $g_{\rm eff}^2
\sim 1/N$ is then very small. We see that the $1/N$ corrections
are simply corrections to the radius of the compactified boson, which
correspond to a marginal operator. This is very different to the case of the
Ising or Ashkin-Teller CFTs, and does not produce the wild IR divergences
that are necessary to be able to define a double scaling limit. 

\acknowledgments
I would like to acknowledge interesting discussions with Duncan Haldane.
\appendix
\section{General solitary waves solutions}
As discussed in Section 2.3.3, the most general static, finite
energy solutions to the field equations (\ref{fieldeq}) satisfy
\begin{equation}
\label{neumanneq}
{d^2\Phi _i\over dx^2} = (\alpha -h_i)\, \Phi _i
\end{equation}
with
\begin{equation}
\label{bcsol}
\lim_{x\rightarrow\pm\infty}\Phi _i^2 = {\delta_{i,N}\over g^2}
\end{equation}
when $h_N=\max_{1\leq i\leq N}h_i$, which we shall assume in the
following.  $\alpha$ is a Lagrange multiplier implementing the constraint
\begin{equation}
\label{cst}
\sum _{i=1}^N \Phi _i^2 = {1\over g^2}\cdotp
\end{equation}
The set of equations (\ref{neumanneq},\ref{cst}) can be interpreted as the 
Newton equations for the motion of a
particle constrained to move on the sphere $S^{N-1}$ of radius $1/g$
in the quadratic potential
\begin{equation}
\label{mecpot}
U(\Phi) = + {1\over 2}\sum_{i=1}^N h_i \Phi_i^2 = -V(\Phi).
\end{equation}
The r\^ole of the time in the mechanical problem is played by $x$,
and $V(\Phi)=-U(\Phi)$ is the potential of the corresponding field theory.
That such a mechanical analogy is possible for solitons in two dimensions
is of course banal (see e.g.~\cite{raja}). The peculiarity of the soliton
problem is that we are interested only in the motions satisfying 
(\ref{bcsol}).

We show in this Appendix that all the solitonic solutions to our model can
be found, and are expressed in terms of elementary functions only. This
generalizes the similar result for the sine-Gordon theory. This is possible
because the classical
mechanical problem (\ref{neumanneq}, \ref{cst}) is integrable in the
Liouville sense. It was first studied by C. Neumann in the case $N=3$,
who shows that one can separate the variables in the Hamilton-Jacobi
equation \cite{neumann}, and discussed much later in the general case
\cite{uhlenbeck}. In the following, we present a self-contained elementary
analysis in the case where all the $h_i$s are distinct.
The more symmetric cases where some of the $h_i$s would coincide can
be easily obtained by taking suitable limits,
but we will not spell out these details.
In A.1 we explain how to solve the equations of motion in general, 
and then we specialize to the soliton problem.
\subsection{General analysis}
The starting point are the formulas for the constants of the motion
\cite{uhlenbeck},
\begin{equation}
\label{integrals}
{\cal I}_i = {\Phi_i^2 \over g^2} + \sum_{\scriptstyle
j=1\atop\scriptstyle j\neq i}^N {(\Phi_j \Phi_i' - \Phi_j' \Phi_i)^2\over
h_i-h_j} \raise 2pt\hbox{,} \quad 1\leq i\leq N,
\end{equation}
where $\Phi_i' = d\Phi_i /dx$.
Only $N-1$ of the ${\cal I}_{i}$s are independent because of the identity
\begin{equation}
\label{identityint}
\sum _{i=1}^N {\cal I}_i = {1\over g^4}\cdotp
\end{equation}
It is straightforward to check that $d{\cal I}_i/dx =0$ by using
(\ref{neumanneq}) and (\ref{cst}). These $N-1$ independent integrals of
the motion actually 
form an involutive system with respect to the Poisson bracket
(by identifying $\Phi_i'$ with the conjugate momentum), which prove
Liouville integrability. The energy is simply expressed in terms of the
${\cal I}_i$s,
\begin{equation}
\label{energy}
{\cal E} = {1\over 2} \sum_{i=1}^N \Bigl( \Phi_i'^2 + h_i\Phi_i^2\Bigr) =
{g^2\over 2}\sum_{i=1}^N h_i{\cal I}_i.
\end{equation}
We are thus left with $N-1$ independent {\it first order} 
differential equations, instead of the second order equations 
(\ref{neumanneq}).
In order to solve these equations, it is very useful to rewrite the
definition of the ${\cal I}_i$s in a more compact, elegant way. For this
purpose, we introduce an arbitrary complex parameter $z$ and define
\begin{equation}
\label{defQs}
q(x,y;z) = \sum_{i=1}^N {x_i y_i\over z-h_i}\, \raise 2pt\hbox{,}\quad
q(x;z) = q(x,x;z),
\end{equation}
\begin{equation}
{\cal I}(\Phi ,\Phi ';z) = \sum_{i=1}^N {{\cal I}_i(\Phi ,\Phi ')\over
z-h_i}\, \cdotp
\end{equation}
The formulas (\ref{integrals}) are equivalent to the equation
\begin{equation}
\label{identityqs}
{\cal I}(\Phi ,\Phi ';z) = q(\Phi ;z)\Bigl( q(\Phi ';z) + {1\over g^2}
\Bigr) - q(\Phi ,\Phi ';z)^2,
\end{equation}
and the differential equations we want to solve can be written
\begin{equation}
\label{eqdiffsol}
{\cal I} (\Phi ,\Phi ';z) = \sum_{i=1}^{N} {{\cal I}_{i}\over 
z-h_{i}}={1\over g^{4}}\, {b(z)\over a(z)}\, \raise 2pt\hbox{,}
\end{equation}
where
\begin{equation}
\label{defa}
a(z) = \prod_{i=1}^N (z-h_i),
\end{equation}
and $b(z)$ is defined by (\ref{eqdiffsol}). The arbitrary parameter $z$ is 
then chosen to make the equations (\ref{eqdiffsol})
as simple as possible. (\ref{identityqs}) suggests that a good choice is
\begin{equation}
\label{ellipc1}
q(\Phi ;z=\mu ) = 0.
\end{equation}
This equation has generically $N-1$ solutions $z=\mu _1 , \ldots ,\mu
_{N-1}$ that can be expressed in terms of the $N-1$ independent
coordinates $\Phi _i$ on the sphere, and vice-versa. The $\mu _i$s are
usually called elliptic coordinates. Explicitly, by introducing
\begin{equation}
\label{defm}
m(z;\mu) = \prod_{i=1}^{N-1} (z-\mu _i),
\end{equation}
we have
\begin{equation}
\label{neweq}
q(\Phi ;z) = {1\over g^{2}}\, {m(z;\mu)\over a(z)}
\end{equation}
and
\begin{equation}
\label{xvsmu}
g^{2}\, \Phi _i^2 = {m(z=h_i;\mu)\over a'(z=h_i)} =
{\displaystyle\prod_{j=1}^{N-1} (h_{i}-\mu_{j})\over\displaystyle
\prod_{j=1,j\neq i}^{N} (h_{i}-h_{j})}\,\cdotp
\end{equation}
We now make the change of variable $\Phi\mapsto\mu$ in 
(\ref{eqdiffsol}). By taking the derivative of (\ref{ellipc1}) 
with respect to $x$ we get
\begin{equation}
\label{q1}
q(\Phi,\Phi';\mu _{i}) = {1\over 2} \mu_{i}' \, \sum_{j=1}^{N} {\Phi 
_{j}^{2}\over (\mu _{i}-h_{k})^{2}}\, \raise 2pt\hbox{,}
\end{equation}
and by taking the derivative of (\ref{neweq}) with respect to $z$ we get
\begin{equation}
\label{q2}
\sum_{j=1}^{N} {\Phi_{j}^{2}\over (\mu_{i}-h_{j})^{2}} = -{1\over 
g^{2}}\, {\partial_{z} m (z=\mu_{i};\mu)\over a(z=\mu_{i})}\, \cdotp
\end{equation}
From these relations and (\ref{identityqs}, \ref{eqdiffsol}, \ref{ellipc1}) 
we finally obtain
\begin{equation}
\label{sol1}
\mu_{i}' = 2 \epsilon_{i} \, {\sqrt{\mathstrut -a(\mu_{i}) b(\mu_{i})}\over
\displaystyle\prod_{\scriptstyle j=1\atop \scriptstyle j\neq i}^{N-1} 
(\mu_{i}-\mu_{j})}\, \raise 2pt\hbox{,}
\end{equation}
with $\epsilon_{i}=\pm 1$.
We can put (\ref{sol1}) in a more suggestive form by using the

\noindent {\it Lemma}: if $P(z)=\prod _{i=1}^{n} (z-r_{i})$ is an arbitrary 
polynomial, then
\begin{equation}
\label{lemma}
\sum_{i=1}^{n} {r_{i}^{n-j}\over P'(r_{i})} = \delta _{j,1}, \quad 
j\geq 1.
\end{equation}
This is proven by calculating
\begin{equation}
\oint _{C} {dz\over 2i\pi } \, {z^{n-j}\over P(z)}
\end{equation}
with the help of the residue theorem and by taking $C$ to be a circle 
centered at $z=0$ of radius $R\rightarrow\infty$.

The lemma implies that (\ref{sol1}) is equivalent to
\begin{equation}
\label{sol2}
\sum_{i=1}^{N-1} \epsilon_{i}\, {\mu_{i}^{N-1-j}\mu_{i}'\over
2\sqrt{-a(\mu_{i})b(\mu_{i})}} = \delta_{j,1}, \quad 1\leq j\leq 
N-1,\quad \epsilon_{i} =\pm 1.
\end{equation}
The general solution of the equations of motion are then given by 
elliptic integrals associated with the genus $N-1$ hyperelliptic curve
\begin{equation}
\label{curve}
y^{2}=-4 a(x) b(x).
\end{equation}
However, as already stressed, we are not interested with the most 
general solutions, but with the ones corresponding to the solitons of our 
field theory model. We now focus on this special case.
\subsection{The solitons for general $N$}
\subsubsection{Basic equations}
The boundary conditions (\ref{bcsol}) were imposed in order to have a
finite mass $M$ for the solitons. $M$ is simply given by the action of
the mechanical problem,
\begin{equation}
\label{solitonmass}
M = {1\over 2}\, \int_{-\infty}^{+\infty}dx 
\Bigl( \sum_{i=1}^N \Phi_i'^2 + v_i\Phi_i^2 \Bigr),
\end{equation}
where $v_{i}=h_{N}-h_{i}$ as usual.
For $M$ to be finite, it is necessary to have
\begin{equation}
\label{bcint}
{\cal I}_i = {\delta_{i,N}\over g^4}\raise 2pt\hbox{,}
\end{equation}
that is to say,
\begin{equation}
\label{bsol}
b(z)=\prod_{i=1}^{N-1} (z-h_{i}).
\end{equation}
We will see below that (\ref{bcint}) is actually sufficient to have a
finite $M$. The conditions (\ref{bcint}) thus characterize the particular
motions that can be interpreted as solitary waves in our original theory.

With this particular choice for $b(z)$, (\ref{sol2}) yields
\begin{equation}
\label{sol3}
\sum_{i=1}^{N-1} \epsilon_{i} \, {\mu_{i}^{N-1-j} \, d\mu_{i}\over
\displaystyle 2\sqrt{h_{N}-\mu_{i}} \,\prod_{j=1}^{N-1}(\mu_{i}-h_{j})} =
\delta_{j,1}\, dx,\quad 1\leq j\leq N-1, \quad \epsilon_{i}=\pm 1.
\end{equation}
We introduce the new coordinates
\begin{equation}
\label{coordy}
y_{i}=\epsilon_{i}\, \sqrt{h_{N}-\mu_{i}},
\end{equation}
and we finally get our basic equation that can be written in two equivalent
forms corresponding respectively to (\ref{sol1}) and (\ref{sol2}),
\begin{equation}
\label{solA}
{\displaystyle\prod_{j=1,j\neq i}^{N-1} 
(y_{j}^{2}-y_{i}^{2})\over\displaystyle\prod_{j=1}^{N-1} (v_{j}-y_{i}^{2})}
\, dy_{i} = - dx\, ,\quad 1\leq i\leq N-1,
\end{equation}
\begin{equation}
\label{solB}
\sum_{i=1}^{N-1} {\displaystyle (h_{N}-y_{i}^{2})^{N-1-j} \, dy_{i}\over
\displaystyle
\prod_{j=1}^{N-1} (v_{j}-y_{i}^{2})} = -\delta_{j,1}\, dx\, , \quad 
1\leq j\leq N-1.
\end{equation}
(\ref{solB}) can be integrated immediately in terms of elementary 
functions and $N-1$ constants of integration, which gives the full 
solution of the problem. The fact that only elemetary functions are 
involved is due to the fact that the 
solitonic solutions correspond to a highly 
degenerate hyperelliptic curve (\ref{curve}). Explicit formulas will 
be given for the case $N=3$ in the next subsection. Note that though 
(\ref{solB}) is best suited to an explicit integration, a
qualitative analysis of the solution, to which we now turn,
is most easily performed by using (\ref{solA}).
\subsubsection{Description of the solutions}
We choose without loss of generality $h_{N}>h_{N-1}>\cdots >h_{1}$. 
Equation (\ref{xvsmu}) shows that $\Phi_{j}^{2}\geq 0$ implies that
each interval $[h_{i},h_{i+1}]$ 
contains one and only one of the $\mu_{j}$s, or equivalently that 
each interval $[v_{i},v_{i-1}]$ contains one and only one of the 
$y_{j}$s. We consider a solution such that
\begin{equation}
\label{specialbc}
\lim _{x\rightarrow -\infty} \Phi_{i} = + {\delta_{i,N}\over g}
\raise 2pt\hbox{,}
\end{equation}
and we conventionally choose the ordering of the coordinates such 
that this boundary condition corresponds to
\begin{equation}
\label{bcys}
\lim _{x\rightarrow -\infty} y_{i} = \sqrt{v_{i}}.
\end{equation}
By looking at (\ref{solA}), we see that $dy_{1}<0$ when 
$x\rightarrow -\infty$. More generally, as long as $y_{i}\in 
[\sqrt{v_{i+1}},\sqrt{v_{i}}]$, which is true at least up to some 
finite value of $x$ because one can have only one $y$ in each of these 
intervals, a simple recursive reasoning using (\ref{solA}) shows that 
$dy_{i}<0$ for all $i$. Two $y_{i}^{2}$ can eventually swap their 
intervals. Equation (\ref{solA}) shows that it is possible to
have $y_{i}^{2}=v_{j}$ at some finite $x$ as long as some other 
$y_{k}^{2}=v_{j}$ at the same $x$, the pole in the denominator being 
then compensated by a zero in the numerator. With our particular 
boundary conditions (\ref{bcys}), the only consistent way this can 
happen is by having first $y_{N-1}\rightarrow -\sqrt{v_{N-1}}$ and 
$y_{N-2}\rightarrow\sqrt{v_{N-1}}$ (step 
$1$), then $y_{N-1}\rightarrow -\sqrt{v_{N-2}}$ and $y_{N-3}\rightarrow 
\sqrt{v_{N-2}}$ (step $2$), and so on up to step $N-2$ when 
$y_{N-1}\rightarrow -\sqrt{v_{2}}$ and $y_{1}\rightarrow\sqrt{v_{2}}$. 
This is followed by $N-3$ additional steps during which 
$y_{1}\rightarrow\sqrt{v_{3}}$ and $y_{N-2}\rightarrow -\sqrt{v_{3}}$ 
(step $N-1$),
$y_{1}\rightarrow\sqrt{v_{4}}$ and $y_{N-3}\rightarrow 
-\sqrt{v_{4}}$ (step $N$), and so on up to step $2N-5$ when 
$y_{1}\rightarrow\sqrt{v_{N-1}}$ and $y_{2}\rightarrow 
-\sqrt{v_{N-1}}$. At each intermediate step it can be checked using 
(\ref{solB}) that $dy_{i}<0$ for all $i$, so the $y_{i}$s are monotonic 
decreasing functions of $x$.
At the end of the day we have
\begin{equation}
\label{limplus}
\lim _{x\rightarrow +\infty} y_{i} = -\sqrt{v_{N-i}}.   
\end{equation}
What we have been describing is the generic solution. By restricting 
oneself to motions for which some of the $\Phi_{i}$s are
equal to zero, one can obtain other solutions, corresponding to
the generic solution at a lower value of $N$.

To get a better physical understanding of the nature of the solution, 
let us consider the coordinates $\Phi _{i}$. First, let us note that 
the transition between the different steps correspond to the successive
crossing of the hyperplanes 
$\Phi_{N-1}=0,\Phi_{N-2}=0,\ldots ,\Phi_{2}=0,\Phi_{3}=0,\ldots ,
\Phi_{N-1}=0$. This shows that the trajectory lies entirely on the 
half sphere $\Phi_{1}>0$ or $\Phi_{1}<0$, the two cases being related 
by the symmetry ${Z}_{2(1)}$. Second, and more importantly, as 
$x$ increases we have successively $y_{N-1}=0$, $y_{N-2}=0$, and so on
up to $y_{1}=0$. This shows that
\begin{equation}
\label{phiN}
\Phi_{N}^{2} = {1\over g^{2}}\, {y_{1}^{2}\cdots y_{N-1}^{2}\over
v_{1}\cdots v_{N-1}}
\end{equation}
has $N-1$ zeros when $x$ goes from $-\infty$ to $+\infty$. From this 
we can conclude that when $N$ is odd the solution is topologically 
trivial, while when $N$ is even, it belongs to the topologically 
non-trivial sector of the theory. Equivalently, we have
\begin{equation}
\label{implusphi}
\lim_{x\rightarrow +\infty} \Phi_{i}= {(-1)^{N-1} \delta_{i,N}\over 
g}\cdotp
\end{equation}
The fact that we cross the equator $\Phi_{N}=0$ $N-1$ times suggests 
that the solution actually corresponds to the succession of $N-1$ kinks and 
anti-kinks. The $N-1$ parameters (constants of integration) would be
in this interpretation 
the global center of mass and the relative separations of the 
individual kinks. To prove that this interpretation is indeed correct, we 
are going to show that the total mass (\ref{solitonmass})
of our solution is the sum 
of the masses of the individual sine-Gordon solitons of the model 
(\ref{massgensol}),
\begin{equation}
\label{Mass}
M = {2\over g^{2}}\, \sum_{i=1}^{N-1}\sqrt{v_{i}}.
\end{equation}
Note that $M$ does not depend of the $N-1$ parameters of the 
solution, which is a direct consequence of the fact that $M$ coincide with 
the action of the mechanical problem, and that the boundary conditions 
in the action are independent of those parameters. We have thus obtained 
a {\it static} solution consisting of a succession of sine-Gordon 
kinks and anti-kinks! As pointed out before, we actually have solutions for 
any number $k\leq N-1$ kinks and anti-kinks. Note that all the kinks 
or anti-kinks appearing in the solution have different masses.
A way to prove 
(\ref{Mass}) is to choose the constants of integration in some limit 
that simplify the general solution (kinks and anti-kinks in a limit 
of very large separation). We will present instead a direct 
derivation which turns out to be simpler.

Our starting point is
\begin{equation}
\label{mass1}
M = \int_{-\infty}^{+\infty} dx \sum_{i=1}^{N-1} v_{i}\Phi_{i}^{2},
\end{equation}
which is derived from (\ref{solitonmass}, \ref{bcint}, \ref{energy}). We 
then use the following algebraic identities:
\begin{equation}
\label{identmass1}
\sum_{i=1}^{N-1} v_{i}\Phi_{i}^{2} = {1\over g^{2}}\, 
\sum_{i=1}^{N-1} \bigl(v_{i}-y_{i}^{2}\bigr),
\end{equation}
\begin{equation}
\label{identmass2}
\sum_{i=1}^{N-1} \bigl(v_{i}-y_{i}^{2}\bigr) = \sum_{i=1}^{N-1} 
{\displaystyle\prod_{j=1}^{N-1} (v_{j}-y_{i}^{2})\over\displaystyle
\prod_{j=1,j\neq i}^{N-1} (y_{j}^{2}-y_{i}^{2})}\, \cdotp
\end{equation}
Equation (\ref{identmass1}) is proven by writing (\ref{neweq}) as
\begin{equation}
\label{neweq2}
{1\over g^{2}}\, m(z;\mu)= a(z) q(\Phi ;z)
\end{equation}
and identifying on both sides the coefficient of $z^{N-2}$. Equation
(\ref{identmass2}) is proven by using the fact that the right hand 
side can be viewed as a rational function of $y_{1}^{2}$ of degree $d\leq 
1$, and that it is actually a linear function of $y_{1}^{2}$ since it 
has a finite limit when $y_{1}^{2}\rightarrow y_{i}^{2}$, for all
$i=2,\ldots N-1$. Taking into account the permutation symmetry amongst the 
$y_{i}$s, this leaves us with two unknown coefficients which are 
determined easily by looking at the $y_{1}^{2}\rightarrow\infty$ 
limit up to terms ${\cal O}(1/y_{1}^{2})$. By using successively
(\ref{mass1}), (\ref{identmass1}), (\ref{identmass2}) and (\ref{solA}), we 
get
\begin{equation}
\label{massinterm}
M = -{2\over g^{2}}\, 
\sum_{i=1}^{N-1}\int\! dy_{i}.
\end{equation}
By using the fact that the $y_{i}$s are monotonic functions of $x$, and 
(\ref{bcys}) and (\ref{limplus}), we finally obtain the desired result 
(\ref{Mass}). From our understanding of the solution, we deduce
in particular that the mass density
\begin{equation}
\label{massd}
\rho (x)=\sum_{i=1}^{N-1} v_i \Phi_i^2
\end{equation}
will have $N-1$ maxima, located at the positions of the kinks and
anti-kinks, whose values approach the corresponding
values for sine-Gordon solitons, $v_i/g^2$, at large separations.
\subsection{Formulas for the case N=3}
We give below the explicit formulas used in Section 2.3.3 to make the 
Figure 2. By integrating (\ref{solB}) for $N=3$ we 
get
\begin{equation}
\label{solN3}
{(\sqrt{v_{1}}-y_{1})(\sqrt{v_{1}}-y_{2})\over 
(\sqrt{v_{1}}+y_{1})(\sqrt{v_{1}}+y_{2})} = 
e^{2\sqrt{v_{1}}(x-x_{1})},\quad
{(y_{1}-\sqrt{v_{2}})(\sqrt{v_{2}}-y_{2})\over 
(\sqrt{v_{2}}+y_{1})(\sqrt{v_{2}} + y_{2})} = e^{2\sqrt{v_{2}} 
(x-x_{2})}.
\end{equation}
When $|x_{2}-x_{1}| >> 1/\sqrt{v_{2}}$, we have two well separated
kinks centered at $x=x_{1}$ and $x=x_{2}$. Solving explicitly for 
$y_{1}$ and $y_{2}$ amounts to solving a degree 2 polynomial 
equation. The relevant root is picked up by using (\ref{bcys}); the 
resulting formulas are very complicated and will not be listed here.
The spherical coordinates $(\theta ,\phi )$ are then given by
\begin{equation}
\label{solN3tp}
\theta = \arccos {y_{1}y_{2}\over \sqrt{v_{1}v_{2}}} \raise 
2pt\hbox{,}\quad \phi = \sign (y_{1}-\sqrt{v_{2}}) \, \arctan 
\sqrt{ {v_{1}\over v_{2}}\, {(y_{1}^{2}-v_{2})(v_{2}-y_{2}^{2})\over
(v_{1}-y_{1}^{2})(v_{1}-y_{2}^{2})}} \cdotp
\end{equation}
\vfill\eject
\section{The functional $s\lbrack f\rbrack $} 
In this Appendix, we study the functional $s[f]$ defined by
\begin{equation}
\label{sfunctionalmink}
s[f]= {i\over 2}\, \tr\ln (-\partial_{\mu}\partial^{\mu} - f + i\epsilon)
+ {1\over 4\pi}\ln {\Lambda_{0}\over\mu }\,\int\! d^{2}x\, f(x),
\end{equation}
or equivalently in the euclidean by
\begin{equation}
\label{sfunctional}
s[f]= {1\over 2}\, \tr\ln (-\partial ^{2} + f) -
{1\over 4\pi}\ln {\Lambda_{0}\over\mu }\,\int\! d^{2}x\, f(x).
\end{equation}
This functional is ubiquitous in the study of the large $N$ limit of 
our model. $s[f]$ depends on an arbitrary renormalization scale 
$\mu$.

All UV divergent integrals are regulated by a momentum cutoff
$|p|\leq\Lambda_{0}$. In finite quantities, we will take the limit 
$\Lambda_{0}\rightarrow\infty$.
\subsection{$s\lbrack f\rbrack $ for constant $f$}
For constant $f$, $s[f]$ can be written in terms of a potential 
$v(f)$ as
\begin{equation}
\label{svsv}
s[f={\rm cst}]=\int\! d^{2}x \, v(f).
\end{equation}
When $f$ is a {\it positive} constant, the euclidean formula 
(\ref{sfunctional}) yields
\begin{equation}
\label{v1}
v(f)={1\over 2}\,\int\! {d^{2}p\over (2\pi)^{2}}\, \ln (p^{2}+f) -
{f\over 4\pi}\ln {\Lambda_{0}\over\mu }\raise 2pt\hbox{,}
\end{equation}
which is, up to $f$-independent terms,
\begin{equation}
\label{v}
v(f)=-{f\over 8\pi}\,\ln {f\over e\mu ^{2}}\, \raise 2pt\hbox{,}
\quad {\rm for}\ f>0.
\end{equation}
The correct formula for $f<0$ is obtained by using
Feynman's $i\epsilon$ prescription (\ref{sfunctionalmink}),
\begin{equation}
\label{v2}
v(f)= -{f\over 8\pi}\,\Bigl( \ln {-f\over e\mu ^{2}} -i\pi\Bigr),
\quad {\rm for\ } f<0.
\end{equation}
Equation (\ref{v2}) is the analytic
continuation of (\ref{v}) by going to $\im f <0$.
\subsection{$s\lbrack f\rbrack$ for arbitrary $f$}
When $f$ is arbitrary, one can expand $s[f]$ in terms of ordinary 
one-loop Feynman diagrams. Introducing an arbitrary mass scale $m$
and defining $\phi (x) = f(x)-m^{2}$, we have
\begin{equation}
\label{expand1}
s[f]=\sum_{n=0}^{\infty} s_{n}[f;m^{2}],
\end{equation}
where
\begin{eqnarray}
\label{expand2}
\hskip -2em s_{n}[f;m^{2}]&=&s_{n}[m^{2}+\phi;m^{2}]\nonumber \\
&=&{1\over n!}\, \int\!\prod_{i=1}^{n}d^{2}x_{i}\, 
s^{(n)}(x_{1},\ldots,x_{n};m^{2})\, \phi(x_{1})\cdots\phi(x_{n})\nonumber \\
&=&{1\over n!}\, \int\!\prod_{i=1}^{n}
{d^{2}p_{i}\over (2\pi)^{2}}\, (2\pi)^{2}\delta^{(2)}
\Bigl(\sum_{i=1}^{n} p_{i}\Bigr)
\, \tilde s^{(n)}(p_{1},\ldots ,p_{n};m^{2})\, \tilde\phi 
(p_{1})\cdots\tilde\phi (p_{n}),\\ \nonumber
\end{eqnarray}
with
\begin{eqnarray}
&&\tilde\phi (p)=\int\! d^{2}x \, e^{-ipx}\phi(x),\label{phivsphit}\\
&&(2\pi)^{2}\delta^{(2)}\Bigl(\sum_{i=1}^{n} p_{i}\Bigr)\,
\tilde s^{(n)}(p_{1},\ldots ,p_{n};m^{2}) =\nonumber\\
&&\qquad\qquad\qquad\qquad
\int\!\prod_{i=1}^{n}d^{2}x_{i}\, e^{i\sum_{j=1}^{n}p_{j}x_{j}}
s^{(n)}(x_{1},\ldots ,x_{n};m^{2}).\label{svsst}\\ \nonumber
\end{eqnarray}
The euclidean formula for $\tilde s^{(n)}$ is
\begin{equation}
\label{sn}
\tilde s^{(n)}(p_{1},\ldots p_{n};m^{2})= (-1)^{n-1}\, {(n-1)!\over 
8\pi^{2}}\,\int {\displaystyle d^{2}k\over\displaystyle
\prod_{i=0}^{n-1}\biggl( \Bigl(k+\sum_{j=1}^{i}p_{j}\Bigr)^{2}+
m^{2}\biggr)} - {\delta_{n,1}\over 4\pi}\, \ln {\Lambda_0\over\mu}
\cdotp
\end{equation}
Explicitly, the linear term is given by
\begin{equation}
\label{slinear}
\tilde s^{(1)} = {1\over 8\pi}\,\ln {\mu^2\over m^2}
\end{equation}
or equivalently
\begin{equation}
\label{slinearbis}
s_{1}[f;m^{2}]={1\over 8\pi}\,\ln {\mu^{2}\over m^{2}}\,
\int\! d^{2}x\,\phi(x).
\end{equation}
In the euclidean regime $p^2 >0$, the quadratic term is given by 
\begin{eqnarray}
\tilde s^{(2)}(p,-p;m^{2}) &=&\tilde s^{(2)}(p^{2},m^{2}) \\
\nonumber
&=& -{1\over 4\pi}\, {1\over p^{2}\sqrt{1+4m^{2}/p^{2}}}\, 
\ln {\sqrt{1+4m^{2}/p^{2}} + 1\over\sqrt{1+4m^{2}/p^{2}} 
-1}\, \raise 2pt\hbox{,}\quad {\rm for\ } p^2 >0.\label{squad}
\end{eqnarray}
At low momentum we can use the expansion
\begin{equation}
\label{squadlp}
\tilde s^{(2)}(p^{2},m^{2}) = -{1\over 8\pi m^{2}}\,\Bigl( 1-{p^{2}\over
6m^{2}}+{\cal O}(p^{4})\Bigr)
\end{equation}
or equivalently
\begin{equation}
\label{squadbis}
s_{2}[f;m^{2}]=-{1\over 16\pi m^{2}}\,\int\! d^{2}x\,\phi ^{2}(x) -
{1\over 96\pi m^{4}}\,\int\! d^{2}x\, \phi(x)\partial^{2}\phi(x) +
{\cal O}(\partial ^{4}).
\end{equation}
For physical momenta below the pair production threshold,
$0< -p^{2}< 4m^{2}$, the correct analytic continuation is given by
\begin{equation}
\label{squadtert}
\tilde s^{(2)}(p^{2};m^{2})= {1\over 2\pi}\,
{1\over p^{2}\sqrt{-1-4m^{2}/p^{2}}}
\, \arctan {1\over\sqrt{-1-4m^{2}/p^{2}}}\, \raise 2pt\hbox{,}\quad {\rm 
for} \ 0< -p^{2}< 4m^{2}.
\end{equation}
Finally, for $-p^2 > 4m^2$, which is the region relevant for two-particle
scattering, we have
\begin{equation}
\label{squadscat}
\tilde s^{(2)}(p^{2},m^{2})=
 -{1\over 4\pi}\, {1\over p^{2}\sqrt{1+4m^{2}/p^{2}}}\,\left( 
\ln {1+ \sqrt{1+4m^{2}/p^{2}}\over 1- \sqrt{1+4m^{2}/p^{2}}} - i\pi \right) 
\, ,\quad {\rm for\ } -p^2 > 4 m^2.
\end{equation}
In the latter case, it is convenient to introduce the rapidity parameter
$\theta >0$ such that
\begin{equation}
\label{rapidity}
-p^2 = 4 m^2 \cosh ^2 (\theta/ 2)
\end{equation}
and in terms of which
\begin{equation}
\label{squadscat2}
\tilde s^{(2)}(p^{2},m^{2}) = {1\over 8\pi m^2}\,
{\theta -i\pi \over\sinh\theta}
\cdotp
\end{equation}
Let us quote to 
finish the asymptotic behaviour of $\tilde s^{(2)}$ at large 
euclidean $p^{2}$, which is 
useful to investigate the UV properties of the $1/N$ expansion,
\begin{equation}
\label{s2asymp}
\tilde s^{(2)}(p^{2};m^{2})=-{1\over 4\pi p^{2}}\, \biggl(
\ln {p^{2}\over m^{2}} + {2m^{2}\over p^{2}}\,\Bigl( 1-\ln{p^{2}\over 
m^{2}}\Bigr) + {\cal O}\Bigl( {m^{4}\over p^{4}}\ln {p^{2}\over m^{2}}
\Bigr) \biggr).
\end{equation}
\section{The 1/N corrections to the critical hypersurface}
We discuss in this Appendix the $1/N$ corrections to the equation
\begin{equation}
\label{Heqa}
{1\over N} \sum_{i=1}^{N-1}\ln {v_{i}\over\Lambda ^{2}}=0
\end{equation}
for the hypersurface of singularities $\HQ$ derived in Section 3.4. The
main goal of this calculation is to understand the r\^ ole of the 
infrared divergences, which generically
plague the $1/N$ expansion near $\HQ$. As the very existence of $\HQ$ was 
derived within the large $N$ expansion, the consistency of the analysis 
requires that the equation for $\HQ$ itself gets only small corrections 
when we go beyond the leading approximation. We show below that the 
first corrections are of order $N^{-1}\ln N$, larger than the naive
$N^{-1}$, but nevertheless smaller that the leading $N^{0}$ term.
The calculation also illustrates nicely how renormalization works
within the $1/N$ expansion, consistently with the discussion of Section 3.1.

\subsection{The effective potential}
The starting point is the effective action
\begin{equation}
\label{Seffa}
S_{\rm eff}[\alpha ,\Phi _{N}]=(N-1)\, s_{\rm eff}[\alpha, \varphi]
\end{equation}
where
\begin{equation}
\label{seffa}
s_{\rm eff}[\alpha ,\varphi]=\int\! d^{2}x\, \left[
{1\over 2}\, \partial _{\alpha}\varphi\partial_{\alpha}\varphi +
{\alpha - h_{N,0}\over 2}\, \varphi^{2} - {\alpha\over 4\pi}\,
\ln{\mu\over\lambda}\right]
+ {1\over N-1}\, \sum_{i=1}^{N-1}s[\alpha -h_{i,0}].
\end{equation}
We were careful in keeping subleading terms in (\ref{seffa}), since we 
are now willing to take them into account. We have defined
\begin{equation}
\label{varphidef}
\varphi = \Phi_{N}/\sqrt{N-1}
\end{equation}
and $\lambda$ by the equation
\begin{equation}
\label{lambdadef}
{1\over g_{0}^{2}}={N-1\over 2\pi}\, \ln 
{\Lambda_{0}\over\lambda}\cdotp
\end{equation}
To order $N^{0}$, we have $\lambda =\Lambda$, but to order $N^{-1}$, 
$\lambda$ must pick an infinite, $\Lambda_{0}$-dependent 
contribution, coming from the $1/N$ corrections to the $\beta$ 
function. Equations (\ref{renNg}), (\ref{renNz2}) and (\ref{renNzs22})
of Section 3.1 gives the form of this 
contribution, as well as the renormalization of $h_{i,0}=h_{i}Z_{(s,2)}/Z$:
\begin{eqnarray}
\lambda^{2}&=&\Lambda^{2}\left( 1+ {1\over N}\, \ln 
{\Lambda_{0}^{2}\over\Lambda^{2}} - {2\over N}\, \ln\ln 
{\Lambda_{0}\over\Lambda} + {c_{1}\over N} \right),\label{lambdaren}\\
h_{i,0}&=&h_{i}\left( 1+{2\over N}\, \ln\ln {\Lambda_{0}\over\Lambda} +
{c_{2}\over N}\right),\label{hren}\\ \nonumber
\end{eqnarray}
where $c_{1}$ and $c_{2}$ are finite constants to be determined by some 
renormalization conditions.

The $1/N$ corrections correspond to the one-loop diagrams derived from
the non-local action $s_{\rm eff}$. In particular,
the effective potential $v_{\rm eff}$ to order $N^{-1}$ is such that
\begin{eqnarray}
\hskip -1.5cm\int\! d^{2}x\, v_{\rm eff}(\alpha ,\varphi) &=& 
\int\! d^{2}x\, \Biggl( {1\over 2}\, (\alpha 
-h_{N,0})\,\varphi^{2} - {\alpha\over 4\pi}\,\ln{\mu\over\lambda} \nonumber\\
&&\hskip .5cm - {1\over 8\pi (N-1)}\,\sum_{i=1}^{N-1} (\alpha -h_{i,0})\,
\ln {\alpha - h_{i,0}\over e\mu^{2}}\Biggr)
+{1\over 2N}\tr\ln {\cal J}[\alpha ,\varphi].\label{veff}\\
\nonumber
\end{eqnarray}
$\cal J$ is the functional hessian of $s_{\rm eff}$, for which
\begin{equation}
\label{matJ}
{1\over 2}\,\tr\ln {\cal J}[\alpha ,\varphi] =
{1\over 8\pi ^{2}}\,\int\! d^{2}x\,\int\! d^{2}p\,\tr\ln\pmatrix{
p^{2}+\alpha-h_{N,0}&\varphi\cr
\varphi & \tilde S(p^{2};\alpha ,h_{1,0},\ldots ,h_{N-1,0})\cr}
\end{equation}
where we have defined
\begin{equation}
\label{Sdef}
\tilde S(p^{2};\alpha ,h_{1},\ldots ,h_{N-1}) =
{1\over N-1}\sum_{i=1}^{N-1} \tilde s^{(2)}(p^{2};\alpha -h_{i}).
\end{equation}
From (\ref{veff}, \ref{matJ}) we get the basic formulas, valid up to 
terms of order $1/N^{2}$,
\begin{eqnarray}
\!\!\!\!\!\!\!\!\!\!\!\!\!\!\!\!\!\!\!\!\!\!\!\!
{\partial v_{\rm eff}\over\partial\varphi} &=&
(\alpha -h_{N,0})\,\varphi - {1\over N}\, {\varphi\over 4\pi^{2}}\,
\int\! d^{2}p\, {1\over (p^{2}+\alpha -h_{N})\tilde S(p^{2};\alpha 
,h_{1},\ldots ,h_{N-1}) - \varphi^{2}}\raise 2pt\hbox{,}\label{dvdp}\\
\!\!\!\!\!\!\!\!\!\!\!\!\!\!\!\!\!\!\!\!\!\!\!\!
{\partial v_{\rm eff}\over\partial\alpha} &=&
{\varphi^{2}\over 2} - {1\over 8\pi (N-1)}\,\sum_{i=1}^{N-1}
\ln{\alpha -h_{i,0}\over\lambda^{2}} + {1\over 8\pi^{2} N}\,
\int\! d^{2}p\, {\tilde S +(p^{2}+\alpha -h_{N})\partial_{\alpha}
\tilde S \over (p^{2}+\alpha -h_{N}) \tilde S
-\varphi^{2}}\cdotp\label{dvda}\\ \nonumber
\end{eqnarray}
\subsection{The equation for $\HQ$}
The naive strategy to get the equation for $\HQ$ up to order $1/N$ 
would be to solve
\begin{equation}
\label{Hcond}
{\partial v_{\rm eff}\over\partial\varphi}=
{\partial v_{\rm eff}\over\partial\alpha}=0
\end{equation}
by substituting $\alpha$ and $\varphi$ in the terms of order $1/N$
by their respective values 
at order $N^{0}$, that is $\alpha =h_{N}$ and $\varphi =0$ (see 
Section 3). However, this yields IR divergent integrals, and thus is 
not correct. Instead, we will identify the terms responsible for the 
IR divergences, and set $\alpha =h_{N}$ and $\varphi =0$ in the other 
terms only. The IR divergences can be analysed easily by using
\begin{eqnarray}
\tilde S(p^{2};\alpha =h_{N},h_{1},\ldots ,h_{N-1}) =
-{1\over 8\pi V} + {\cal O}(p^{2}),\label{Sasym}\\
\partial_{\alpha} \tilde S(p^{2};\alpha =h_{N},h_{1},\ldots ,h_{N-1}) =
{1\over 8\pi (N-1)}\,\sum_{i=1}^{N-1} {1\over v_{i}^{2}} 
+ {\cal O}(p^{2}),\label{dSasym}\\ \nonumber
\end{eqnarray}
where $V$ is defined in (\ref{Vdef}).
The term containing $\partial_{\alpha}\tilde S$ in
$\partial v_{\rm eff}/\partial\alpha$ is not IR divergent, while 
the other term can be written
\vfill\eject
\begin{eqnarray}
\label{IRdiv}
&&\int\! {d^{2}p\over 8\pi^{2}}\, \Biggl[
{\tilde S(p^{2};\alpha ,h_{1},\ldots ,h_{N-1})\over 
(p^{2}+\alpha -h_{N})\tilde S(p^{2};\alpha ,h_{1},\ldots ,h_{N-1})
-\varphi^{2}} - \nonumber\\
&&\qquad\qquad\qquad {\tilde S(p^{2}=0;\alpha ,h_{1},\ldots ,h_{N-1})\over 
(p^{2}+\alpha -h_{N})\tilde S(p^{2}=0;\alpha ,h_{1},\ldots ,h_{N-1})
-\varphi^{2}} \Biggr] \nonumber \\
+&&\int\! {d^{2}p\over 8\pi^{2}}\,
{\tilde S(p^{2}=0;\alpha ,h_{1},\ldots ,h_{N-1})\over 
(p^{2}+\alpha -h_{N})\tilde S(p^{2}=0;\alpha ,h_{1},\ldots ,h_{N-1})
-\varphi^{2}}\cdotp\\ \nonumber
\end{eqnarray}
The first term is now IR convergent, and goes actually to zero when 
$\alpha\rightarrow h_{N}$ and $\varphi\rightarrow 0$, while the second 
term is easily calculated. The condition 
$\partial v_{\rm eff}/\partial\alpha =0$ is thus equivalent to
\begin{equation}
\label{dvdazero}
0={\varphi^{2}\over 2} - {1\over 8\pi (N-1)}\,\sum_{i=1}^{N-1}
\ln{\alpha -h_{i,0}\over\lambda^{2}} + 
{1\over 8\pi N}\, \ln{\Lambda_{0}^{2}\over \alpha - h_{N} + 8\pi 
V\varphi^{2}} + {I_{\alpha}(v_{1},\ldots ,v_{N-1})\over 8\pi N}
\raise 2pt\hbox{,}
\end{equation}
where
\begin{equation}
\label{Iadef}
I_{\alpha}(v_{1},\ldots ,v_{N-1}) = {1\over\pi}\,\int\! d^{2}p\,
{\partial_{\alpha} \tilde S(p^{2};\alpha =h_{N},h_{1},\ldots ,h_{N-1})\over
\tilde S(p^{2};\alpha =h_{N},h_{1},\ldots ,h_{N-1})}\cdotp
\end{equation}
We treat $\partial v_{\rm eff}/\partial\varphi$ is a similar way. We 
write the $1/N$ contribution as
\begin{eqnarray}
\label{IRdiv2}
&&-\int\! {d^{2}p\over 4\pi^{2}}\, \Biggl[
{1 \over 
(p^{2}+\alpha -h_{N})\tilde S(p^{2};\alpha ,h_{1},\ldots ,h_{N-1})
-\varphi^{2}} - \nonumber\\
&&\qquad\qquad\qquad {1 \over 
(p^{2}+\alpha -h_{N})\tilde S(p^{2}=0;\alpha ,h_{1},\ldots ,h_{N-1})
-\varphi^{2}} \Biggr] \nonumber \\
&&-\int\! {d^{2}p\over 4\pi^{2}}\,
{1 \over 
(p^{2}+\alpha -h_{N})\tilde S(p^{2}=0;\alpha ,h_{1},\ldots ,h_{N-1})
-\varphi^{2}}\,\raise 2pt\hbox{,}\\ \nonumber
\end{eqnarray}
check that the first term is now IR convergent while the second 
term is easy to compute, and conclude that the condition
$\partial v_{\rm eff}/\partial\varphi =0$ is equivalent to
\begin{equation}
\label{dvdpzero}
0=\alpha -h_{N,0} +{V\over N}\,\left( 
2 \ln{\Lambda_{0}^{2}\over \alpha - h_{N} + 8\pi 
V\varphi^{2}} - I_{\varphi}(v_{1},\ldots ,v_{N-1})\right),
\end{equation}
where
\begin{eqnarray}
\label{Ipdef}
I_{\varphi}(v_{1},\ldots ,v_{N-1}) &=& {1\over 4\pi^{2} V}\,\int\! 
{d^{2}p\over p^{2}}\,
\Biggl( {1\over \tilde S(p^{2};\alpha =h_{N},h_{1},\ldots ,h_{N-1})} - 
\nonumber \\ 
&&\qquad\qquad\qquad\qquad
{1\over \tilde S(p^{2}=0;\alpha =h_{N},h_{1},\ldots ,h_{N-1})}\Biggr) \cdotp
\nonumber \\
\end{eqnarray}
We have chosen to approach $\HQ$ from weak coupling, and thus we have
discarded the solution $\varphi =0$ to the equation 
$\partial v_{\rm eff}/\partial\varphi =0$. $\HQ$ is precisely the locus in
parameter space where $\varphi =0$ becomes the true minimum of the
effective potential for $\varphi$, see Section 3.4.

The equations (\ref{dvdazero}) and (\ref{dvdpzero}) are our new starting
point. They are badly UV divergent, as discussed further in the next
subsection, and the renormalizations
(\ref{lambdaren}) and (\ref{hren}) are not enough to make them finite.
This is not surprising: as explained in Section 2.1, formulas containing
the field $\alpha$ need an infinite number of counterterms. 
To obtain a finite, physically sensible, formula, we must eliminate
$\alpha$ using (\ref{dvdazero}), and then check that the result is finite.
To make the presentation as clear as possible, let us
introduce
\begin{equation}
\label{betadef}
\tilde\alpha = \alpha - h_{N,0}.
\end{equation}
We know from Section 3.4 that 
$\tilde\alpha =0$ on $\HQ$ to order $N^0$ (this is
a direct consequence of (\ref{dvdpzero}). We will thus always set 
$\tilde\alpha =0$ in $1/N$ corrections, as long as this does not lead
to an IR divergence. Moreover, to leading order, (\ref{dvdazero})
shows that $\tilde\alpha = \tilde\alpha _{N^0} 
(v_{1,0},\ldots ,v_{N-1,0};\lambda)$ with
\begin{equation}
\label{beq}
\sum_{i=1}^{N-1} \ln {\tilde\alpha_{N^0} + v_{i,0}\over\lambda^2} =0.
\end{equation}
In order to solve (\ref{dvdazero}), we then substitute $\tilde\alpha =
\tilde\alpha_{N^0} + {\cal O}(1/N)$ and use (\ref{beq}) to obtain
\begin{equation}
\label{betaeq}
\tilde\alpha = \tilde\alpha_{N^0} + {V\over N}\,\left( I_{\alpha} - \ln
{\tilde\alpha_{N^0}\over\Lambda_0^2}\right).
\end{equation}
Finally, using (\ref{dvdpzero}),
we obtain the $\alpha$-independent equation
\begin{equation}
\label{Heqapp1}
\tilde\alpha_{N^0} (v_{1,0},\ldots ,v_{N-1,0};\lambda) =
{V\over N}\,\left( I_{\varphi}
 - I_{\alpha} + 3\ln {\tilde\alpha_{N^0}
(v_{1,0},\ldots ,v_{N-1,0};\lambda)\over\Lambda_0^2}\right).
\end{equation}
This is the equation for $\HQ$ at order $1/N$. A highly non-trivial check
is that it is actually $\Lambda _0$-independent. A crucial ingredient for
this to be possible is that only the combination $I_{\varphi} -
I_{\alpha}$ appears, so that most of the UV divergences in $I_{\varphi}$ and
$I_{\alpha}$ cancel each other (see
(\ref{ipmia}) for an explicit formula). 
Moreover, by using (\ref{lambdaren}) and (\ref{hren}), it is
straightforward to show that up to terms of order $1/N^2$,
\begin{equation}
\label{betaren}
\tilde\alpha_{N^0}(v_{1,0},\ldots ,v_{N-1,0};\lambda ) =
\tilde\alpha_{N^0}(v_{1},\ldots ,v_{N-1};\Lambda ) + 
{V\over N}\,\left( \ln{\Lambda_0^2\over\Lambda^2} - 4\ln\ln
{\Lambda_0\over\Lambda} + c_1 - c_2\right).
\end{equation}
This turns out to be exactly what is required to cancel the divergences
in (\ref{Heqapp1}). In terms of finite quantities only, the equation for
$\HQ$ can then be written
\begin{equation}
\label{Heq2}
\tilde\alpha_{N^0} = {V\over N}\,\biggl( I -4\ln 2 + 3\ln
{\tilde\alpha_{N^0}\over\Lambda^2} + c_2 - c_1\biggr).
\end{equation}
The finite integral $I(v_1,\ldots v_{N-1})$ is defined by (\ref{Idef}). 
It remains to enforce a renormalization condition in order to determine
$c_2 - c_1$. The study of this seemingly secondary point is actually
important, because the correction in (\ref{Heq2}) are small
only if we can choose $c_2-c_1$ to be much smaller than $N$. 
In particular, the naive guess that $c_2-c_1$ can be an arbitrary
finite constant independent of $N$ is not correct.
Our renormalization condition will be written at $v_1=\cdots
=v_{N-1}=v$, values for which $I=0$ and $\tilde\alpha_{N^0} = -v+\Lambda^2$.
We impose that
\begin{equation}
\label{rencond}
v = \Lambda^2 \biggl(1- {\Delta\over N}\biggr),
\end{equation}
$\Delta$ being an $N$-independent, strictly positive, constant. It is
impossible to choose $\Delta =0$, due to the IR divergences, very much like
it is impossible to use renormalization conditions at
zero momentum in massless theories. Equation (\ref{rencond}) implies
\begin{equation}
\label{delta}
c_2 - c_1 = 3\ln N + \Delta - 3\ln\Delta + 4\ln 2, 
\end{equation}
and the final equation for $\HQ$ is
\begin{equation}
\label{Heq}
\tilde\alpha _{N^0} = {V\over N}\,\biggl( 3\ln N + I+\Delta
+ 3\ln {\tilde\alpha_{N^0}\over\Delta\Lambda^2}\biggr).
\end{equation}
We recall that $\tilde\alpha _{N^0}$ is the unique solution to 
\begin{equation}
\label{beq2}
\sum_{i=1}^{N-1} \ln {\tilde\alpha_{N^0} + v_{i}\over\Lambda^2} =0,
\end{equation}
$I$ is defined by (\ref{Idef}), and $\Delta >0$ is a renormalization
constant independent of $N$. We thus see that the IR instability is
responsible for a correction of order $(\ln N)/N$, larger than the expected
$1/N$ but much smaller than the leading $N^0$ term, as required.
\subsection{Formulas for $I_{\varphi}$, $I_{\alpha}$ and 
$I_{\varphi} - I_{\alpha}$}
In view of (B.14), it is natural to introduce the variables
\begin{equation}
\label{xdef}
x_{i}(p^{2})= {\sqrt{1+ 4v_{i}/p^{2}} +1\over\sqrt{1+ 4v_{i}/p^{2}} -1}
\cdotp
\end{equation}
This set of variables is particularly well suited to study the integrals
$I_{\varphi}$ and $I_{\alpha}$ when the $v_{i}$s are equal or nearly 
equal. The UV behaviour is also easily studied in this representation.
One has
\begin{equation}
{x_{j}-1\over x_{j}+1} = {x_{i}-1\over x_{i}+1} g_{ji}(x_{i})
\end{equation}
with
\begin{equation}
g_{ji}(x_{i}) = 1\Biggm/ \sqrt{1+{4(v_{j}-v_{i})\over v_{i}}\,
{x_{i}\over (x_{i}+1)^{2}}}\ \raise 2pt\hbox{,}
\end{equation}
so that
\begin{equation}
g_{ij}(x_{k})g_{ji}(x_{i})=g_{jk}(x_{k}).
\end{equation}
We have the explicit formulas
\begin{equation}
\label{Iaex}
I_{\alpha} = -\sum_{i=1}^{N-1}\int_{1}^{x_{i}(\Lambda_{0}^{2})}\! 
dx_{i}\, {\displaystyle {(x_{i}-1)^{3}\over x_{i}^{2}(x_{i}+1)} +
{2\over x_{i}}\, \left({x_{i}-1\over x_{i}+1}\right)^{2}\, \ln x_{i}
\over\displaystyle\sum_{j=1}^{N-1} {x_{j}(x_{i})-1\over x_{j}(x_{i}) + 
1} \, \ln x_{j}(x_{i})}\ \raise 2pt\hbox{,}
\end{equation}
\begin{equation}
\label{Ipex}
I_{\varphi}=-\sum_{i=1}^{N-1}\int_{1}^{x_{i}(\Lambda_{0}^{2})}\!
dx_{i}\, \left[ {\displaystyle {x_{i}^{2}-1\over x_{i}^{2}}\over
\displaystyle\sum_{j=1}^{N-1} {x_{j}(x_{i})-1\over x_{j}(x_{i}) + 
1} \, \ln x_{j}(x_{i})} - {2\over N-1}\, {x_{i}+1\over x_{i}(x_{i}-1)}
\right]\cdotp
\end{equation}
$I_{\varphi}$ and $I_{\alpha}$ have separately untamable UV 
divergencies. These divergencies can be studied for example 
for $v_{1}=\cdots =v_{N-1}$. In that case one has
\begin{equation}
\label{Ipvvv}
I_{\varphi}(v_{1}=\cdots =v_{N-1}=v)= 2\ln {\Lambda_{0}^{2}\over v} -
2\ln\ln {\Lambda_{0}^{2}\over v} + 2\gamma - \li x(\Lambda_{0}^{2}),
\end{equation}
with 
\begin{equation}
x(\Lambda_{0}^{2}) = {\Lambda _{0}^{2}\over v}\,\left(
1 + {v\over\Lambda_{0}^{2}} + {\cal 
O}({v^{2}\over\Lambda_{0}^{4}})\right)
\end{equation}
and $\li$ is the logarithmic integral, whose expansion is
\begin{equation}
\li x ={\rm P}\int_{0}^{x}{dx\over\ln x}=
 \gamma + \ln\ln x +\sum_{k=1}^{\infty} {(\ln x)^{k}\over k!\,k}
\quad {\rm for}\quad x>1.
\end{equation}
On the other hand, $I_{\varphi}-I_{\alpha}$ has simple UV 
divergencies that are renormalizable with a finite number of 
counterterms. The general formula is
\begin{equation}
\label{ipmia}
I_{\varphi}-I_{\alpha} = I -4 \ln\ln {\Lambda_{0}\over\Lambda} + 4
\ln {\Lambda_{0}^{2}\over\Lambda^{2}} -4 \ln 2
+{4\over N-1}\,\sum_{i=1}^{N-1}\ln {\Lambda^{2}\over v_{i}}
\raise 2pt\hbox{,}
\end{equation}
where $I$ is a finite integral (which vanishes when $v_{1}=\cdots 
=v_{N-1}$),
\begin{eqnarray}
\label{Idef}
I(v_{1},\ldots ,v_{N-1}) &=&\sum_{i=1}^{N-1}\int_{1}^{\infty}\! dx_{i}\,
\Biggl[ {4\over N-1}\, {1\over x_{i}\ln x_{i}} - {2\over N-1}\,
{x_{i}-1\over x_{i}(x_{i}+1)} - \nonumber\\
&&\qquad\qquad\qquad\qquad {\displaystyle
{4(x_{i}-1)\over x_{i}(x_{i}+1)}\, \left( 1 - {1\over 2}\, 
{x_{i}-1\over x_{i}+1}\, \ln x_{i}\right)\over\displaystyle
\sum_{j=1}^{N-1} {x_{j}(x_{i})-1\over x_{j}(x_{i}) + 1}\,\ln 
x_{j}(x_{i})}\Biggr] \cdotp\\ \nonumber
\end{eqnarray}
On the critical hypersurface $\HQ$, the last term in (\ref{ipmia}) is 
of order $1/N$ and can be neglected.

\end{document}